%% file: main.tex
\RequirePackage{amsmath}

\documentclass{jfphack} 

\usepackage{amsmath}
\usepackage{url}
\usepackage{hyperref} 
\usepackage{amssymb}
\usepackage{longtable}
\usepackage{tikz}
\usetikzlibrary{cd, automata, arrows, positioning, backgrounds}

\usepackage{subcaption}
\usepackage{natbib}

\usepackage[utf8]{inputenc}

\input{macros.TeX}

\input{PIKPalette}

\hypersetup{
  pdfauthor    = {Nuria Brede, Nicola Botta},
  pdftitle     = {Brede, Botta -- On the Correctness of Monadic Backward Induction},
  colorlinks   = true,             
  urlcolor     = PIKorangeDark3,   
  linkcolor    = PIKcyanDark3,     
  citecolor    = PIKgreenDark3     
}

\makeatletter
\@ifundefined{lhs2tex.lhs2tex.sty.read}%
  {\@namedef{lhs2tex.lhs2tex.sty.read}{}%
   \newcommand\SkipToFmtEnd{}%
   \newcommand\EndFmtInput{}%
   \long\def\SkipToFmtEnd#1\EndFmtInput{}%
  }\SkipToFmtEnd

\newcommand\ReadOnlyOnce[1]{\@ifundefined{#1}{\@namedef{#1}{}}\SkipToFmtEnd}
\usepackage{amstext}
\usepackage{amssymb}
\usepackage{stmaryrd}
\DeclareFontFamily{OT1}{cmtex}{}
\DeclareFontShape{OT1}{cmtex}{m}{n}
  {<5><6><7><8>cmtex8
   <9>cmtex9
   <10><10.95><12><14.4><17.28><20.74><24.88>cmtex10}{}
\DeclareFontShape{OT1}{cmtex}{m}{it}
  {<-> ssub * cmtt/m/it}{}

\DeclareFontShape{OT1}{cmtt}{bx}{n}
  {<5><6><7><8>cmtt8
   <9>cmbtt9
   <10><10.95><12><14.4><17.28><20.74><24.88>cmbtt10}{}
\DeclareFontShape{OT1}{cmtex}{bx}{n}
  {<-> ssub * cmtt/bx/n}{}

\newcommand{\Conid}[1]{\mathit{#1}}
\newcommand{\Varid}[1]{\mathit{#1}}
\newcommand{\anonymous}{\kern0.06em \vbox{\hrule\@width.5em}}

\renewcommand{\leq}{\leqslant}

\usepackage{polytable}

\@ifundefined{mathindent}%
  {\newdimen\mathindent\mathindent\leftmargini}%
  {}%

\def\resethooks{%
  \global\let\SaveRestoreHook\empty
  \global\let\ColumnHook\empty}
\newcommand*{\savecolumns}[1][default]%
  {\g@addto@macro\SaveRestoreHook{\savecolumns[#1]}}
\newcommand*{\restorecolumns}[1][default]%
  {\g@addto@macro\SaveRestoreHook{\restorecolumns[#1]}}
\newcommand*{\aligncolumn}[2]%
  {\g@addto@macro\ColumnHook{\column{#1}{#2}}}

\resethooks

\newcommand{\onelinecommentchars}{\quad-{}- }
\newcommand{\commentbeginchars}{\enskip\{-}
\newcommand{\commentendchars}{-\}\enskip}

\newcommand{\visiblecomments}{%
  \let\onelinecomment=\onelinecommentchars
  \let\commentbegin=\commentbeginchars
  \let\commentend=\commentendchars}

\newcommand{\invisiblecomments}{%
  \let\onelinecomment=\empty
  \let\commentbegin=\empty
  \let\commentend=\empty}

\visiblecomments

\newlength{\blanklineskip}
\newcommand{\hsindent}[1]{\quad}
\let\hspre\empty
\let\hspost\empty

\EndFmtInput
\makeatother
\def\doubleequals{\mathrel{\unitlength 0.01em
  \begin{picture}(78,40)
    \put(7,34){\line(1,0){25}} \put(45,34){\line(1,0){25}}
    \put(7,14){\line(1,0){25}} \put(45,14){\line(1,0){25}}
  \end{picture}}}
\ReadOnlyOnce{polycode.fmt}%
\makeatletter

\newcommand{\hsnewpar}[1]%
  {{\parskip=0pt\parindent=0pt\par\vskip #1\noindent}}

\newcommand{\hscodestyle}{}

\newcommand{\sethscode}[1]%
  {\expandafter\let\expandafter\hscode\csname #1\endcsname
   \expandafter\let\expandafter\endhscode\csname end#1\endcsname}

  {\par\noindent
   \advance\leftskip\mathindent
   \hscodestyle
   \let\\=\@normalcr
   \let\hspre\(\let\hspost\)%
   \pboxed}%
  {\endpboxed\)%
   \par\noindent
   \ignorespacesafterend}

  {\hsnewpar\abovedisplayskip
   \advance\leftskip\mathindent
   \hscodestyle
   \let\hspre\(\let\hspost\)%
   \pboxed}%
  {\endpboxed%
   \hsnewpar\belowdisplayskip
   \ignorespacesafterend}

  {\hsnewpar\abovedisplayskip
   \advance\leftskip\mathindent
   \hscodestyle
   \let\\=\@normalcr
   \(\pboxed}%
  {\endpboxed\)%
   \hsnewpar\belowdisplayskip
   \ignorespacesafterend}

\newcommand{\plainhs}{\sethscode{plainhscode}}

\plainhs

  {\hsnewpar\abovedisplayskip
   \advance\leftskip\mathindent
   \hscodestyle
   \let\\=\@normalcr
   \(\parray}%
  {\endparray\)%
   \hsnewpar\belowdisplayskip
   \ignorespacesafterend}

  {\parray}{\endparray}

  {\(\parray}{\endparray\)}

\def\codeframewidth{\arrayrulewidth}
\RequirePackage{calc}

  {\parskip=\abovedisplayskip\par\noindent
   \hscodestyle
   \arrayrulewidth=\codeframewidth
   \tabular{@{}|p{\linewidth-2\arraycolsep-2\arrayrulewidth-2pt}|@{}}%
   \hline\framedhslinecorrect\\{-1.5ex}%
   \let\endoflinesave=\\
   \let\\=\@normalcr
   \(\pboxed}%
  {\endpboxed\)%
   \framedhslinecorrect\endoflinesave{.5ex}\hline
   \endtabular
   \parskip=\belowdisplayskip\par\noindent
   \ignorespacesafterend}

\newcommand{\framedhslinecorrect}[2]%
  {#1[#2]}

  {\(\def\column##1##2{}%
   \let\>\undefined\let\<\undefined\let\\\undefined
   \newcommand\>[1][]{}\newcommand\<[1][]{}\newcommand\\[1][]{}%
   \def\fromto##1##2##3{##3}%
   }{\) }%

  {\let\orighscode=\hscode
   \let\origendhscode=\endhscode
   \def\endhscode{\def\hscode{\endgroup\def\@currenvir{hscode}\\}\begingroup}
   \orighscode\def\hscode{\endgroup\def\@currenvir{hscode}}}%
  {\origendhscode
   \global\let\hscode=\orighscode
   \global\let\endhscode=\origendhscode}%

\makeatother
\EndFmtInput

\usepackage{enumitem}
\setlist[itemize,1]{leftmargin=10pt}
\setlist[enumerate,1]{leftmargin=10pt}

\begin{document}

\journaltitle{JFP}
\cpr{The Author(s),}
\doival{10.1017/xxxxx}

\lefttitle{Brede, Botta}
\righttitle{Correctness of Monadic BI}

\totalpg{\pageref{lastpage01}}
\jnlDoiYr{2021}

\title[Correctness of Monadic BI]{On the Correctness of\\ Monadic Backward Induction}

\begin{authgrp}
\author{NURIA BREDE}
\affiliation{University of Potsdam,
             Potsdam, Germany }\vspace{-1em}
\affiliation{Potsdam Institute for Climate Impact Research,
             Potsdam, Germany\\[0.2em]
             (\email{brede@uni-potsdam.de})
            }
\author{NICOLA BOTTA}
\affiliation{Potsdam Institute for Climate Impact Research,
             Potsdam, Germany
           }\vspace{-1em}
\affiliation{Chalmers University of Technology, Göteborg, Sweden\\[0.2em]
             (\email{botta@pik-potsdam.de}) }
\end{authgrp}

\begin{abstract}
  In control theory, to solve a finite-horizon sequential decision
  problem (SDP) commonly means to find a list of decision rules that
  result in an optimal expected total reward (or cost) when taking a
  given number of decision steps. SDPs
  are routinely solved using Bellman's backward induction.
  Textbook authors (e.g. Bertsekas or Puterman) typically give more or less formal
  proofs to show that the backward induction algorithm is correct as
  solution method for deterministic and stochastic SDPs.
  
  Botta, Jansson and Ionescu
  propose a generic framework for finite horizon, \emph{monadic} SDPs
  together with a monadic version of backward induction for
  solving such SDPs. In monadic SDPs, the monad captures a generic
  notion of uncertainty, while a generic measure function aggregates
  rewards.

  In the present paper we define a notion of correctness for monadic SDPs and identify
  three conditions that allow us to prove a correctness result for monadic
  backward induction that is comparable to textbook correctness proofs
  for ordinary backward induction.
  The conditions that we impose are fairly general and can be cast in
  category-theoretical terms using the notion of
  Eilenberg-Moore-algebra.
  They hold in familiar settings like those of
  deterministic or stochastic SDPs but we also give examples in which they fail.
  Our results show that backward induction can safely be employed for a
  broader class of SDPs than usually treated in textbooks. However, they also rule out
  certain instances that were considered admissible in the context of
  \bottaetal's generic framework.
  
  Our development is formalised in Idris as an extension
  of the \bottaetal framework and the sources are available as
  supplementary material.

\end{abstract}


\maketitle
\setlength\mathindent{0.5cm}
\renewcommand{\hscodestyle}{\small\setlength{\belowdisplayskip}{6pt plus 0pt minus 0pt}}


\section{Introduction}
\label{section:introduction}

Backward induction is a method introduced by \cite{bellman1957} that
is routinely used to solve \emph{finite-horizon sequential decision
  problems (SDP)}. Such problems lie
at the core of many applications in economics, logistics,
and computer science \citep{finus+al2003, helm2003,
  heitzig2012, gintis2007, botta+al2013b,
  de_moor1995, de_moor1999}.
Examples include inventory, scheduling and shortest path
problems, but also the search for optimal strategies in
games~\citep{bertsekas1995, diederich01}.

Botta, Jansson and Ionescu (\citeyear{2017_Botta_Jansson_Ionescu})
propose a generic framework for \emph{monadic} finite-horizon SDPs as
generalisation of the deterministic, non-deterministic and stochastic SDPs
treated in control theory textbooks \citep{bertsekas1995,
  puterman2014markov}. This framework allows to
specify such problems and to solve them with a generic version of
backward induction that we will refer to as \emph{monadic backward
  induction}.

The Botta-Jansson-Ionescu-framework, subsequently referred to as
\emph{BJI-framework}, \emph{BJI-theory} or simply \emph{framework},
already includes a verification of monadic
backward induction with respect to a certain underlying \emph{value}
function (see Sec.~\ref{subsection:solution_components}). However,
in the literature on stochastic SDPs this formulation of the function
is itself part of the backward induction algorithm and needs to be
verified against an optimisation criterion, the \emph{expected total
  reward} \citep[Ch.~4.2]{puterman2014markov}.
For stochastic SDPs semi-formal proofs can be found in textbooks -- but
monadic SDPs are substantially more general than the stochastic SDPs
for which these results are established.
This observation raises a number of questions:
\begin{itemize}
\item What exactly should ``correctness'' mean for a solution of
  monadic SDPs?
\item Does monadic backward induction provide
  correct solutions in this sense for monadic SDPs in their full
  generality?
\item And if not, is there a class of monadic SDPs for which
  monadic backward induction does provide provably correct
  solutions?
\end{itemize}
In the present paper we address these questions and make the
following contributions to answering them:
\begin{itemize}
\item We put forward a formal specification that monadic backward
  induction should meet in order
  to be considered ``correct'' as solution method for monadic SDPs.
  This specification uses an optimisation criterion that is
  a generic version of the \emph{expected total reward} of standard control
  theory textbooks.\footnote{Note that in control theory
  backward  
  induction is often referred to as \emph{the dynamic programming
    algorithm} where the term \emph{dynamic programming} is used in
  the original sense of \cite{bellman1957}.} In analogy, we call this
  criterion \emph{measured total reward}.
\item We consider the value function underlying monadic backward
  induction as ``correct'' if it computes the \emph{measured total reward}.
\item If the value function is correct, monadic backward induction can
  be proven to be correct in our sense by extending the result of
  \cite {2017_Botta_Jansson_Ionescu}.
  However, we show that this is not necessarily the case, i.e.
  the value function does not compute the \emph{measured total reward}
  for arbitrary monadic SDPs.
\item We therefore formulate conditions that identify a class of monadic SDPs
  for which the value function and thus monadic backward induction
  can be shown to be correct. The conditions are fairly simple and allow for
  a neat description in category-theoretical terms using the notion of
  Eilenberg-Moore-algebra.
\item We give a formalised proof that monadic backward induction fulfils
  the correctness criterion if the conditions hold. This correctness result can
  be seen as a generic version of correctness results for standard backward induction
like \citep[Prop.~1.3.1]{bertsekas1995} and
\citep[Th.~4.5.1.c]{puterman2014markov}.
\end{itemize}

Our results rule out the application of backward induction to certain monadic SDPs that
were previously considered as admissible in the BJI-framework. Thus, they complement
the verification result of \bottaetal and provide a necessary clarification. 
Still, the new conditions are simple enough to be checked for
non-standard instantiations of the framework. This
allows to broaden the applicability of backward induction to settings
which are not commonly discussed in the literature and to obtain a
formalised proof of correctness with considerably less effort.
It is worth stressing that our conditions can
be useful for anyone interested in applying monadic backward induction in
non-standard situations -- completely independent of the BJI-framework. 

Finally, the value function underlying monadic backward induction is
also interesting in itself. Given the conditions hold, it can be used to compute
the measured total reward efficiently, using a method reminiscent of a \emph{thinning}
algorithm \cite[Ch.~10]{adwh}.

For the reader unfamiliar with SDPs, we provide a brief informal
overview and two simple examples in the next section. We recap the
BJI-framework and its (partial) verification result for monadic
backward induction in Sec.~\ref{section:framework}.
In Sec.~\ref{section:preparation}
we specify correctness for monadic backward induction
and the BJI-value function. We also
show that in the general monadic case the value function does not
necessarily meet the specification. To resolve this problem, we
identify conditions under which the value function does meet the
specification. These conditions are stated and analysed in
Sec.~\ref{section:conditions}. In Sec.~\ref{section:valval} we
prove that, given the conditions hold, the BJI-value function and monadic
backward induction are correct in the sense defined in
Sec.~\ref{section:preparation}. We discuss the conditions from a more
abstract perspective in Sec.~\ref{section:discussion} and 
conclude in Sec.~\ref{section:conclusion}.

Throughout the paper we use Idris as our host language
\citep{JFP:9060502,idrisbook}. We assume some familiarity
with Haskell-like syntax and notions like \emph{functor} and
\emph{monad} as used in functional programming. We tacitly consider
types as logical statements and programs as proofs, justified by the
propositions-as-types correspondence \citep[for an accessible
introduction see][]{DBLP:journals/cacm/Wadler15}.
\paragraph*{Source code.} \hspace{0.1cm}
Our development is
formalised in Idris as an extension of a lightweight version of the
BJI-framework. The proofs are machine-checked and the source code is
available as supplementary material attached to this paper.
The sources of this document have been written in literal Idris and are
available at \citep{IdrisLibsValVal}, together with some example code.
All source files can be type checked with Idris~1.3.2.


\section{Finite-horizon Sequential Decision Problems}
\label{section:SDPs}

In deterministic, non-deterministic and stochastic finite-horizon
SDPs, a decision maker seeks
to control the evolution of a \emph{(dynamical) system} at a finite number of
\emph{decision steps} by selecting certain \emph{controls} in sequence,
one after the other. The controls
available to the decision maker at a given decision step typically
depend on the \emph{state} of the system at that step.

In \emph{deterministic} problems, selecting a control in a state at
decision step \ensuremath{\Varid{t} \mathop{:} \mathbb{N}}, determines a unique next state at decision step
\ensuremath{\Varid{t}\mathbin{+}\mathrm{1}} through a given \emph{transition function}.
In \emph{non-deterministic} problems, the transition function yields a
whole set of \emph{possible} states at the next decision step.
In \emph{stochastic} problems, the transition function yields a
\emph{probability distribution} on states at the next decision step.

The notion of \emph{monadic} problem generalises that of deterministic,
non-deterministic and stochastic problem through a transition
function that yields an \ensuremath{\Conid{M}}-structure of next states where \ensuremath{\Conid{M}} is a
monad.
For example, the identity monad can be applied to model deterministic
systems. Non-deterministic systems can be represented in terms of
transition functions that return lists (or some other representations
of sets) of next states. Stochastic systems can be represented in
terms of probability distribution monads \citep{giry1981,
  DBLP:journals/jfp/ErwigK06, DBLP:journals/scp/AudebaudP09,
  DBLP:journals/tcs/Jacobs11}.
The uncertainty monad, the states, the controls and the next function
define what is often called a \emph{decision process}.

The idea of sequential decision problems is that each single decision
yields a \emph{reward} and these rewards add up to a \emph{total
  reward} over all decision steps. Rewards are often represented by
  values of a numeric type, and added up using the canonical addition.
  If the transition function and thus
the evolution of the system is not deterministic, then the resulting
possible total rewards need to be aggregated to yield a single outcome
value.
In stochastic SDPs, evolving the
underlying stochastic system leads to a probability distribution
on total rewards which is usually aggregated using the familiar
\emph{expected value} measure. The value thus obtained is called the
\emph{expected total reward} \citep[ch.~4.1.2]{puterman2014markov} and
its role is central: It is the quantity that is to be optimised in
an SDP.

In monadic SDPs, the measure is generic, i.e. it is not fixed in advance
but has to be given as part of the specification of a concrete problem. 
Therefore we will generalise the notion of \emph{expected total reward} to
a corresponding notion for monadic SDPs that we call
\emph{measured total reward} in analogy (see Sec.~\ref{section:preparation}).

\emph{Solving a stochastic SDP} consists in \emph{finding a list of rules
  for selecting controls that maximises the expected total reward for
  \ensuremath{\Varid{n}} decision steps when starting at
  decision step \ensuremath{\Varid{t}}}.
Similarly, we define that \emph{solving a monadic SDP} consists in
\emph{finding a list of rules for selecting controls that maximises
  the measured total reward}.
This means that when starting from any initial state at decision step
\ensuremath{\Varid{t}}, following the computed list of rules for selecting controls will
result in a value that is maximal as measure of the sum of rewards
along all possible trajectories rooted in that initial state.

Equivalently, rewards can instead be considered as \emph{costs}
that need to be \emph{minimised}. This dual perspective is taken e.g.
in \citep{bertsekas1995}. In the subsequent sections we will follow
the terminology of the BJI-framework and \citep{puterman2014markov}
and speak of ``rewards'', but our second example below will illustrate
the ``cost'' perspective. 

In mathematical theories of optimal control, the rules for selecting
controls are called \emph{policies}. A \emph{policy} for a decision
step is simply a function that maps each possible state to a
control. As mentioned above, the controls available in a given
state typically depend on that state, thus policies are dependently typed
functions. A list of such policies is called a \emph{policy sequence}.

The central idea underlying backward induction is to compute a globally
optimal solution of a multi-step SDP incrementally by solving
local optimisation problems at each decision step. This is captured
by \emph{Bellman's principle}: \emph{Extending an optimal
  policy sequence with an optimal policy yields again an optimal policy
  sequence}. However, as we will see in Sec.~\ref{subsection:counterEx},
  one has to carefully check whether for a given SDP backward
  induction is indeed applicable. 

Two features are crucial for finite-horizon, monadic SDPs to be
solvable with the BJI-framework that we study in this  
paper: (1) the number of decision steps has to be given explicitly as
input to the backward induction and (2) at each decision step, the
number of possible next states has to be \emph{finite}.
While (2) is a necessary condition for backward induction to be
computable, (1) is a genuine limitation of the BJI-framework: in many SDPs,
for example in a game of tic-tac-toe, the number of decision steps is
bounded but not known a priori.
  
Before we discuss the BJI-framework in the next section, we illustrate
the notion of sequential decision problem with two simple examples,
one in which the purpose is to maximise rewards and one in which the
purpose is to minimise costs. Rewards and costs in these examples are
just natural numbers and are summed up with ordinary addition. The
first example is a non-deterministic SDP. Although it is somewhat
oversimplified, it has the advantage of being tractable for
computations by hand while still being sufficient as basis for
illustrations in
sections~\ref{section:framework}--\ref{section:conditions}. The second
example is a deterministic SDP that stands for the important class of
scheduling SDPs. This problem highlights why dependent types are necessary
to model state and control spaces accurately. As in these simple
examples state and control spaces are finite, the transition functions
can be described by directed graphs. These are given in
Fig.~\ref{fig:examplesGraph}.

\begin{exm}[\emph{A toy climate problem}]
\label{subsection:example1SDPs}

\begin{figure}
    \centering
    \framebox{
    \parbox{\textwidth-0.5cm}{
    \vspace{0.2cm}

    \begin{subfigure}[b]{.35\textwidth-0.55cm}
      \centering

        \includegraphics[height=0.26\textheight]{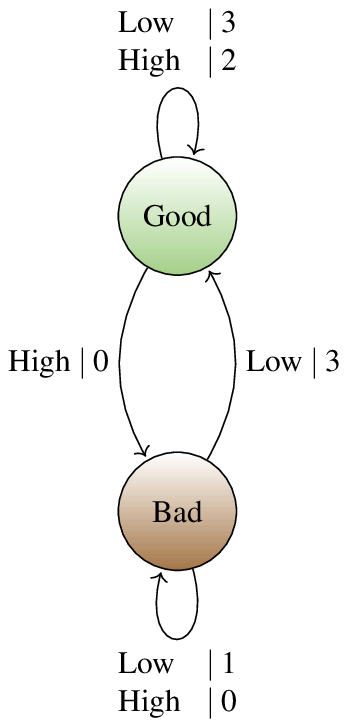}
        

            \caption{Example~1}
      \label{fig:example1}
    \end{subfigure}
    \begin{subfigure}[b]{.65\textwidth}
      \centering

        \includegraphics[height=0.25\textheight]{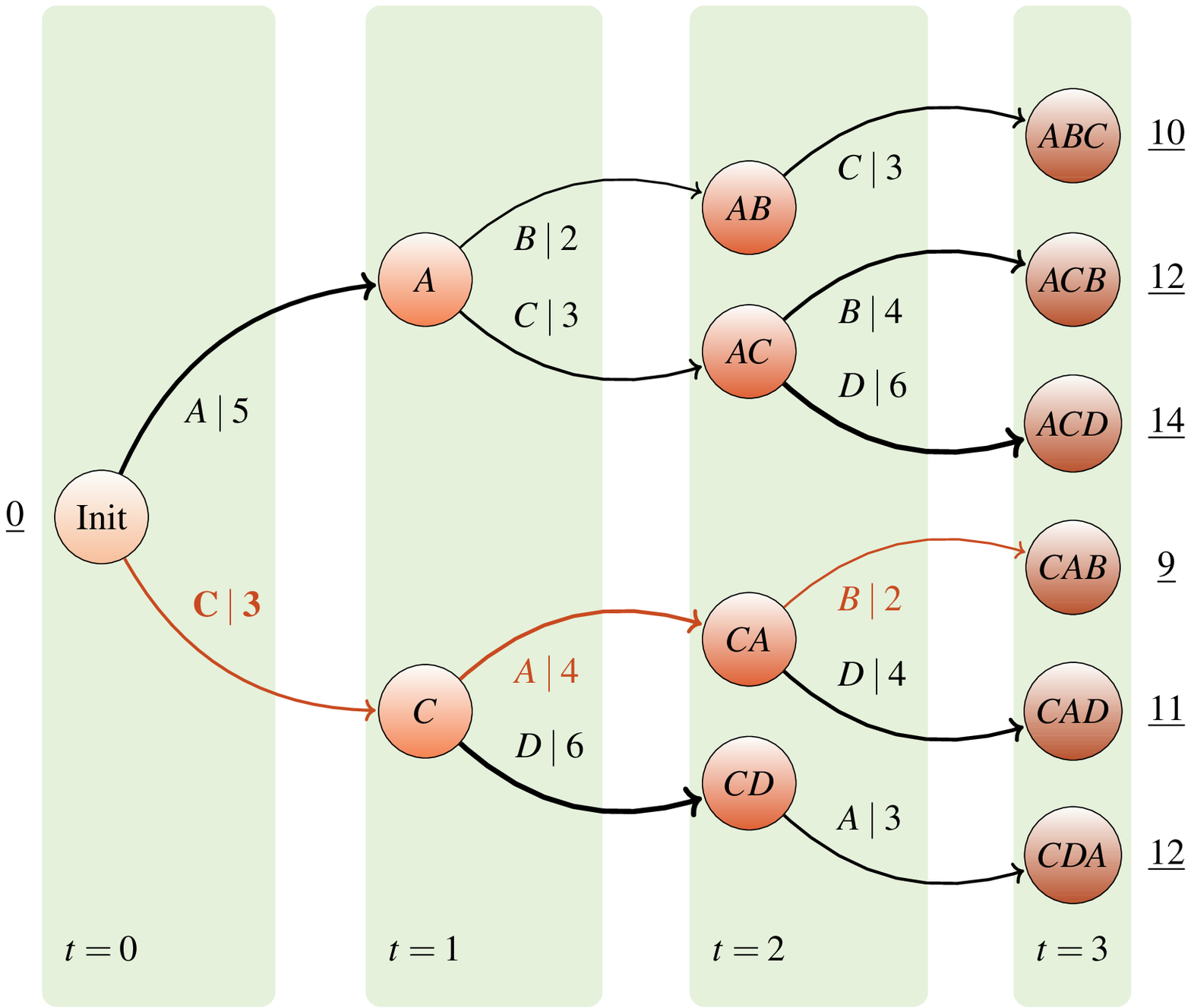}
        

      \caption{Example~2}
      \label{fig:example2}
    \end{subfigure}

     \begin{small}
       \caption{Transition graphs for the example SDPs described in
        Sec.~\ref{section:SDPs}. The edge labels denote pairs
        $\textit{control} \mid  \textit{reward}$ for the associated
        transitions. In the first example, state and control spaces
        are constant over time therefore we have omitted the temporal
        dimension.
        \label{fig:examplesGraph}
      }
      \end{small}
    }
  }

\end{figure}

Our first example is a variant of a
stochastic climate science SDP studied in \citep{esd-9-525-2018},
stripped down to a simple non-deterministic SDP. At every decision
step, the world may be in
one of two \emph{states}, namely \emph{Good} or \emph{Bad}, and the
\emph{controls} determine whether a \emph{Low} or a \emph{High} amount
of green house gases is emitted into the
atmosphere. If the world is
in the \emph{Good} state, choosing \emph{Low} emissions will definitely
keep the world in the \emph{Good} state, but the result of choosing high
emissions is non-deterministic: either the world may stay in the
\emph{Good} or tip to the \emph{Bad} state. Similarly, in the \emph{Bad}
state, \emph{High} emissions will definitely keep the world in
\emph{Bad}, while with \emph{Low} emissions it might either stay in
\emph{Bad} or recover and return to the \emph{Good} state. The
transitions just described define a non-deterministic \emph{transition
  function}. The \emph{rewards}
associated with each transition are determined by the respective control
and reached state. Now we can formulate an SDP: \emph{``Which
policy sequence will maximise the worst-case sum of rewards along all
possible trajectories when taking \ensuremath{\Varid{n}} decisions starting at decision
step \ensuremath{\Varid{t}}?''}. In this simple example the question is not hard to
answer: always choose \emph{Low} emissions, independent of decision step and
state. The \emph{optimal policy sequence} for any \ensuremath{\Varid{n}} and \ensuremath{\Varid{t}} would
thus consist of \ensuremath{\Varid{n}} constant \emph{Low} functions. But in a more
realistic example the
situation will be more involved: every option will have its benefits and
drawbacks encoded in a more complicated reward function, uncertainties
might come with different probabilities, there might be intermediate
states, different combinations of control options etc. For more along
these lines we refer the interested reader to~\citep{esd-9-525-2018}.

\end{exm}

\begin{exm}[\emph{Scheduling}]
\label{subsection:example2SDPs}

Scheduling
problems serve as canonical examples in control theory textbooks. The
one we present here is a slightly modified version of
\citep[Example~1.1.2]{bertsekas1995}.

Think of some machine in a factory that can perform different
operations, say $A, B, C$ and $D$. Each of these operations is supposed
to be performed once. The machine can only perform one operation at a
time, thus an order has to be fixed in which to perform the
operations. Setting the machine up for each operation incurs a specific
cost that might vary according to which operation has been performed
before. Moreover, operation $B$ can only be performed after operation
$A$ has already been completed, and operation $D$ only after operation
$C$. It suffices to fix the order in which the first three operations
are to be performed as this uniquely determines which will be the
fourth task. The aim is now to choose an order that minimises
  the total cost of performing the four operations.

This situation can be modelled as follows as a problem with three
decision steps: The \emph{states at each step} are the sequences of
operations already performed, with the empty sequence at step 0. The
\emph{controls at a decision step and in a state} are the operations which have
not already been performed at previous steps and which are permitted in
that state. For example, at decision step 0, only controls $A$ and $C$
are available because of the above constraint on performing $B$ and
$D$.
The \emph{transition} and \emph{cost} functions of the problem are depicted by
the graph in Fig.~\ref{fig:example2}. As the problem is deterministic,
picking a control will result in a unique next state and each sequence
of policies will result in a unique trajectory. In this setting,
solving the SDP
reduces to finding a control sequence that \emph{minimises the sum of costs
along the single resulting trajectory}. In Fig.~\ref{fig:example2} this
is the sequence $CAB(D)$.

\end{exm}

\section{The BJI-framework}
\label{section:framework}

The BJI-framework is a dependently typed formalisation of optimal
control theory for finite-horizon, discrete-time SDPs. It extends
mathematical formulations for stochastic
SDPs \citep{bertsekas1995, bertsekasShreve96, puterman2014markov}
to the general problem of optimal decision making under \emph{monadic}
uncertainty.

For monadic SDPs, the framework provides a
generic implementation of backward induction. It has been applied to
study the impact of uncertainties on optimal emission policies
\citep{esd-9-525-2018} and is currently used to investigate solar
radiation management problems under tipping point
uncertainty \citep{TiPES::Website}.

In a nutshell, the framework consists of two sets of components: one
for the \emph{specification} of an SDP and one for its \emph{solution}
with monadic backward induction.

\subsection{Problem specification components}
\label{subsection:specification_components}

In the following we discuss the components necessary to specify a
monadic SDP.\\
The first one is the monad \ensuremath{\Conid{M}}:
\begin{hscode}\SaveRestoreHook
\column{B}{@{}>{\hspre}l<{\hspost}@{}}%
\column{3}{@{}>{\hspre}l<{\hspost}@{}}%
\column{13}{@{}>{\hspre}c<{\hspost}@{}}%
\column{13E}{@{}l@{}}%
\column{16}{@{}>{\hspre}l<{\hspost}@{}}%
\column{E}{@{}>{\hspre}l<{\hspost}@{}}%
\>[3]{}\Conid{M}{}\<[13]%
\>[13]{} \mathop{:} {}\<[13E]%
\>[16]{}\Conid{Type} \to \Conid{Type}{}\<[E]%
\ColumnHook
\end{hscode}\resethooks
As discussed in the previous
section, \ensuremath{\Conid{M}} accounts for the uncertainties that affect the decision
problem. For our first example, we could for instance define \ensuremath{\Conid{M}} to be
\ensuremath{\Conid{List}}. For the second example it suffices to use \ensuremath{\Conid{M}\mathrel{=}\Conid{Id}} as the
problem is deterministic.

Further, the BJI-framework supports the specification of the
\emph{states}, the \emph{controls} and the \emph{transition function} of
an SDP through
\begin{hscode}\SaveRestoreHook
\column{B}{@{}>{\hspre}l<{\hspost}@{}}%
\column{3}{@{}>{\hspre}l<{\hspost}@{}}%
\column{13}{@{}>{\hspre}c<{\hspost}@{}}%
\column{13E}{@{}l@{}}%
\column{16}{@{}>{\hspre}l<{\hspost}@{}}%
\column{E}{@{}>{\hspre}l<{\hspost}@{}}%
\>[3]{}\Conid{X}{}\<[13]%
\>[13]{} \mathop{:} {}\<[13E]%
\>[16]{}(\Varid{t} \mathop{:} \mathbb{N}) \to \Conid{Type}{}\<[E]%
\\
\>[3]{}\Conid{Y}{}\<[13]%
\>[13]{} \mathop{:} {}\<[13E]%
\>[16]{}(\Varid{t} \mathop{:} \mathbb{N}) \to \Conid{X}\;\Varid{t} \to \Conid{Type}{}\<[E]%
\\
\>[3]{}\Varid{next}{}\<[13]%
\>[13]{} \mathop{:} {}\<[13E]%
\>[16]{}(\Varid{t} \mathop{:} \mathbb{N}) \to (\Varid{x} \mathop{:} \Conid{X}\;\Varid{t}) \to \Conid{Y}\;\Varid{t}\;\Varid{x} \to \Conid{M}\;(\Conid{X}\;(\Conid{S}\;\Varid{t})){}\<[E]%
\ColumnHook
\end{hscode}\resethooks
The interpretation is that \ensuremath{\Conid{X}\;\Varid{t}} represents the states at decision step
\ensuremath{\Varid{t}}.\footnote{Note that in Idris, $S$ and $Z$ are the familiar
  constructors of the data type \ensuremath{\mathbb{N}}.} In the first example of
Sec.~\ref{section:SDPs}, there are just two states
(\ensuremath{\Conid{Good}} and \ensuremath{\Conid{Bad}}) such that \ensuremath{\Conid{X}} is a constant family:
\begin{hscode}\SaveRestoreHook
\column{B}{@{}>{\hspre}l<{\hspost}@{}}%
\column{3}{@{}>{\hspre}l<{\hspost}@{}}%
\column{15}{@{}>{\hspre}c<{\hspost}@{}}%
\column{15E}{@{}l@{}}%
\column{18}{@{}>{\hspre}l<{\hspost}@{}}%
\column{E}{@{}>{\hspre}l<{\hspost}@{}}%
\>[3]{}\mathbf{data}\;\Conid{State}{}\<[15]%
\>[15]{}\mathrel{=}{}\<[15E]%
\>[18]{}\Conid{Good}\mid \Conid{Bad}{}\<[E]%
\\
\>[3]{}\Conid{X}\;\Varid{\char95 t}{}\<[15]%
\>[15]{}\mathrel{=}{}\<[15E]%
\>[18]{}\Conid{State}{}\<[E]%
\ColumnHook
\end{hscode}\resethooks
But in the second example the possible states depend on the decision
step \ensuremath{\Varid{t}}. Taking for example step \ensuremath{\Varid{t}\mathrel{=}\mathrm{2}}, we could simply define
\begin{hscode}\SaveRestoreHook
\column{B}{@{}>{\hspre}l<{\hspost}@{}}%
\column{3}{@{}>{\hspre}l<{\hspost}@{}}%
\column{6}{@{}>{\hspre}l<{\hspost}@{}}%
\column{16}{@{}>{\hspre}c<{\hspost}@{}}%
\column{16E}{@{}l@{}}%
\column{19}{@{}>{\hspre}l<{\hspost}@{}}%
\column{E}{@{}>{\hspre}l<{\hspost}@{}}%
\>[3]{}\mathbf{data}\;\Conid{State2}{}\<[16]%
\>[16]{}\mathrel{=}{}\<[16E]%
\>[19]{}\Conid{AB}\mid \Conid{AC}\mid \Conid{CA}\mid \Conid{CD}{}\<[E]%
\\
\>[3]{}\Conid{X}\;{}\<[6]%
\>[6]{}\mathrm{2}{}\<[16]%
\>[16]{}\mathrel{=}{}\<[16E]%
\>[19]{}\Conid{State2}{}\<[E]%
\ColumnHook
\end{hscode}\resethooks

Alternatively, we could employ type dependency in a more systematic way
to express that in Ex.~2 states are admissible sequences of
actions
\begin{hscode}\SaveRestoreHook
\column{B}{@{}>{\hspre}l<{\hspost}@{}}%
\column{3}{@{}>{\hspre}l<{\hspost}@{}}%
\column{E}{@{}>{\hspre}l<{\hspost}@{}}%
\>[3]{}\mathbf{data}\;\Conid{Act}\mathrel{=}\Conid{A}\mid \Conid{B}\mid \Conid{C}\mid \Conid{D}{}\<[E]%
\ColumnHook
\end{hscode}\resethooks
Recall that actions could require that another action was performed before,
no action was to be carried out twice and the problem was limited to
3 steps. These conditions might be captured by a type-valued predicate
\begin{hscode}\SaveRestoreHook
\column{B}{@{}>{\hspre}l<{\hspost}@{}}%
\column{3}{@{}>{\hspre}l<{\hspost}@{}}%
\column{20}{@{}>{\hspre}l<{\hspost}@{}}%
\column{E}{@{}>{\hspre}l<{\hspost}@{}}%
\>[3]{}\Conid{AdmissibleState}{}\<[20]%
\>[20]{} \mathop{:} \{\mskip1.5mu \Varid{t} \mathop{:} \mathbb{N}\mskip1.5mu\} \to \Conid{Vect}\;\Varid{t}\;\Conid{Act} \to \Conid{Type}{}\<[E]%
\ColumnHook
\end{hscode}\resethooks
and the type of states might then be expressed as dependent pair
of a vector of actions and a proof that it is admissible.
\begin{hscode}\SaveRestoreHook
\column{B}{@{}>{\hspre}l<{\hspost}@{}}%
\column{3}{@{}>{\hspre}l<{\hspost}@{}}%
\column{E}{@{}>{\hspre}l<{\hspost}@{}}%
\>[3]{}\Conid{X}\;\Varid{t}\mathrel{=}(\Varid{as} \mathop{:} \Conid{Vect}\;\Varid{t}\;\Conid{Act}  \mathbin{*\!*} \Conid{AdmissibleState}\;\Varid{as}){}\<[E]%
\ColumnHook
\end{hscode}\resethooks
Similarly, \ensuremath{\Conid{Y}\;\Varid{t}\;\Varid{x}} represents the controls available at decision step
\ensuremath{\Varid{t}} and in state \ensuremath{\Varid{x}} and \ensuremath{\Varid{next}\;\Varid{t}\;\Varid{x}\;\Varid{y}} represents the states that can
be obtained by selecting control \ensuremath{\Varid{y}} in state \ensuremath{\Varid{x}} at decision step
\ensuremath{\Varid{t}}. In our first example, the available controls remain constant over
time (\ensuremath{\Conid{High}} or \ensuremath{\Conid{Low}}) like the states, but for the second example,
the type dependency is relevant: e.g. we might define (again at step
\ensuremath{\Varid{t}\mathrel{=}\mathrm{2}})
\begin{hscode}\SaveRestoreHook
\column{B}{@{}>{\hspre}l<{\hspost}@{}}%
\column{3}{@{}>{\hspre}l<{\hspost}@{}}%
\column{6}{@{}>{\hspre}l<{\hspost}@{}}%
\column{9}{@{}>{\hspre}l<{\hspost}@{}}%
\column{17}{@{}>{\hspre}c<{\hspost}@{}}%
\column{17E}{@{}l@{}}%
\column{20}{@{}>{\hspre}l<{\hspost}@{}}%
\column{E}{@{}>{\hspre}l<{\hspost}@{}}%
\>[3]{}\mathbf{data}\;\Conid{CtrlAC}{}\<[17]%
\>[17]{}\mathrel{=}{}\<[17E]%
\>[20]{}\Conid{B}\mid \Conid{D}{}\<[E]%
\\
\>[3]{}\Conid{Y}\;{}\<[6]%
\>[6]{}\mathrm{2}\;{}\<[9]%
\>[9]{}\Conid{AC}{}\<[17]%
\>[17]{}\mathrel{=}{}\<[17E]%
\>[20]{}\Conid{CtrlAC}{}\<[E]%
\ColumnHook
\end{hscode}\resethooks

or more elegantly use dependent pairs to define the type of controls,
using the observation that an action is an
admissible control for some state represented by a vector of actions,
if adding the action to the vector results again in an admissible state :
\begin{hscode}\SaveRestoreHook
\column{B}{@{}>{\hspre}l<{\hspost}@{}}%
\column{3}{@{}>{\hspre}l<{\hspost}@{}}%
\column{22}{@{}>{\hspre}c<{\hspost}@{}}%
\column{22E}{@{}l@{}}%
\column{25}{@{}>{\hspre}l<{\hspost}@{}}%
\column{27}{@{}>{\hspre}c<{\hspost}@{}}%
\column{27E}{@{}l@{}}%
\column{30}{@{}>{\hspre}l<{\hspost}@{}}%
\column{E}{@{}>{\hspre}l<{\hspost}@{}}%
\>[3]{}\Conid{AdmissibleControl}{}\<[22]%
\>[22]{} \mathop{:} {}\<[22E]%
\>[25]{}\{\mskip1.5mu \Varid{t} \mathop{:} \mathbb{N}\mskip1.5mu\} \to \Conid{Vect}\;\Varid{t}\;\Conid{Act} \to \Conid{Act} \to \Conid{Type}{}\<[E]%
\\
\>[3]{}\Conid{AdmissibleControl}\;\Varid{as}\;\Varid{a}{}\<[27]%
\>[27]{}\mathrel{=}{}\<[27E]%
\>[30]{}\Conid{AdmissibleState}\;(\Varid{a}\mathbin{::}\Varid{as}){}\<[E]%
\ColumnHook
\end{hscode}\resethooks
\begin{hscode}\SaveRestoreHook
\column{B}{@{}>{\hspre}l<{\hspost}@{}}%
\column{3}{@{}>{\hspre}l<{\hspost}@{}}%
\column{E}{@{}>{\hspre}l<{\hspost}@{}}%
\>[3]{}\Conid{Y}\;\Varid{t}\;\Varid{x}\mathrel{=}(\Varid{a} \mathop{:} \Conid{Act}  \mathbin{*\!*} \Conid{AdmissibleControl}\;(\Varid{fst}\;\Varid{x})\;\Varid{a}){}\<[E]%
\ColumnHook
\end{hscode}\resethooks

Recall from Sec.~\ref{section:SDPs} that the monad, the states,
the controls and the next function together define a decision process.
In order to fully
specify a decision problem, one also has to define the rewards obtained
at each decision step and the operation that is used to add up rewards.
In the BJI-framework, this is done in terms of
\begin{hscode}\SaveRestoreHook
\column{B}{@{}>{\hspre}l<{\hspost}@{}}%
\column{3}{@{}>{\hspre}l<{\hspost}@{}}%
\column{13}{@{}>{\hspre}c<{\hspost}@{}}%
\column{13E}{@{}l@{}}%
\column{16}{@{}>{\hspre}l<{\hspost}@{}}%
\column{E}{@{}>{\hspre}l<{\hspost}@{}}%
\>[3]{}\Conid{Val}{}\<[13]%
\>[13]{} \mathop{:} {}\<[13E]%
\>[16]{}\Conid{Type}{}\<[E]%
\\
\>[3]{}\Varid{reward}{}\<[13]%
\>[13]{} \mathop{:} {}\<[13E]%
\>[16]{}(\Varid{t} \mathop{:} \mathbb{N}) \to (\Varid{x} \mathop{:} \Conid{X}\;\Varid{t}) \to \Conid{Y}\;\Varid{t}\;\Varid{x} \to \Conid{X}\;(\Conid{S}\;\Varid{t}) \to \Conid{Val}{}\<[E]%
\\
\>[3]{}( \mathbin{\oplus} ){}\<[13]%
\>[13]{} \mathop{:} {}\<[13E]%
\>[16]{}\Conid{Val} \to \Conid{Val} \to \Conid{Val}{}\<[E]%
\ColumnHook
\end{hscode}\resethooks
Here, \ensuremath{\Conid{Val}} is the type of rewards and \ensuremath{\Varid{reward}\;\Varid{t}\;\Varid{x}\;\Varid{y}\;\Varid{x'}} is the reward
obtained by selecting control \ensuremath{\Varid{y}} in state \ensuremath{\Varid{x}} if the next state is
\ensuremath{\Varid{x'}}, an element of the state space at step \ensuremath{\Varid{t}\mathbin{+}\mathrm{1}}.
Note that for deterministic problems it is unnecessary to parameterise
the reward function over the next state as it is unique and can thus be
obtained from the current state and control. But for non-deterministic
problems it is useful to be able to assign rewards depending on the
(uncertain) outcome of a transition.

A few remarks are at place here.

\begin{itemize}
\item In many applications, \ensuremath{\Conid{Val}} is a numerical type and the controls of the
  SDP represent resources (fuel, water, etc.) that come at a cost. In these
cases, the reward function encodes the costs and perhaps also the
benefits associated with a decision step. Often, the latter also depends
both on the current state \ensuremath{\Varid{x}} and on the next state \ensuremath{\Varid{x'}}. The
BJI-framework nicely copes with all these situations.
\item The operation \ensuremath{ \mathbin{\oplus} } determines how rewards are added up. It
  could be a simple arithmetic operation, but it could also be defined
  in terms of problem-specific parameters, e.g.\ discount factors to
  give more weight to current rewards as compared to future rewards.
\item Mapping \ensuremath{\Varid{reward}\;\Varid{t}\;\Varid{x}\;\Varid{y}} onto \ensuremath{\Varid{next}\;\Varid{t}\;\Varid{x}\;\Varid{y}} (remember
that \ensuremath{\Conid{M}} is a monad and thus a functor) yields a value of type \ensuremath{\Conid{M}\;\Conid{Val}}. These are the \emph{possible} rewards obtained by selecting
control \ensuremath{\Varid{y}} in state \ensuremath{\Varid{x}} at decision step \ensuremath{\Varid{t}}.
In mathematical theories of optimal control, the implicit assumption
often is that \ensuremath{\Conid{Val}} is equal to \ensuremath{ \mathbb{R} } and that the \ensuremath{\Conid{M}}-structure is a
probability distribution on real numbers  which can be evaluated with
the \emph{expected value} measure.
However, in many practical applications, measuring uncertainty of rewards in
terms of the expected value is inadequate \citep{mercure2020risk}.
The BJI-framework therefore takes a generic approach and allows the
specification of SDPs in terms of problem-specific measures.
\end{itemize}
As just discussed, in SDPs with uncertainty a measure is required to
aggregate multiple possible rewards. The BJI-framework supports
the specification of the measure by:
\begin{hscode}\SaveRestoreHook
\column{B}{@{}>{\hspre}l<{\hspost}@{}}%
\column{3}{@{}>{\hspre}l<{\hspost}@{}}%
\column{20}{@{}>{\hspre}c<{\hspost}@{}}%
\column{20E}{@{}l@{}}%
\column{23}{@{}>{\hspre}l<{\hspost}@{}}%
\column{E}{@{}>{\hspre}l<{\hspost}@{}}%
\>[3]{}\Varid{meas}{}\<[20]%
\>[20]{} \mathop{:} {}\<[20E]%
\>[23]{}\Conid{M}\;\Conid{Val} \to \Conid{Val}{}\<[E]%
\ColumnHook
\end{hscode}\resethooks
In our first example we could use the minimum of a list as worst-case
measure, while in the second example the measure
would just be the identity (as the problem is deterministic).

Before we get to the solution components of the BJI-framework, one
more ingredient needs to be specified. In the next section we will
formalise a notion of optimality for which it is necessary to be able
to compare elements of \ensuremath{\Conid{Val}}.
The framework allows users to compare \ensuremath{\Conid{Val}}-values in terms of a problem
specific comparison operator:
\begin{hscode}\SaveRestoreHook
\column{B}{@{}>{\hspre}l<{\hspost}@{}}%
\column{3}{@{}>{\hspre}l<{\hspost}@{}}%
\column{20}{@{}>{\hspre}c<{\hspost}@{}}%
\column{20E}{@{}l@{}}%
\column{23}{@{}>{\hspre}l<{\hspost}@{}}%
\column{E}{@{}>{\hspre}l<{\hspost}@{}}%
\>[3]{}( \,\sqsubseteq\, ){}\<[20]%
\>[20]{} \mathop{:} {}\<[20E]%
\>[23]{}\Conid{Val} \to \Conid{Val} \to \Conid{Type}{}\<[E]%
\ColumnHook
\end{hscode}\resethooks
The operator \ensuremath{( \,\sqsubseteq\, )} is required to define a total preorder on \ensuremath{\Conid{Val}}.
In our two examples, we simply have:
\begin{hscode}\SaveRestoreHook
\column{B}{@{}>{\hspre}l<{\hspost}@{}}%
\column{3}{@{}>{\hspre}l<{\hspost}@{}}%
\column{20}{@{}>{\hspre}c<{\hspost}@{}}%
\column{20E}{@{}l@{}}%
\column{23}{@{}>{\hspre}l<{\hspost}@{}}%
\column{E}{@{}>{\hspre}l<{\hspost}@{}}%
\>[3]{}\Conid{Val}{}\<[20]%
\>[20]{}\mathrel{=}{}\<[20E]%
\>[23]{}\mathbb{N}{}\<[E]%
\\
\>[3]{}( \mathbin{\oplus} ){}\<[20]%
\>[20]{}\mathrel{=}{}\<[20E]%
\>[23]{}(\mathbin{+}){}\<[E]%
\\
\>[3]{}( \,\sqsubseteq\, ){}\<[20]%
\>[20]{}\mathrel{=}{}\<[20E]%
\>[23]{}(\leq){}\<[E]%
\ColumnHook
\end{hscode}\resethooks
Three more ingredients are necessary to fully specify a monadic SDP,
but we defer discussing them to when they come
up in the next subsection.
For illustration, a formalisation of Ex.~\ref{subsection:example1SDPs} can be found in
Fig.~\ref{fig:example1Formal}. A formalisation of Ex.~\ref{subsection:example2SDPs} is
included in the supplementary material.




\begin{figure}

\centering
  \framebox{

    \parbox{\textwidth-0.5cm}{

      \scalebox{0.8}{

        \small
        \begin{tabular}{m{7.3cm} m{6.7cm}}

          \multicolumn{2}{l}{ \parbox[t]{13cm}{
          \textbf{Formalisation of Ex.~\ref{subsection:example1SDPs}:}\\

          We use the monoid and preorder structure on \ensuremath{\mathbb{N}}, i.e.
          \ensuremath{\Conid{Val}\mathrel{=}\mathbb{N}}, \ensuremath{( \mathbin{\oplus} )\mathrel{=}(\mathbin{+})}, \ensuremath{\Varid{zero}\mathrel{=}\mathrm{0}}, \ensuremath{(\leq)\mathrel{=}( \,\sqsubseteq\, )}.

          }
          }
          \\

          \parbox{6.9cm}{
\begin{hscode}\SaveRestoreHook
\column{B}{@{}>{\hspre}l<{\hspost}@{}}%
\column{3}{@{}>{\hspre}l<{\hspost}@{}}%
\column{E}{@{}>{\hspre}l<{\hspost}@{}}%
\>[3]{}\Conid{M}\mathrel{=}\Conid{List}{}\<[E]%
\ColumnHook
\end{hscode}\resethooks
Measure:
\begin{hscode}\SaveRestoreHook
\column{B}{@{}>{\hspre}l<{\hspost}@{}}%
\column{3}{@{}>{\hspre}l<{\hspost}@{}}%
\column{12}{@{}>{\hspre}c<{\hspost}@{}}%
\column{12E}{@{}l@{}}%
\column{15}{@{}>{\hspre}l<{\hspost}@{}}%
\column{22}{@{}>{\hspre}c<{\hspost}@{}}%
\column{22E}{@{}l@{}}%
\column{25}{@{}>{\hspre}l<{\hspost}@{}}%
\column{E}{@{}>{\hspre}l<{\hspost}@{}}%
\>[3]{}\Varid{minList}{}\<[12]%
\>[12]{} \mathop{:} {}\<[12E]%
\>[15]{}\Conid{List}\;\mathbb{N} \to \mathbb{N}{}\<[E]%
\\
\>[3]{}\Varid{minList}\;[\mskip1.5mu \mskip1.5mu]{}\<[22]%
\>[22]{}\mathrel{=}{}\<[22E]%
\>[25]{}\mathrm{0}{}\<[E]%
\\
\>[3]{}\Varid{minList}\;(\Varid{x}\mathbin{::}[\mskip1.5mu \mskip1.5mu]){}\<[22]%
\>[22]{}\mathrel{=}{}\<[22E]%
\>[25]{}\Varid{x}{}\<[E]%
\\
\>[3]{}\Varid{minList}\;(\Varid{x}\mathbin{::}\Varid{xs}){}\<[22]%
\>[22]{}\mathrel{=}{}\<[22E]%
\>[25]{}\Varid{x}\mathbin{`\Varid{minimum}`}\Varid{xs}{}\<[E]%
\\[\blanklineskip]%
\>[3]{}\Varid{meas}\mathrel{=}\Varid{minList}{}\<[E]%
\ColumnHook
\end{hscode}\resethooks
States and Controls:
\begin{hscode}\SaveRestoreHook
\column{B}{@{}>{\hspre}l<{\hspost}@{}}%
\column{3}{@{}>{\hspre}l<{\hspost}@{}}%
\column{12}{@{}>{\hspre}c<{\hspost}@{}}%
\column{12E}{@{}l@{}}%
\column{15}{@{}>{\hspre}l<{\hspost}@{}}%
\column{18}{@{}>{\hspre}c<{\hspost}@{}}%
\column{18E}{@{}l@{}}%
\column{21}{@{}>{\hspre}l<{\hspost}@{}}%
\column{E}{@{}>{\hspre}l<{\hspost}@{}}%
\>[3]{}\mathbf{data}\;\Conid{States}{}\<[18]%
\>[18]{}\mathrel{=}{}\<[18E]%
\>[21]{}\Conid{Good}\mid \Conid{Bad}{}\<[E]%
\\
\>[3]{}\mathbf{data}\;\Conid{Controls}{}\<[18]%
\>[18]{}\mathrel{=}{}\<[18E]%
\>[21]{}\Conid{High}\mid \Conid{Low}{}\<[E]%
\\[\blanklineskip]%
\>[3]{}\Conid{X}\;\Varid{\char95 t}{}\<[12]%
\>[12]{}\mathrel{=}{}\<[12E]%
\>[15]{}\Conid{States}{}\<[E]%
\\
\>[3]{}\Conid{Y}\;\Varid{\char95 t}\;\Varid{\char95 x}{}\<[12]%
\>[12]{}\mathrel{=}{}\<[12E]%
\>[15]{}\Conid{Controls}{}\<[E]%
\ColumnHook
\end{hscode}\resethooks
          }
                    &

                      \parbox{6.5cm}{

                      Transition function:
\begin{hscode}\SaveRestoreHook
\column{B}{@{}>{\hspre}l<{\hspost}@{}}%
\column{3}{@{}>{\hspre}l<{\hspost}@{}}%
\column{17}{@{}>{\hspre}l<{\hspost}@{}}%
\column{23}{@{}>{\hspre}c<{\hspost}@{}}%
\column{23E}{@{}l@{}}%
\column{26}{@{}>{\hspre}l<{\hspost}@{}}%
\column{E}{@{}>{\hspre}l<{\hspost}@{}}%
\>[3]{}\Varid{next}\;\Varid{\char95 t}\;\Conid{Good}\;{}\<[17]%
\>[17]{}\Conid{Low}{}\<[23]%
\>[23]{}\mathrel{=}{}\<[23E]%
\>[26]{}[\mskip1.5mu \Conid{Good}\mskip1.5mu]{}\<[E]%
\\
\>[3]{}\Varid{next}\;\Varid{\char95 t}\;\Conid{Bad}\;{}\<[17]%
\>[17]{}\Conid{High}{}\<[23]%
\>[23]{}\mathrel{=}{}\<[23E]%
\>[26]{}[\mskip1.5mu \Conid{Bad}\mskip1.5mu]{}\<[E]%
\\
\>[3]{}\Varid{next}\;\Varid{\char95 t}\;\Varid{\char95 x}\;{}\<[17]%
\>[17]{}\Varid{\char95 y}{}\<[23]%
\>[23]{}\mathrel{=}{}\<[23E]%
\>[26]{}[\mskip1.5mu \Conid{Good},\Conid{Bad}\mskip1.5mu]{}\<[E]%
\ColumnHook
\end{hscode}\resethooks
Rewards:
\begin{hscode}\SaveRestoreHook
\column{B}{@{}>{\hspre}l<{\hspost}@{}}%
\column{3}{@{}>{\hspre}l<{\hspost}@{}}%
\column{22}{@{}>{\hspre}l<{\hspost}@{}}%
\column{28}{@{}>{\hspre}c<{\hspost}@{}}%
\column{28E}{@{}l@{}}%
\column{31}{@{}>{\hspre}l<{\hspost}@{}}%
\column{E}{@{}>{\hspre}l<{\hspost}@{}}%
\>[3]{}\Varid{reward}\;\Varid{\char95 t}\;\Varid{\char95 x}\;\Conid{Low}\;{}\<[22]%
\>[22]{}\Conid{Good}{}\<[28]%
\>[28]{}\mathrel{=}{}\<[28E]%
\>[31]{}\mathrm{3}{}\<[E]%
\\
\>[3]{}\Varid{reward}\;\Varid{\char95 t}\;\Varid{\char95 x}\;\Conid{High}\;{}\<[22]%
\>[22]{}\Conid{Good}{}\<[28]%
\>[28]{}\mathrel{=}{}\<[28E]%
\>[31]{}\mathrm{2}{}\<[E]%
\\
\>[3]{}\Varid{reward}\;\Varid{\char95 t}\;\Varid{\char95 x}\;\Conid{Low}\;{}\<[22]%
\>[22]{}\Conid{Bad}{}\<[28]%
\>[28]{}\mathrel{=}{}\<[28E]%
\>[31]{}\mathrm{1}{}\<[E]%
\\
\>[3]{}\Varid{reward}\;\Varid{\char95 t}\;\Varid{\char95 x}\;\Conid{High}\;{}\<[22]%
\>[22]{}\Conid{Bad}{}\<[28]%
\>[28]{}\mathrel{=}{}\<[28E]%
\>[31]{}\mathrm{0}{}\<[E]%
\ColumnHook
\end{hscode}\resethooks

          }
        \end{tabular}
      }

      \caption{A formalisation of Ex.~\ref{subsection:example1SDPs} from
        Sec.~\ref{section:SDPs}
        \label{fig:example1Formal}.
      }
    }
  }
\end{figure}

\subsection{Problem solution components}
\label{subsection:solution_components}

The second set of components of the BJI-framework is an extension of the
mathematical theory of optimal control for stochastic sequential
decision problems to monadic problems. Here, we provide a summary of the
theory. Motivation and full details can be found in
\citep{2014_Botta_et_al, 2017_Botta_Jansson_Ionescu, esd-9-525-2018}.

The theory formalises the notions of policy (decision
rule) from Sec.~\ref{section:SDPs} as
\begin{hscode}\SaveRestoreHook
\column{B}{@{}>{\hspre}l<{\hspost}@{}}%
\column{3}{@{}>{\hspre}l<{\hspost}@{}}%
\column{13}{@{}>{\hspre}c<{\hspost}@{}}%
\column{13E}{@{}l@{}}%
\column{16}{@{}>{\hspre}l<{\hspost}@{}}%
\column{E}{@{}>{\hspre}l<{\hspost}@{}}%
\>[3]{}\Conid{Policy}{}\<[13]%
\>[13]{} \mathop{:} {}\<[13E]%
\>[16]{}(\Varid{t} \mathop{:} \mathbb{N}) \to \Conid{Type}{}\<[E]%
\\
\>[3]{}\Conid{Policy}\;\Varid{t}{}\<[13]%
\>[13]{}\mathrel{=}{}\<[13E]%
\>[16]{}(\Varid{x} \mathop{:} \Conid{X}\;\Varid{t}) \to \Conid{Y}\;\Varid{t}\;\Varid{x}{}\<[E]%
\ColumnHook
\end{hscode}\resethooks
Policy sequences are then essentially vectors
of policies\footnote{The curly brackets in the types of \ensuremath{\Conid{Nil}} and
  \ensuremath{(\mathbin{::})} indicate that \ensuremath{\Varid{t}} and \ensuremath{\Varid{n}} are implicit arguments.}.
\begin{hscode}\SaveRestoreHook
\column{B}{@{}>{\hspre}l<{\hspost}@{}}%
\column{3}{@{}>{\hspre}l<{\hspost}@{}}%
\column{5}{@{}>{\hspre}l<{\hspost}@{}}%
\column{11}{@{}>{\hspre}c<{\hspost}@{}}%
\column{11E}{@{}l@{}}%
\column{14}{@{}>{\hspre}l<{\hspost}@{}}%
\column{19}{@{}>{\hspre}c<{\hspost}@{}}%
\column{19E}{@{}l@{}}%
\column{22}{@{}>{\hspre}l<{\hspost}@{}}%
\column{E}{@{}>{\hspre}l<{\hspost}@{}}%
\>[3]{}\mathbf{data}\;\Conid{PolicySeq}{}\<[19]%
\>[19]{} \mathop{:} {}\<[19E]%
\>[22]{}(\Varid{t},\Varid{n} \mathop{:} \mathbb{N}) \to \Conid{Type}\;\mathbf{where}{}\<[E]%
\\
\>[3]{}\hsindent{2}{}\<[5]%
\>[5]{}\Conid{Nil}{}\<[11]%
\>[11]{} \mathop{:} {}\<[11E]%
\>[14]{}\{\mskip1.5mu \Varid{t} \mathop{:} \mathbb{N}\mskip1.5mu\} \to \Conid{PolicySeq}\;\Varid{t}\;\Conid{Z}{}\<[E]%
\\
\>[3]{}\hsindent{2}{}\<[5]%
\>[5]{}(\mathbin{::}){}\<[11]%
\>[11]{} \mathop{:} {}\<[11E]%
\>[14]{}\{\mskip1.5mu \Varid{t},\Varid{n} \mathop{:} \mathbb{N}\mskip1.5mu\} \to \Conid{Policy}\;\Varid{t} \to \Conid{PolicySeq}\;(\Conid{S}\;\Varid{t})\;\Varid{n} \to \Conid{PolicySeq}\;\Varid{t}\;(\Conid{S}\;\Varid{n}){}\<[E]%
\ColumnHook
\end{hscode}\resethooks
Notice the role of the step (time) index \ensuremath{\Varid{t}} and of the length index \ensuremath{\Varid{n}}
in the constructors of policy sequences:
For a policy sequence to make sense, policies for taking decisions at
step \ensuremath{\Varid{t}} can only be prepended to policy sequences for taking \emph{first}
decisions at step \ensuremath{\Varid{t}\mathbin{+}\mathrm{1}} and the operation yields
policy sequences for taking \emph{first} decisions at step \ensuremath{\Varid{t}}.
Thus the time index allows to ensure a consistency property of policy
sequences by construction.
As for plain vectors and lists, prepending a policy to a policy sequence of
length \ensuremath{\Varid{n}} yields a policy sequence of length \ensuremath{\Varid{n}\mathbin{+}\mathrm{1}}.
Both the time and the
length index will be useful below: they allow to express that the backward
induction algorithm computes policy sequences starting at a specific time
and having a specific length depending on its inputs.   

The perhaps most important ingredient of backward induction is a
\emph{value function} that incrementally measures and adds up rewards.
For a given decision problem,
the value function takes two arguments: a policy sequence \ensuremath{\Varid{ps}} for making
\ensuremath{\Varid{n}} decision steps starting from decision step \ensuremath{\Varid{t}} and an initial state
\ensuremath{\Varid{x} \mathop{:} \Conid{X}\;\Varid{t}}. It computes the value of taking \ensuremath{\Varid{n}} decision steps
according to the policies \ensuremath{\Varid{ps}} when starting in \ensuremath{\Varid{x}}. In the
BJI-framework, the value function is defined as

\begin{hscode}\SaveRestoreHook
\column{B}{@{}>{\hspre}l<{\hspost}@{}}%
\column{3}{@{}>{\hspre}l<{\hspost}@{}}%
\column{8}{@{}>{\hspre}c<{\hspost}@{}}%
\column{8E}{@{}l@{}}%
\column{11}{@{}>{\hspre}l<{\hspost}@{}}%
\column{12}{@{}>{\hspre}l<{\hspost}@{}}%
\column{21}{@{}>{\hspre}l<{\hspost}@{}}%
\column{24}{@{}>{\hspre}c<{\hspost}@{}}%
\column{24E}{@{}l@{}}%
\column{27}{@{}>{\hspre}l<{\hspost}@{}}%
\column{36}{@{}>{\hspre}c<{\hspost}@{}}%
\column{36E}{@{}l@{}}%
\column{39}{@{}>{\hspre}l<{\hspost}@{}}%
\column{E}{@{}>{\hspre}l<{\hspost}@{}}%
\>[3]{}\Varid{val}{}\<[8]%
\>[8]{} \mathop{:} {}\<[8E]%
\>[11]{}\{\mskip1.5mu \Varid{t},\Varid{n} \mathop{:} \mathbb{N}\mskip1.5mu\} \to \Conid{PolicySeq}\;\Varid{t}\;\Varid{n} \to \Conid{X}\;\Varid{t} \to \Conid{Val}{}\<[E]%
\\
\>[3]{}\Varid{val}\;\{\mskip1.5mu \Varid{t}\mskip1.5mu\}\;{}\<[12]%
\>[12]{}\Conid{Nil}\;{}\<[21]%
\>[21]{}\Varid{x}{}\<[24]%
\>[24]{}\mathrel{=}{}\<[24E]%
\>[27]{}\Varid{zero}{}\<[E]%
\\
\>[3]{}\Varid{val}\;\{\mskip1.5mu \Varid{t}\mskip1.5mu\}\;(\Varid{p}\mathbin{::}\Varid{ps})\;\Varid{x}{}\<[24]%
\>[24]{}\mathrel{=}{}\<[24E]%
\>[27]{}\mathbf{let}\;\Varid{y}{}\<[36]%
\>[36]{}\mathrel{=}{}\<[36E]%
\>[39]{}\Varid{p}\;\Varid{x}\;\mathbf{in}{}\<[E]%
\\
\>[27]{}\mathbf{let}\;\Varid{mx'}{}\<[36]%
\>[36]{}\mathrel{=}{}\<[36E]%
\>[39]{}\Varid{next}\;\Varid{t}\;\Varid{x}\;\Varid{y}\;\mathbf{in}{}\<[E]%
\\
\>[27]{}\Varid{meas}\;(\Varid{map}\;(\Varid{reward}\;\Varid{t}\;\Varid{x}\;\Varid{y} \mathbin{\medoplus} \Varid{val}\;\Varid{ps})\;\Varid{mx'}){}\<[E]%
\ColumnHook
\end{hscode}\resethooks
Notice that, independently of the initial state \ensuremath{\Varid{x}}, the value of the
empty policy sequence is \ensuremath{\Varid{zero}}.
This is a problem-specific reference
value
\begin{hscode}\SaveRestoreHook
\column{B}{@{}>{\hspre}l<{\hspost}@{}}%
\column{3}{@{}>{\hspre}l<{\hspost}@{}}%
\column{11}{@{}>{\hspre}c<{\hspost}@{}}%
\column{11E}{@{}l@{}}%
\column{14}{@{}>{\hspre}l<{\hspost}@{}}%
\column{E}{@{}>{\hspre}l<{\hspost}@{}}%
\>[3]{}\Varid{zero}{}\<[11]%
\>[11]{} \mathop{:} {}\<[11E]%
\>[14]{}\Conid{Val}{}\<[E]%
\ColumnHook
\end{hscode}\resethooks
that has to be provided as part of a problem
specification.\footnote{The name might
  suggest that \ensuremath{\Varid{zero}} is supposed to be a neutral element relative to
  \ensuremath{ \mathbin{\oplus} }. However, this is not required by the framework.} It is one of
the specification components that we have not
discussed in Sec.~\ref{subsection:specification_components}.
The value of a policy sequence consisting of a first policy \ensuremath{\Varid{p}} and of a
tail policy sequence \ensuremath{\Varid{ps}} is defined inductively as the measure of an
\ensuremath{\Conid{M}}-structure of \ensuremath{\Conid{Val}}-values. These values are obtained by first
computing the control \ensuremath{\Varid{y}} dictated by the policy \ensuremath{\Varid{p}} in state \ensuremath{\Varid{x}} and
the \ensuremath{\Conid{M}}-structure of possible next states \ensuremath{\Varid{mx'}} dictated by the
transition function \ensuremath{\Varid{next}}. Then, for all \ensuremath{\Varid{x'}} in \ensuremath{\Varid{mx'}} the current reward 
\ensuremath{\Varid{reward}\;\Varid{t}\;\Varid{x}\;\Varid{y}\;\Varid{x'}} is added to the aggregated outcome for the tail policy sequence
\ensuremath{\Varid{val}\;\Varid{ps}\;\Varid{x'}} . Finally, the result of
this functorial mapping is aggregated with the problem-specific
measure \ensuremath{\Varid{meas}} to obtain a result of type \ensuremath{\Conid{Val}} as outcome
for the policy sequence \ensuremath{(\Varid{p}\mathbin{::}\Varid{ps})}. The function which is
mapped onto \ensuremath{\Varid{mx'}} is just a lifted version of \ensuremath{ \mathbin{\oplus} }:
\begin{hscode}\SaveRestoreHook
\column{B}{@{}>{\hspre}l<{\hspost}@{}}%
\column{3}{@{}>{\hspre}l<{\hspost}@{}}%
\column{11}{@{}>{\hspre}c<{\hspost}@{}}%
\column{11E}{@{}l@{}}%
\column{14}{@{}>{\hspre}l<{\hspost}@{}}%
\column{E}{@{}>{\hspre}l<{\hspost}@{}}%
\>[3]{}( \mathbin{\medoplus} ){}\<[11]%
\>[11]{} \mathop{:} {}\<[11E]%
\>[14]{}\{\mskip1.5mu \Conid{A} \mathop{:} \Conid{Type}\mskip1.5mu\} \to (\Varid{f},\Varid{g} \mathop{:} \Conid{A} \to \Conid{Val}) \to \Conid{A} \to \Conid{Val}{}\<[E]%
\\
\>[3]{}\Varid{f} \mathbin{\medoplus} \Varid{g}\mathrel{=}\lambda \Varid{a}\Rightarrow \Varid{f}\;\Varid{a} \mathbin{\oplus} \Varid{g}\;\Varid{a}{}\<[E]%
\ColumnHook
\end{hscode}\resethooks
We will come back to the value function of the BJI-theory in
Sec.~\ref{section:preparation} where we will contrast it
with a function \ensuremath{\Varid{val'}} that, for a policy sequence \ensuremath{\Varid{ps}} and an initial state
\ensuremath{\Varid{x}}, computes the measure of the sum of the
rewards along all possible trajectories starting at \ensuremath{\Varid{x}} under
\ensuremath{\Varid{ps}} (the \emph{measured total reward} that we anticipated in
Sec.~\ref{section:SDPs}).
For the time being, though, we accept the notion of value of a policy
sequence as put forward in the BJI-theory and we show how the
definition of \ensuremath{\Varid{val}} can be employed to compute policy sequences that
are provably optimal in the sense of
\begin{hscode}\SaveRestoreHook
\column{B}{@{}>{\hspre}l<{\hspost}@{}}%
\column{3}{@{}>{\hspre}l<{\hspost}@{}}%
\column{17}{@{}>{\hspre}c<{\hspost}@{}}%
\column{17E}{@{}l@{}}%
\column{20}{@{}>{\hspre}l<{\hspost}@{}}%
\column{28}{@{}>{\hspre}c<{\hspost}@{}}%
\column{28E}{@{}l@{}}%
\column{31}{@{}>{\hspre}l<{\hspost}@{}}%
\column{E}{@{}>{\hspre}l<{\hspost}@{}}%
\>[3]{}\Conid{OptPolicySeq}{}\<[17]%
\>[17]{} \mathop{:} {}\<[17E]%
\>[20]{}\{\mskip1.5mu \Varid{t},\Varid{n} \mathop{:} \mathbb{N}\mskip1.5mu\} \to \Conid{PolicySeq}\;\Varid{t}\;\Varid{n} \to \Conid{Type}{}\<[E]%
\\
\>[3]{}\Conid{OptPolicySeq}\;\{\mskip1.5mu \Varid{t}\mskip1.5mu\}\;\{\mskip1.5mu \Varid{n}\mskip1.5mu\}\;\Varid{ps}{}\<[28]%
\>[28]{}\mathrel{=}{}\<[28E]%
\>[31]{}(\Varid{ps'} \mathop{:} \Conid{PolicySeq}\;\Varid{t}\;\Varid{n}) \to (\Varid{x} \mathop{:} \Conid{X}\;\Varid{t}) \to \Varid{val}\;\Varid{ps'}\;\Varid{x} \,\sqsubseteq\, \Varid{val}\;\Varid{ps}\;\Varid{x}{}\<[E]%
\ColumnHook
\end{hscode}\resethooks
Notice the universal quantification in this definition:
A policy sequence \ensuremath{\Varid{ps}} is defined to be optimal iff \ensuremath{\Varid{val}\;\Varid{ps'}\;\Varid{x} \,\sqsubseteq\, \Varid{val}\;\Varid{ps}\;\Varid{x}} for any policy sequence \ensuremath{\Varid{ps'}} and for any state \ensuremath{\Varid{x}}.

The generic implementation of backward induction in the
BJI-framework is an application of \emph{Bellman's principle of
optimality} mentioned in Sec.~\ref{section:SDPs}.
In control theory textbooks, this principle is often
referred to as \emph{Bellman's equation}. It can be suitably
formulated in terms of the notion of \emph{optimal extension}. We say
that a policy \ensuremath{\Varid{p} \mathop{:} \Conid{Policy}\;\Varid{t}} is an optimal extension of a policy
sequence \ensuremath{\Varid{ps} \mathop{:} \Conid{Policy}\;(\Conid{S}\;\Varid{t})\;\Varid{n}} if it is
the case that the value of \ensuremath{\Varid{p}\mathbin{::}\Varid{ps}} is at least as good as the value of
\ensuremath{\Varid{p'}\mathbin{::}\Varid{ps}} for any policy \ensuremath{\Varid{p'}} and for any state \ensuremath{\Varid{x} \mathop{:} \Conid{X}\;\Varid{t}}:
\begin{hscode}\SaveRestoreHook
\column{B}{@{}>{\hspre}l<{\hspost}@{}}%
\column{3}{@{}>{\hspre}l<{\hspost}@{}}%
\column{11}{@{}>{\hspre}c<{\hspost}@{}}%
\column{11E}{@{}l@{}}%
\column{14}{@{}>{\hspre}l<{\hspost}@{}}%
\column{20}{@{}>{\hspre}c<{\hspost}@{}}%
\column{20E}{@{}l@{}}%
\column{23}{@{}>{\hspre}l<{\hspost}@{}}%
\column{E}{@{}>{\hspre}l<{\hspost}@{}}%
\>[3]{}\Conid{OptExt}{}\<[11]%
\>[11]{} \mathop{:} {}\<[11E]%
\>[14]{}\{\mskip1.5mu \Varid{t},\Varid{n} \mathop{:} \mathbb{N}\mskip1.5mu\} \to \Conid{PolicySeq}\;(\Conid{S}\;\Varid{t})\;\Varid{n} \to \Conid{Policy}\;\Varid{t} \to \Conid{Type}{}\<[E]%
\\
\>[3]{}\Conid{OptExt}\;\{\mskip1.5mu \Varid{t}\mskip1.5mu\}\;\Varid{ps}\;\Varid{p}{}\<[20]%
\>[20]{}\mathrel{=}{}\<[20E]%
\>[23]{}(\Varid{p'} \mathop{:} \Conid{Policy}\;\Varid{t}) \to (\Varid{x} \mathop{:} \Conid{X}\;\Varid{t}) \to \Varid{val}\;(\Varid{p'}\mathbin{::}\Varid{ps})\;\Varid{x} \,\sqsubseteq\, \Varid{val}\;(\Varid{p}\mathbin{::}\Varid{ps})\;\Varid{x}{}\<[E]%
\ColumnHook
\end{hscode}\resethooks
With the notion of optimal extension in place, Bellman's principle can
be formulated as
\begin{hscode}\SaveRestoreHook
\column{B}{@{}>{\hspre}l<{\hspost}@{}}%
\column{3}{@{}>{\hspre}l<{\hspost}@{}}%
\column{12}{@{}>{\hspre}c<{\hspost}@{}}%
\column{12E}{@{}l@{}}%
\column{15}{@{}>{\hspre}l<{\hspost}@{}}%
\column{21}{@{}>{\hspre}c<{\hspost}@{}}%
\column{21E}{@{}l@{}}%
\column{24}{@{}>{\hspre}l<{\hspost}@{}}%
\column{E}{@{}>{\hspre}l<{\hspost}@{}}%
\>[3]{}\Conid{Bellman}{}\<[12]%
\>[12]{} \mathop{:} {}\<[12E]%
\>[15]{}\{\mskip1.5mu \Varid{t},\Varid{n} \mathop{:} \mathbb{N}\mskip1.5mu\} \to {}\<[E]%
\\
\>[15]{}(\Varid{ps}{}\<[21]%
\>[21]{} \mathop{:} {}\<[21E]%
\>[24]{}\Conid{PolicySeq}\;(\Conid{S}\;\Varid{t})\;\Varid{n}) \to \Conid{OptPolicySeq}\;\Varid{ps} \to {}\<[E]%
\\
\>[15]{}(\Varid{p}{}\<[21]%
\>[21]{} \mathop{:} {}\<[21E]%
\>[24]{}\Conid{Policy}\;\Varid{t}) \to \Conid{OptExt}\;\Varid{ps}\;\Varid{p} \to {}\<[E]%
\\
\>[15]{}\Conid{OptPolicySeq}\;(\Varid{p}\mathbin{::}\Varid{ps}){}\<[E]%
\ColumnHook
\end{hscode}\resethooks
In words: \emph{extending an optimal policy sequence with an optimal
extension (of that policy sequence) yields an optimal policy sequence}
or shorter \emph{prefixing with optimal extensions preserves
  optimality}.
Proving Bellman's optimality principle is almost straightforward and
relies on \ensuremath{ \,\sqsubseteq\, } being reflexive and transitive and two
\emph{monotonicity} properties:
\begin{hscode}\SaveRestoreHook
\column{B}{@{}>{\hspre}l<{\hspost}@{}}%
\column{3}{@{}>{\hspre}l<{\hspost}@{}}%
\column{17}{@{}>{\hspre}c<{\hspost}@{}}%
\column{17E}{@{}l@{}}%
\column{20}{@{}>{\hspre}l<{\hspost}@{}}%
\column{E}{@{}>{\hspre}l<{\hspost}@{}}%
\>[3]{}\Varid{plusMonSpec}{}\<[17]%
\>[17]{} \mathop{:} {}\<[17E]%
\>[20]{}\{\mskip1.5mu \Varid{v1},\Varid{v2},\Varid{v3},\Varid{v4} \mathop{:} \Conid{Val}\mskip1.5mu\} \to \Varid{v1} \,\sqsubseteq\, \Varid{v2} \to \Varid{v3} \,\sqsubseteq\, \Varid{v4} \to (\Varid{v1} \mathbin{\oplus} \Varid{v3}) \,\sqsubseteq\, (\Varid{v2} \mathbin{\oplus} \Varid{v4}){}\<[E]%
\\[\blanklineskip]%
\>[3]{}\Varid{measMonSpec}{}\<[17]%
\>[17]{} \mathop{:} {}\<[17E]%
\>[20]{}\{\mskip1.5mu \Conid{A} \mathop{:} \Conid{Type}\mskip1.5mu\} \to (\Varid{f},\Varid{g} \mathop{:} \Conid{A} \to \Conid{Val}) \to ((\Varid{a} \mathop{:} \Conid{A}) \to \Varid{f}\;\Varid{a} \,\sqsubseteq\, \Varid{g}\;\Varid{a}) \to {}\<[E]%
\\
\>[20]{}(\Varid{ma} \mathop{:} \Conid{M}\;\Conid{A}) \to \Varid{meas}\;(\Varid{map}\;\Varid{f}\;\Varid{ma}) \,\sqsubseteq\, \Varid{meas}\;(\Varid{map}\;\Varid{g}\;\Varid{ma}){}\<[E]%
\ColumnHook
\end{hscode}\resethooks
The second condition is a special case of the measure monotonicity
requirement originally formulated by \cite{ionescu2009} in
the context of a theory of vulnerability and monadic dynamical
systems. It is a natural property and the expected value measure, the
worst (best) case measure and any sound statistical measure fulfil it.
Like the reference value \ensuremath{\Varid{zero}} discussed above, \ensuremath{\Varid{plusMonSpec}} and
\ensuremath{\Varid{measMonSpec}} are specification components of the BJI-framework that
we have not discussed in Sec.~\ref{subsection:specification_components}.
We provide a proof of \ensuremath{\Conid{Bellman}} in Appendix~\ref{appendix:Bellman}. As one
would expect, the proof makes essential use of the recursive definition of
the function \ensuremath{\Varid{val}} discussed above.
As a consequence, this precise definition of \ensuremath{\Varid{val}} plays a crucial
role for the verification of backward induction in
\citep{2017_Botta_Jansson_Ionescu}.

Apart from the increased level of generality, the
definition of \ensuremath{\Varid{val}} and \ensuremath{\Conid{Bellman}} are in fact just an
Idris formalisation of Bellman's equation as formulated in control
theory textbooks. With \ensuremath{\Conid{Bellman}} and provided that we can compute
optimal extensions of arbitrary policy sequences
\begin{hscode}\SaveRestoreHook
\column{B}{@{}>{\hspre}l<{\hspost}@{}}%
\column{3}{@{}>{\hspre}l<{\hspost}@{}}%
\column{15}{@{}>{\hspre}c<{\hspost}@{}}%
\column{15E}{@{}l@{}}%
\column{18}{@{}>{\hspre}l<{\hspost}@{}}%
\column{E}{@{}>{\hspre}l<{\hspost}@{}}%
\>[3]{}\Varid{optExt}{}\<[15]%
\>[15]{} \mathop{:} {}\<[15E]%
\>[18]{}\{\mskip1.5mu \Varid{t},\Varid{n} \mathop{:} \mathbb{N}\mskip1.5mu\} \to \Conid{PolicySeq}\;(\Conid{S}\;\Varid{t})\;\Varid{n} \to \Conid{Policy}\;\Varid{t}{}\<[E]%
\\[\blanklineskip]%
\>[3]{}\Varid{optExtSpec}{}\<[15]%
\>[15]{} \mathop{:} {}\<[15E]%
\>[18]{}\{\mskip1.5mu \Varid{t},\Varid{n} \mathop{:} \mathbb{N}\mskip1.5mu\} \to (\Varid{ps} \mathop{:} \Conid{PolicySeq}\;(\Conid{S}\;\Varid{t})\;\Varid{n}) \to \Conid{OptExt}\;\Varid{ps}\;(\Varid{optExt}\;\Varid{ps}){}\<[E]%
\ColumnHook
\end{hscode}\resethooks
it is easy to derive an implementation of monadic backward
induction that computes provably optimal policy sequences with
respect to \ensuremath{\Varid{val}}: first, notice that the empty policy sequence is
optimal:
\begin{hscode}\SaveRestoreHook
\column{B}{@{}>{\hspre}l<{\hspost}@{}}%
\column{3}{@{}>{\hspre}l<{\hspost}@{}}%
\column{20}{@{}>{\hspre}c<{\hspost}@{}}%
\column{20E}{@{}l@{}}%
\column{23}{@{}>{\hspre}l<{\hspost}@{}}%
\column{E}{@{}>{\hspre}l<{\hspost}@{}}%
\>[3]{}\Varid{nilOptPolicySeq}{}\<[20]%
\>[20]{} \mathop{:} {}\<[20E]%
\>[23]{}\Conid{OptPolicySeq}\;\Conid{Nil}{}\<[E]%
\\
\>[3]{}\Varid{nilOptPolicySeq}\;\Conid{Nil}\;\Varid{x}\mathrel{=}\Varid{reflexive}\;\Varid{lteTP}\;\Varid{zero}{}\<[E]%
\ColumnHook
\end{hscode}\resethooks
This is the base case for constructing optimal policy sequences by
backward induction, starting from the empty policy sequence:
\begin{hscode}\SaveRestoreHook
\column{B}{@{}>{\hspre}l<{\hspost}@{}}%
\column{3}{@{}>{\hspre}l<{\hspost}@{}}%
\column{7}{@{}>{\hspre}c<{\hspost}@{}}%
\column{7E}{@{}l@{}}%
\column{9}{@{}>{\hspre}l<{\hspost}@{}}%
\column{11}{@{}>{\hspre}l<{\hspost}@{}}%
\column{15}{@{}>{\hspre}c<{\hspost}@{}}%
\column{15E}{@{}l@{}}%
\column{18}{@{}>{\hspre}l<{\hspost}@{}}%
\column{E}{@{}>{\hspre}l<{\hspost}@{}}%
\>[3]{}\Varid{bi}{}\<[7]%
\>[7]{} \mathop{:} {}\<[7E]%
\>[11]{}(\Varid{t},\Varid{n} \mathop{:} \mathbb{N}) \to \Conid{PolicySeq}\;\Varid{t}\;\Varid{n}{}\<[E]%
\\
\>[3]{}\Varid{bi}\;\Varid{t}\;{}\<[9]%
\>[9]{}\Conid{Z}{}\<[15]%
\>[15]{}\mathrel{=}{}\<[15E]%
\>[18]{}\Conid{Nil}{}\<[E]%
\\
\>[3]{}\Varid{bi}\;\Varid{t}\;(\Conid{S}\;\Varid{n}){}\<[15]%
\>[15]{}\mathrel{=}{}\<[15E]%
\>[18]{}\mathbf{let}\;\Varid{ps}\mathrel{=}\Varid{bi}\;(\Conid{S}\;\Varid{t})\;\Varid{n}\;\mathbf{in}\;\Varid{optExt}\;\Varid{ps}\mathbin{::}\Varid{ps}{}\<[E]%
\ColumnHook
\end{hscode}\resethooks
That \ensuremath{\Varid{bi}} computes optimal policy sequences with respect to \ensuremath{\Varid{val}}
is then proved by induction on \ensuremath{\Varid{n}}, the input that determines the
length of the resulting policy sequence:
\begin{hscode}\SaveRestoreHook
\column{B}{@{}>{\hspre}l<{\hspost}@{}}%
\column{3}{@{}>{\hspre}l<{\hspost}@{}}%
\column{5}{@{}>{\hspre}l<{\hspost}@{}}%
\column{10}{@{}>{\hspre}c<{\hspost}@{}}%
\column{10E}{@{}l@{}}%
\column{13}{@{}>{\hspre}l<{\hspost}@{}}%
\column{15}{@{}>{\hspre}l<{\hspost}@{}}%
\column{16}{@{}>{\hspre}l<{\hspost}@{}}%
\column{21}{@{}>{\hspre}c<{\hspost}@{}}%
\column{21E}{@{}l@{}}%
\column{24}{@{}>{\hspre}l<{\hspost}@{}}%
\column{32}{@{}>{\hspre}l<{\hspost}@{}}%
\column{36}{@{}>{\hspre}l<{\hspost}@{}}%
\column{41}{@{}>{\hspre}c<{\hspost}@{}}%
\column{41E}{@{}l@{}}%
\column{44}{@{}>{\hspre}l<{\hspost}@{}}%
\column{E}{@{}>{\hspre}l<{\hspost}@{}}%
\>[3]{}\Varid{biOptVal}{}\<[13]%
\>[13]{} \mathop{:} {}\<[16]%
\>[16]{}(\Varid{t},\Varid{n} \mathop{:} \mathbb{N}) \to \Conid{OptPolicySeq}\;(\Varid{bi}\;\Varid{t}\;\Varid{n}){}\<[E]%
\\
\>[3]{}\Varid{biOptVal}\;\Varid{t}\;{}\<[15]%
\>[15]{}\Conid{Z}{}\<[21]%
\>[21]{}\mathrel{=}{}\<[21E]%
\>[24]{}\Varid{nilOptPolicySeq}{}\<[E]%
\\
\>[3]{}\Varid{biOptVal}\;\Varid{t}\;(\Conid{S}\;\Varid{n}){}\<[21]%
\>[21]{}\mathrel{=}{}\<[21E]%
\>[24]{}\Conid{Bellman}\;\Varid{ps}\;\Varid{ops}\;\Varid{p}\;\Varid{oep}\;\mathbf{where}{}\<[E]%
\\
\>[3]{}\hsindent{2}{}\<[5]%
\>[5]{}\Varid{ps}{}\<[10]%
\>[10]{} \mathop{:} {}\<[10E]%
\>[13]{}\Conid{PolicySeq}\;(\Conid{S}\;\Varid{t})\;\Varid{n}{}\<[32]%
\>[32]{}\quad;\quad {}\<[36]%
\>[36]{}\Varid{ps}{}\<[41]%
\>[41]{}\mathrel{=}{}\<[41E]%
\>[44]{}\Varid{bi}\;(\Conid{S}\;\Varid{t})\;\Varid{n}{}\<[E]%
\\
\>[3]{}\hsindent{2}{}\<[5]%
\>[5]{}\Varid{ops}{}\<[10]%
\>[10]{} \mathop{:} {}\<[10E]%
\>[13]{}\Conid{OptPolicySeq}\;\Varid{ps}{}\<[32]%
\>[32]{}\quad;\quad {}\<[36]%
\>[36]{}\Varid{ops}{}\<[41]%
\>[41]{}\mathrel{=}{}\<[41E]%
\>[44]{}\Varid{biOptVal}\;(\Conid{S}\;\Varid{t})\;\Varid{n}{}\<[E]%
\\
\>[3]{}\hsindent{2}{}\<[5]%
\>[5]{}\Varid{p}{}\<[10]%
\>[10]{} \mathop{:} {}\<[10E]%
\>[13]{}\Conid{Policy}\;\Varid{t}{}\<[32]%
\>[32]{}\quad;\quad {}\<[36]%
\>[36]{}\Varid{p}{}\<[41]%
\>[41]{}\mathrel{=}{}\<[41E]%
\>[44]{}\Varid{optExt}\;\Varid{ps}{}\<[E]%
\\
\>[3]{}\hsindent{2}{}\<[5]%
\>[5]{}\Varid{oep}{}\<[10]%
\>[10]{} \mathop{:} {}\<[10E]%
\>[13]{}\Conid{OptExt}\;\Varid{ps}\;\Varid{p}{}\<[32]%
\>[32]{}\quad;\quad {}\<[36]%
\>[36]{}\Varid{oep}{}\<[41]%
\>[41]{}\mathrel{=}{}\<[41E]%
\>[44]{}\Varid{optExtSpec}\;\Varid{ps}{}\<[E]%
\ColumnHook
\end{hscode}\resethooks
This is the verification result for \ensuremath{\Varid{bi}} of
\citep{2017_Botta_Jansson_Ionescu}.\footnote{Note that \ensuremath{\Varid{biOptVal}} is
  called \ensuremath{\Varid{biLemma}} in \citep{2017_Botta_Jansson_Ionescu}. We chose the
  new name to emphasise that \ensuremath{\Varid{bi}} computes optimal policy sequences
  with respect to \ensuremath{\Varid{val}}.}

\subsection{BJI-framework wrap-up.}
\label{subsection:wrap-up}

The specification and solution components discussed in the last two
sections are all we need to formulate precisely the problem of
correctness for monadic backward induction in the BJI-framework.
This is done in the next section.
Before turning to it, two further remarks are necessary:

\begin{itemize}

\item The theory proposed in \citep{2017_Botta_Jansson_Ionescu} is
slightly more general than the one summarised above. Here, policies are
just functions from states to controls:
\begin{hscode}\SaveRestoreHook
\column{B}{@{}>{\hspre}l<{\hspost}@{}}%
\column{3}{@{}>{\hspre}l<{\hspost}@{}}%
\column{11}{@{}>{\hspre}c<{\hspost}@{}}%
\column{11E}{@{}l@{}}%
\column{14}{@{}>{\hspre}l<{\hspost}@{}}%
\column{E}{@{}>{\hspre}l<{\hspost}@{}}%
\>[3]{}\Conid{Policy}{}\<[11]%
\>[11]{} \mathop{:} {}\<[11E]%
\>[14]{}(\Varid{t} \mathop{:} \mathbb{N}) \to \Conid{Type}{}\<[E]%
\\
\>[3]{}\Conid{Policy}\;\Varid{t}\mathrel{=}(\Varid{x} \mathop{:} \Conid{X}\;\Varid{t}) \to \Conid{Y}\;\Varid{t}\;\Varid{x}{}\<[E]%
\ColumnHook
\end{hscode}\resethooks
By contrast, in \citep{2017_Botta_Jansson_Ionescu}, policies are indexed
over a number of decision steps \ensuremath{\Varid{n}}
\begin{hscode}\SaveRestoreHook
\column{B}{@{}>{\hspre}l<{\hspost}@{}}%
\column{3}{@{}>{\hspre}l<{\hspost}@{}}%
\column{11}{@{}>{\hspre}c<{\hspost}@{}}%
\column{11E}{@{}l@{}}%
\column{14}{@{}>{\hspre}l<{\hspost}@{}}%
\column{19}{@{}>{\hspre}c<{\hspost}@{}}%
\column{19E}{@{}l@{}}%
\column{22}{@{}>{\hspre}l<{\hspost}@{}}%
\column{E}{@{}>{\hspre}l<{\hspost}@{}}%
\>[3]{}\Conid{Policy}{}\<[11]%
\>[11]{} \mathop{:} {}\<[11E]%
\>[14]{}(\Varid{t},\Varid{n} \mathop{:} \mathbb{N}) \to \Conid{Type}{}\<[E]%
\\
\>[3]{}\Conid{Policy}\;\Varid{t}\;\Conid{Z}{}\<[19]%
\>[19]{}\mathrel{=}{}\<[19E]%
\>[22]{}\Conid{Unit}{}\<[E]%
\\
\>[3]{}\Conid{Policy}\;\Varid{t}\;(\Conid{S}\;\Varid{m}){}\<[19]%
\>[19]{}\mathrel{=}{}\<[19E]%
\>[22]{}(\Varid{x} \mathop{:} \Conid{X}\;\Varid{t}) \to \Conid{Reachable}\;\Varid{x} \to \Conid{Viable}\;(\Conid{S}\;\Varid{m})\;\Varid{x} \to \Conid{GoodCtrl}\;\Varid{t}\;\Varid{x}\;\Varid{m}{}\<[E]%
\ColumnHook
\end{hscode}\resethooks
and their domain for \ensuremath{\Varid{n}\mathbin{>}\mathrm{0}} is restricted to states that are
\emph{reachable} and \emph{viable} for \ensuremath{\Varid{n}} steps. This allows
to cope with states whose control set is empty and with transition
functions that return empty \ensuremath{\Conid{M}}-structures of next states.
(For a discussion of reachability and viability
see \citep[Sec.~3.7~and~3.8]{2017_Botta_Jansson_Ionescu}.)

This generality, however, comes at a cost: Compare e.g.\
the proof of Bellman's principle from the last
subsection with the corresponding proof in
\cite[Appendix B]{2017_Botta_Jansson_Ionescu}. The impact of the
reachability and viability
constraints on other parts of the theory is even more severe.

Here, we have decided to trade some generality for better readability
and opted for a simplified version of the original theory.
Still, for the generic backward induction algorithm we need to make
sure that it is possible to define
policy sequences of the length required for a specific SDP. 
This can e.g. be done by postulating
controls to be non-empty:
\begin{hscode}\SaveRestoreHook
\column{B}{@{}>{\hspre}l<{\hspost}@{}}%
\column{3}{@{}>{\hspre}l<{\hspost}@{}}%
\column{14}{@{}>{\hspre}c<{\hspost}@{}}%
\column{14E}{@{}l@{}}%
\column{17}{@{}>{\hspre}l<{\hspost}@{}}%
\column{E}{@{}>{\hspre}l<{\hspost}@{}}%
\>[3]{}\Varid{notEmptyY}{}\<[14]%
\>[14]{} \mathop{:} {}\<[14E]%
\>[17]{}(\Varid{t} \mathop{:} \mathbb{N}) \to (\Varid{x} \mathop{:} \Conid{X}\;\Varid{t}) \to \Conid{Y}\;\Varid{t}\;\Varid{x}{}\<[E]%
\ColumnHook
\end{hscode}\resethooks
We also impose a non-emptiness requirement on
the transition function \ensuremath{\Varid{next}} that will be discussed in
Sec.~\ref{section:discussion}.
\begin{hscode}\SaveRestoreHook
\column{B}{@{}>{\hspre}l<{\hspost}@{}}%
\column{3}{@{}>{\hspre}l<{\hspost}@{}}%
\column{17}{@{}>{\hspre}c<{\hspost}@{}}%
\column{17E}{@{}l@{}}%
\column{20}{@{}>{\hspre}l<{\hspost}@{}}%
\column{E}{@{}>{\hspre}l<{\hspost}@{}}%
\>[3]{}\Varid{nextNotEmpty}{}\<[17]%
\>[17]{} \mathop{:} {}\<[17E]%
\>[20]{}\{\mskip1.5mu \Varid{t} \mathop{:} \mathbb{N}\mskip1.5mu\} \to (\Varid{x} \mathop{:} \Conid{X}\;\Varid{t}) \to (\Varid{y} \mathop{:} \Conid{Y}\;\Varid{t}\;\Varid{x}) \to \Conid{NotEmpty}\;(\Varid{next}\;\Varid{t}\;\Varid{x}\;\Varid{y}){}\<[E]%
\ColumnHook
\end{hscode}\resethooks
\item In section \ref{subsection:solution_components}, we have not
discussed under which conditions one can implement optimal extensions of
arbitrary policy sequences. This is an interesting topic that is however
orthogonal to the purpose of the current paper.
For the same reason we have not addressed the question of how to make \ensuremath{\Varid{bi}}
more efficient by tabulation.
We briefly discuss the specification and implementation of optimal extensions
in the BJI-framework in Appendix~\ref{appendix:optimal_extension}.
We refer the reader interested in tabulation of \ensuremath{\Varid{bi}} to
\href{https://gitlab.pik-potsdam.de/botta/IdrisLibs/-/blob/master/SequentialDecisionProblems/TabBackwardsInduction.lidr}
{SequentialDecisionProblems.TabBackwardsInduction}
of \citep{botta20162018}.

\end{itemize}

\vfill
\pagebreak

\section{Correctness for monadic backward induction}
\label{section:preparation}

In this section we formally specify the notions of correctness for
monadic backward induction \ensuremath{\Varid{bi}} and the value function \ensuremath{\Varid{val}} of the
BJI-framework that we will study in the remainder of this paper.
We develop these notions as generic variants of the corresponding
notions for stochastic SDPs.

\subsection{Extension of the BJI-framework}
\label{subsection:frameworkExtension}

In the previous section, we have seen that a monadic SDP can be
specified in terms of nine components: \ensuremath{\Conid{M}}, \ensuremath{\Conid{X}}, \ensuremath{\Conid{Y}},
\ensuremath{\Varid{next}}, \ensuremath{\Conid{Val}}, \ensuremath{\Varid{zero}}, \ensuremath{ \mathbin{\oplus} }, \ensuremath{ \,\sqsubseteq\, } and \ensuremath{\Varid{reward}}.

Given a policy sequence (optimal or not) and an initial state
for an SDP, we can compute the \ensuremath{\Conid{M}}-structure of possible trajectories
starting at that state:
\begin{hscode}\SaveRestoreHook
\column{B}{@{}>{\hspre}l<{\hspost}@{}}%
\column{3}{@{}>{\hspre}l<{\hspost}@{}}%
\column{5}{@{}>{\hspre}l<{\hspost}@{}}%
\column{11}{@{}>{\hspre}c<{\hspost}@{}}%
\column{11E}{@{}l@{}}%
\column{14}{@{}>{\hspre}l<{\hspost}@{}}%
\column{22}{@{}>{\hspre}c<{\hspost}@{}}%
\column{22E}{@{}l@{}}%
\column{25}{@{}>{\hspre}l<{\hspost}@{}}%
\column{E}{@{}>{\hspre}l<{\hspost}@{}}%
\>[3]{}\mathbf{data}\;\Conid{StateCtrlSeq}{}\<[22]%
\>[22]{} \mathop{:} {}\<[22E]%
\>[25]{}(\Varid{t},\Varid{n} \mathop{:} \mathbb{N}) \to \Conid{Type}\;\mathbf{where}{}\<[E]%
\\
\>[3]{}\hsindent{2}{}\<[5]%
\>[5]{}\Conid{Last}{}\<[11]%
\>[11]{} \mathop{:} {}\<[11E]%
\>[14]{}\{\mskip1.5mu \Varid{t} \mathop{:} \mathbb{N}\mskip1.5mu\} \to \Conid{X}\;\Varid{t} \to \Conid{StateCtrlSeq}\;\Varid{t}\;(\Conid{S}\;\Conid{Z}){}\<[E]%
\\
\>[3]{}\hsindent{2}{}\<[5]%
\>[5]{}(  \mathbin{\#\!\#} ){}\<[11]%
\>[11]{} \mathop{:} {}\<[11E]%
\>[14]{}\{\mskip1.5mu \Varid{t},\Varid{n} \mathop{:} \mathbb{N}\mskip1.5mu\} \to (\Varid{x} \mathop{:} \Conid{X}\;\Varid{t}  \mathbin{*\!*} \Conid{Y}\;\Varid{t}\;\Varid{x}) \to \Conid{StateCtrlSeq}\;(\Conid{S}\;\Varid{t})\;(\Conid{S}\;\Varid{n}) \to \Conid{StateCtrlSeq}\;\Varid{t}\;(\Conid{S}\;(\Conid{S}\;\Varid{n})){}\<[E]%
\ColumnHook
\end{hscode}\resethooks
\begin{hscode}\SaveRestoreHook
\column{B}{@{}>{\hspre}l<{\hspost}@{}}%
\column{3}{@{}>{\hspre}l<{\hspost}@{}}%
\column{8}{@{}>{\hspre}c<{\hspost}@{}}%
\column{8E}{@{}l@{}}%
\column{11}{@{}>{\hspre}l<{\hspost}@{}}%
\column{12}{@{}>{\hspre}l<{\hspost}@{}}%
\column{21}{@{}>{\hspre}l<{\hspost}@{}}%
\column{24}{@{}>{\hspre}c<{\hspost}@{}}%
\column{24E}{@{}l@{}}%
\column{27}{@{}>{\hspre}l<{\hspost}@{}}%
\column{35}{@{}>{\hspre}l<{\hspost}@{}}%
\column{E}{@{}>{\hspre}l<{\hspost}@{}}%
\>[3]{}\Varid{trj}{}\<[8]%
\>[8]{} \mathop{:} {}\<[8E]%
\>[11]{}\{\mskip1.5mu \Varid{t},\Varid{n} \mathop{:} \mathbb{N}\mskip1.5mu\} \to \Conid{PolicySeq}\;\Varid{t}\;\Varid{n} \to \Conid{X}\;\Varid{t} \to \Conid{M}\;(\Conid{StateCtrlSeq}\;\Varid{t}\;(\Conid{S}\;\Varid{n})){}\<[E]%
\\
\>[3]{}\Varid{trj}\;\{\mskip1.5mu \Varid{t}\mskip1.5mu\}\;{}\<[12]%
\>[12]{}\Conid{Nil}\;{}\<[21]%
\>[21]{}\Varid{x}{}\<[24]%
\>[24]{}\mathrel{=}{}\<[24E]%
\>[27]{}\Varid{pure}\;(\Conid{Last}\;\Varid{x}){}\<[E]%
\\
\>[3]{}\Varid{trj}\;\{\mskip1.5mu \Varid{t}\mskip1.5mu\}\;(\Varid{p}\mathbin{::}\Varid{ps})\;\Varid{x}{}\<[24]%
\>[24]{}\mathrel{=}{}\<[24E]%
\>[27]{}\mathbf{let}\;\Varid{y}{}\<[35]%
\>[35]{}\mathrel{=}\Varid{p}\;\Varid{x}\;\mathbf{in}{}\<[E]%
\\
\>[27]{}\mathbf{let}\;\Varid{mx'}\mathrel{=}\Varid{next}\;\Varid{t}\;\Varid{x}\;\Varid{y}\;\mathbf{in}{}\<[E]%
\\
\>[27]{}\Varid{map}\;((\Varid{x}  \mathbin{*\!*} \Varid{y})  \mathbin{\#\!\#} )\;(\Varid{mx'} \mathbin{>\!\!>\!\!=} \Varid{trj}\;\Varid{ps}){}\<[E]%
\ColumnHook
\end{hscode}\resethooks
where we use \ensuremath{\Conid{StateCtrlSeq}} as type of trajectories. Essentially it is
a non-empty list of (dependent) state/control pairs, with the exception of the base case
which is a singleton just containing the last state reached.

Furthermore, we can compute the \emph{total reward} for a single
trajectory, i.e. its sum of rewards:

\begin{hscode}\SaveRestoreHook
\column{B}{@{}>{\hspre}l<{\hspost}@{}}%
\column{3}{@{}>{\hspre}l<{\hspost}@{}}%
\column{9}{@{}>{\hspre}c<{\hspost}@{}}%
\column{9E}{@{}l@{}}%
\column{12}{@{}>{\hspre}l<{\hspost}@{}}%
\column{31}{@{}>{\hspre}c<{\hspost}@{}}%
\column{31E}{@{}l@{}}%
\column{34}{@{}>{\hspre}l<{\hspost}@{}}%
\column{37}{@{}>{\hspre}l<{\hspost}@{}}%
\column{E}{@{}>{\hspre}l<{\hspost}@{}}%
\>[3]{}\Varid{sumR}{}\<[9]%
\>[9]{} \mathop{:} {}\<[9E]%
\>[12]{}\{\mskip1.5mu \Varid{t},\Varid{n} \mathop{:} \mathbb{N}\mskip1.5mu\} \to \Conid{StateCtrlSeq}\;\Varid{t}\;\Varid{n} \to \Conid{Val}{}\<[E]%
\\
\>[3]{}\Varid{sumR}\;\{\mskip1.5mu \Varid{t}\mskip1.5mu\}\;(\Conid{Last}\;\Varid{x}){}\<[34]%
\>[34]{}\mathrel{=}{}\<[37]%
\>[37]{}\Varid{zero}{}\<[E]%
\\
\>[3]{}\Varid{sumR}\;\{\mskip1.5mu \Varid{t}\mskip1.5mu\}\;((\Varid{x}  \mathbin{*\!*} \Varid{y})  \mathbin{\#\!\#} \Varid{xys}){}\<[31]%
\>[31]{}\mathrel{=}{}\<[31E]%
\>[34]{}\Varid{reward}\;\Varid{t}\;\Varid{x}\;\Varid{y}\;(\Varid{head}\;\Varid{xys}) \mathbin{\oplus} \Varid{sumR}\;\Varid{xys}{}\<[E]%
\ColumnHook
\end{hscode}\resethooks
where \ensuremath{\Varid{head}} is the helper function
\begin{hscode}\SaveRestoreHook
\column{B}{@{}>{\hspre}l<{\hspost}@{}}%
\column{3}{@{}>{\hspre}l<{\hspost}@{}}%
\column{9}{@{}>{\hspre}c<{\hspost}@{}}%
\column{9E}{@{}l@{}}%
\column{12}{@{}>{\hspre}l<{\hspost}@{}}%
\column{27}{@{}>{\hspre}c<{\hspost}@{}}%
\column{27E}{@{}l@{}}%
\column{30}{@{}>{\hspre}c<{\hspost}@{}}%
\column{30E}{@{}l@{}}%
\column{33}{@{}>{\hspre}l<{\hspost}@{}}%
\column{E}{@{}>{\hspre}l<{\hspost}@{}}%
\>[3]{}\Varid{head}{}\<[9]%
\>[9]{} \mathop{:} {}\<[9E]%
\>[12]{}\{\mskip1.5mu \Varid{t},\Varid{n} \mathop{:} \mathbb{N}\mskip1.5mu\} \to \Conid{StateCtrlSeq}\;\Varid{t}\;(\Conid{S}\;\Varid{n}) \to \Conid{X}\;\Varid{t}{}\<[E]%
\\
\>[3]{}\Varid{head}\;(\Conid{Last}\;\Varid{x}){}\<[30]%
\>[30]{}\mathrel{=}{}\<[30E]%
\>[33]{}\Varid{x}{}\<[E]%
\\
\>[3]{}\Varid{head}\;((\Varid{x}  \mathbin{*\!*} \Varid{y})  \mathbin{\#\!\#} \Varid{xys}){}\<[27]%
\>[27]{}\mathrel{=}{}\<[27E]%
\>[30]{}\Varid{x}{}\<[30E]%
\ColumnHook
\end{hscode}\resethooks
By mapping \ensuremath{\Varid{sumR}} onto an \ensuremath{\Conid{M}}-structure of trajectories, we obtain an
\ensuremath{\Conid{M}}-structure containing the individual sums of rewards of the
trajectories. Now, using the measure function, we can compute the
generic analogue of the expected total reward for a policy sequence \ensuremath{\Varid{ps}}
and an initial state \ensuremath{\Varid{x}}:
\begin{hscode}\SaveRestoreHook
\column{B}{@{}>{\hspre}l<{\hspost}@{}}%
\column{3}{@{}>{\hspre}l<{\hspost}@{}}%
\column{9}{@{}>{\hspre}c<{\hspost}@{}}%
\column{9E}{@{}l@{}}%
\column{12}{@{}>{\hspre}l<{\hspost}@{}}%
\column{15}{@{}>{\hspre}l<{\hspost}@{}}%
\column{E}{@{}>{\hspre}l<{\hspost}@{}}%
\>[3]{}\Varid{val'}{}\<[9]%
\>[9]{} \mathop{:} {}\<[9E]%
\>[12]{}\{\mskip1.5mu \Varid{t},\Varid{n} \mathop{:} \mathbb{N}\mskip1.5mu\} \to (\Varid{ps} \mathop{:} \Conid{PolicySeq}\;\Varid{t}\;\Varid{n}) \to (\Varid{x} \mathop{:} \Conid{X}\;\Varid{t}) \to \Conid{Val}{}\<[E]%
\\
\>[3]{}\Varid{val'}\;\Varid{ps}{}\<[12]%
\>[12]{}\mathrel{=}{}\<[15]%
\>[15]{}\Varid{meas}\mathbin{\circ}\Varid{map}\;\Varid{sumR}\mathbin{\circ}\Varid{trj}\;\Varid{ps}{}\<[E]%
\ColumnHook
\end{hscode}\resethooks
As anticipated in Sec.~\ref{section:SDPs} we call the value
computed by \ensuremath{\Varid{val'}} the \emph{measured total reward}. Recall that
solving a stochastic SDP commonly means finding a policy sequence that
maximises the \emph{expected total reward}. By analogy, we define that
solving a monadic SDP means to find a policy sequence that maximises
the \emph{measured total reward}. I.e. given \ensuremath{\Varid{t}} and \ensuremath{\Varid{n}}, the solution
of a monadic SDP is a sequence of \ensuremath{\Varid{n}} policies that maximises the measure
of the sum of rewards along all possible trajectories of length \ensuremath{\Varid{n}}
that are rooted in an initial state at step \ensuremath{\Varid{t}}. 

Again by analogy to the stochastic case, we define monadic
backward induction to be correct if, for a given SDP, the policy
sequence computed by \ensuremath{\Varid{bi}} is the solution to the SDP.
I.e., we consider \ensuremath{\Varid{bi}} to be correct if it meets the specification
\begin{hscode}\SaveRestoreHook
\column{B}{@{}>{\hspre}l<{\hspost}@{}}%
\column{3}{@{}>{\hspre}l<{\hspost}@{}}%
\column{25}{@{}>{\hspre}c<{\hspost}@{}}%
\column{25E}{@{}l@{}}%
\column{28}{@{}>{\hspre}l<{\hspost}@{}}%
\column{E}{@{}>{\hspre}l<{\hspost}@{}}%
\>[3]{}\Varid{biOptMeasTotalReward}{}\<[25]%
\>[25]{} \mathop{:} {}\<[25E]%
\>[28]{}(\Varid{t},\Varid{n} \mathop{:} \mathbb{N}) \to \Conid{GenOptPolicySeq}\;\Varid{val'}\;(\Varid{bi}\;\Varid{t}\;\Varid{n}){}\<[E]%
\ColumnHook
\end{hscode}\resethooks
where \ensuremath{\Conid{GenOptPolicySeq}} is a generalised version of the optimality
predicate \ensuremath{\Conid{OptPolicySeq}} from
Sec.~\ref{subsection:solution_components}. It now takes as an
additional parameter the function with respect to which the policy
sequence is to be optimal:
\begin{hscode}\SaveRestoreHook
\column{B}{@{}>{\hspre}l<{\hspost}@{}}%
\column{3}{@{}>{\hspre}l<{\hspost}@{}}%
\column{20}{@{}>{\hspre}c<{\hspost}@{}}%
\column{20E}{@{}l@{}}%
\column{23}{@{}>{\hspre}l<{\hspost}@{}}%
\column{33}{@{}>{\hspre}c<{\hspost}@{}}%
\column{33E}{@{}l@{}}%
\column{36}{@{}>{\hspre}l<{\hspost}@{}}%
\column{40}{@{}>{\hspre}l<{\hspost}@{}}%
\column{74}{@{}>{\hspre}l<{\hspost}@{}}%
\column{E}{@{}>{\hspre}l<{\hspost}@{}}%
\>[3]{}\Conid{GenOptPolicySeq}{}\<[20]%
\>[20]{} \mathop{:} {}\<[20E]%
\>[23]{}\{\mskip1.5mu \Varid{t},\Varid{n} \mathop{:} \mathbb{N}\mskip1.5mu\} \to {}\<[40]%
\>[40]{}(\Conid{PolicySeq}\;\Varid{t}\;\Varid{n} \to \Conid{X}\;\Varid{t} \to \Conid{Val}) \to {}\<[74]%
\>[74]{}\Conid{PolicySeq}\;\Varid{t}\;\Varid{n} \to \Conid{Type}{}\<[E]%
\\[\blanklineskip]%
\>[3]{}\Conid{GenOptPolicySeq}\;\{\mskip1.5mu \Varid{t}\mskip1.5mu\}\;\{\mskip1.5mu \Varid{n}\mskip1.5mu\}\;\Varid{f}\;\Varid{ps}{}\<[33]%
\>[33]{}\mathrel{=}{}\<[33E]%
\>[36]{}(\Varid{ps'} \mathop{:} \Conid{PolicySeq}\;\Varid{t}\;\Varid{n}) \to (\Varid{x} \mathop{:} \Conid{X}\;\Varid{t}) \to \Varid{f}\;\Varid{ps'}\;\Varid{x} \,\sqsubseteq\, \Varid{f}\;\Varid{ps}\;\Varid{x}{}\<[E]%
\ColumnHook
\end{hscode}\resethooks
As recapitulated in Sec.~\ref{subsection:solution_components},
\bottaetal have already shown that if \ensuremath{\Conid{M}} is a monad, \ensuremath{ \,\sqsubseteq\, } a total
preorder and
\ensuremath{ \mathbin{\oplus} } and \ensuremath{\Varid{meas}} fulfil two monotonicity conditions, then \ensuremath{\Varid{bi}\;\Varid{t}\;\Varid{n}}
yields an optimal policy sequence with respect to the value function
\ensuremath{\Varid{val}} in the sense that \ensuremath{\Varid{val}\;\Varid{ps'}\;\Varid{x} \,\sqsubseteq\, \Varid{val}\;(\Varid{bi}\;\Varid{t}\;\Varid{n})\;\Varid{x}} for any
policy sequence \ensuremath{\Varid{ps'}} and initial state \ensuremath{\Varid{x}}, for arbitrary \ensuremath{\Varid{t},\Varid{n} \mathop{:} \mathbb{N}}. Or, expressed using the generalised optimality predicate,
that the type
\begin{hscode}\SaveRestoreHook
\column{B}{@{}>{\hspre}l<{\hspost}@{}}%
\column{3}{@{}>{\hspre}l<{\hspost}@{}}%
\column{E}{@{}>{\hspre}l<{\hspost}@{}}%
\>[3]{}\Conid{GenOptPolicySeq}\;\{\mskip1.5mu \Varid{t}\mskip1.5mu\}\;\{\mskip1.5mu \Varid{n}\mskip1.5mu\}\;\Varid{val}\;(\Varid{bi}\;\Varid{t}\;\Varid{n}){}\<[E]%
\ColumnHook
\end{hscode}\resethooks
is inhabited.
As seen in Sec.~\ref{subsection:solution_components}, the function
\ensuremath{\Varid{val}} measures and adds rewards incrementally. But does it always
compute the measured total reward like \ensuremath{\Varid{val'}}?
Modulo differences in the presentation \citet[Theorem
4.2.1]{puterman2014markov} suggests that for standard
stochastic SDPs, \ensuremath{\Varid{val}} and \ensuremath{\Varid{val'}} are extensionally equal, which in turn
allows the use of backward induction for solving these SDPs.
Generalising, we therefore consider \ensuremath{\Varid{val}} as
correct if it fulfils the specification
\begin{hscode}\SaveRestoreHook
\column{B}{@{}>{\hspre}l<{\hspost}@{}}%
\column{3}{@{}>{\hspre}l<{\hspost}@{}}%
\column{23}{@{}>{\hspre}c<{\hspost}@{}}%
\column{23E}{@{}l@{}}%
\column{26}{@{}>{\hspre}l<{\hspost}@{}}%
\column{E}{@{}>{\hspre}l<{\hspost}@{}}%
\>[3]{}\Varid{valMeasTotalReward}{}\<[23]%
\>[23]{} \mathop{:} {}\<[23E]%
\>[26]{}\{\mskip1.5mu \Varid{t},\Varid{n} \mathop{:} \mathbb{N}\mskip1.5mu\} \to (\Varid{ps} \mathop{:} \Conid{PolicySeq}\;\Varid{t}\;\Varid{n}) \to (\Varid{x} \mathop{:} \Conid{X}\;\Varid{t}) \to \Varid{val}\;\Varid{ps}\;\Varid{x}\mathrel{=}\Varid{val'}\;\Varid{ps}\;\Varid{x}{}\<[E]%
\ColumnHook
\end{hscode}\resethooks
If this equality held for the general monadic SDPs of the BJI-theory,
we could prove the correctness of \ensuremath{\Varid{bi}} as immediate corollary of
\ensuremath{\Varid{valMeasTotalReward}} and \bottaetal\!'s result \ensuremath{\Varid{biOptVal}}.
The statement \ensuremath{\Varid{biOptMeasTotalReward}} can be seen as a generic version
of textbook correctness statements for backward induction as solution
method for stochastic SDPs like \citep[prop.1.3.1]{bertsekas1995} or
\citep[Theorem~4.5.1.c]{puterman2014markov}.
By proving \ensuremath{\Varid{valMeasTotalReward}} we could therefore extend the
verification of \citep{2017_Botta_Jansson_Ionescu} and obtain a
stronger correctness result for monadic backward induction. 

\vspace{0.2cm}
Our main objective in the remainder of the paper is therefore to prove
that \ensuremath{\Varid{valMeasTotalReward}} holds.
But there is a problem.

\subsection{The problem with the BJI-value function}
\label{subsection:counterEx}
A closer look at \ensuremath{\Varid{val}} and \ensuremath{\Varid{val'}} reveals two quite different
computational patterns: applied to a policy sequence \ensuremath{\Varid{ps}} of length \ensuremath{\Varid{n}\mathbin{+}\mathrm{1}}
and a state \ensuremath{\Varid{x}}, the function \ensuremath{\Varid{val}} directly evaluates \ensuremath{\Varid{meas}} on
the \ensuremath{\Conid{M}}-structure of rewards
corresponding to the possible next states after one step. This entails
further evaluations of \ensuremath{\Varid{meas}} for each possible next state.
By contrast, \ensuremath{\Varid{val'}\;\Varid{ps}\;\Varid{x}} entails only one evaluation of \ensuremath{\Varid{meas}},
independently of the length of \ensuremath{\Varid{ps}}. The computation, however, builds up
an intermediate \ensuremath{\Conid{M}}-structure of state-control sequences. The elements of this
\ensuremath{\Conid{M}}-structure, the state-control sequences, are then consumed by \ensuremath{\Varid{sumR}}
and finally the \ensuremath{\Conid{M}}-structure of rewards is reduced by \ensuremath{\Varid{meas}}.

For illustration, let us revisit Ex.~\ref{subsection:example1SDPs} from
Sec.~\ref{section:SDPs} as formalised in
Fig.~\ref{fig:example1Formal}.
To do an example calculation with \ensuremath{\Varid{val}} and \ensuremath{\Varid{val'}} we first need a
concrete policy sequence as input.
The simplest two policies are the two constant policies:
\begin{hscode}\SaveRestoreHook
\column{B}{@{}>{\hspre}l<{\hspost}@{}}%
\column{3}{@{}>{\hspre}l<{\hspost}@{}}%
\column{E}{@{}>{\hspre}l<{\hspost}@{}}%
\>[3]{}\Varid{constH} \mathop{:} (\Varid{t} \mathop{:} \mathbb{N}) \to \Conid{Policy}\;\Varid{t}{}\<[E]%
\\
\>[3]{}\Varid{constH}\;\Varid{\char95 t}\mathrel{=}\Varid{const}\;\Conid{High}{}\<[E]%
\\[\blanklineskip]%
\>[3]{}\Varid{constL} \mathop{:} (\Varid{t} \mathop{:} \mathbb{N}) \to \Conid{Policy}\;\Varid{t}{}\<[E]%
\\
\>[3]{}\Varid{constL}\;\Varid{\char95 t}\mathrel{=}\Varid{const}\;\Conid{Low}{}\<[E]%
\ColumnHook
\end{hscode}\resethooks
From these, we can define a policy sequence
\begin{hscode}\SaveRestoreHook
\column{B}{@{}>{\hspre}l<{\hspost}@{}}%
\column{3}{@{}>{\hspre}l<{\hspost}@{}}%
\column{E}{@{}>{\hspre}l<{\hspost}@{}}%
\>[3]{}\Varid{ps} \mathop{:} \Conid{PolicySeq}\;\mathrm{0}\;\mathrm{3}{}\<[E]%
\\
\>[3]{}\Varid{ps}\mathrel{=}\Varid{constH}\;\mathrm{0}\mathbin{::}(\Varid{constL}\;\mathrm{1}\mathbin{::}(\Varid{constH}\;\mathrm{2}\mathbin{::}\Conid{Nil})){}\<[E]%
\ColumnHook
\end{hscode}\resethooks
It is instructive to compute \ensuremath{\Varid{val}\;\Varid{ps}\;\Conid{Good}} and \ensuremath{\Varid{val'}\;\Varid{ps}\;\Conid{Good}} by hand.  Recall that in this example, we have
\ensuremath{\Conid{M}\mathrel{=}\Conid{List}} with \ensuremath{( \mathbin{>\!\!>\!\!=} )\mathrel{=}\Varid{concatMap}} and \ensuremath{\Conid{Val}\mathrel{=}\mathbb{N}} with \ensuremath{ \mathbin{\oplus} \mathrel{=}\mathbin{+}}.
The measure \ensuremath{\Varid{meas}} thus needs to
have the type \ensuremath{\Conid{List}\;\mathbb{N} \to \mathbb{N}}. Without instantiating \ensuremath{\Varid{meas}} for the
moment, the computations roughly exhibit the structure
\begin{hscode}\SaveRestoreHook
\column{B}{@{}>{\hspre}l<{\hspost}@{}}%
\column{3}{@{}>{\hspre}l<{\hspost}@{}}%
\column{18}{@{}>{\hspre}l<{\hspost}@{}}%
\column{19}{@{}>{\hspre}l<{\hspost}@{}}%
\column{25}{@{}>{\hspre}l<{\hspost}@{}}%
\column{E}{@{}>{\hspre}l<{\hspost}@{}}%
\>[3]{}\Varid{val}\;\Varid{ps}\;\Conid{Good}\mathrel{=}{}\<[18]%
\>[18]{}\Varid{meas}\;[\mskip1.5mu \mathrm{2}\mathbin{+}\Varid{meas}\;[\mskip1.5mu \mathrm{3}\mathbin{+}\Varid{meas}\;[\mskip1.5mu \mathrm{2},\mathrm{0}\mskip1.5mu]\mskip1.5mu],\mathrm{0}\mathbin{+}\Varid{meas}\;[\mskip1.5mu \mathrm{3}\mathbin{+}\Varid{meas}\;[\mskip1.5mu \mathrm{2},\mathrm{0}\mskip1.5mu],\mathrm{1}\mathbin{+}\Varid{meas}\;[\mskip1.5mu \mathrm{0}\mskip1.5mu]\mskip1.5mu]\mskip1.5mu]{}\<[E]%
\\[\blanklineskip]%
\>[3]{}\Varid{val'}\;\Varid{ps}\;\Conid{Good}\mathrel{=}{}\<[19]%
\>[19]{}\Varid{meas}\;{}\<[25]%
\>[25]{}[\mskip1.5mu \mathrm{7},\mathrm{5},\mathrm{5},\mathrm{3},\mathrm{1}\mskip1.5mu]{}\<[E]%
\ColumnHook
\end{hscode}\resethooks
and it is not ``obviously clear'' that \ensuremath{\Varid{val}} and \ensuremath{\Varid{val'}} are
extensionally equal without further knowledge about \ensuremath{\Varid{meas}}.

In the deterministic case, i.e. for \ensuremath{\Conid{M}\mathrel{=}\Conid{Id}} and
\ensuremath{\Varid{meas}\mathrel{=}\Varid{id}}, \ensuremath{\Varid{val}\;\Varid{ps}\;\Varid{x}} and \ensuremath{\Varid{val'}\;\Varid{ps}\;\Varid{x}} are indeed equal for all \ensuremath{\Varid{ps}}
and \ensuremath{\Varid{x}}, without imposing any further 
conditions (as we will see in Sec.~\ref{section:valval}).
For the stochastic case, \cite[Theorem
4.2.1]{puterman2014markov} suggests that the equality
should hold. But for the monadic case, no such result has been
established.
And as it turns out, in general the functions \ensuremath{\Varid{val}} and \ensuremath{\Varid{val'}} are not
unconditionally equal -- consider the following counter-example:
We continue in the setting of Ex.~\ref{subsection:example1SDPs}
from above, but now instantiate the measure to the plain arithmetic sum 
\begin{hscode}\SaveRestoreHook
\column{B}{@{}>{\hspre}l<{\hspost}@{}}%
\column{3}{@{}>{\hspre}l<{\hspost}@{}}%
\column{E}{@{}>{\hspre}l<{\hspost}@{}}%
\>[3]{}\Varid{meas}\mathrel{=}\Varid{foldr}\;(\mathbin{+})\;\mathrm{0}{}\<[E]%
\ColumnHook
\end{hscode}\resethooks

This measure fulfils the monotonicity condition
(\ensuremath{\Varid{measMonSpec}}, Sec.~\ref{subsection:solution_components}) imposed by the
BJI-framework.
But if we instantiate the above computations with it, then we get \ensuremath{\Varid{val}\;\Varid{ps}\;\Conid{Good}\mathrel{=}\mathrm{13}} and \ensuremath{\Varid{val'}\;\Varid{ps}\;\Conid{Good}} = 21!
We thus see that the equality between \ensuremath{\Varid{val}} and \ensuremath{\Varid{val'}} cannot hold
unconditionally in the generic setting of the BJI-framework.
In the next section we therefore present conditions under which the
equality \emph{does} hold.

\section{Correctness conditions}
\label{section:conditions}

We now formulate three conditions on combinations of the monad,
the measure function and the binary operation \ensuremath{ \mathbin{\oplus} } that imply the
extensional equality of \ensuremath{\Varid{val}} and \ensuremath{\Varid{val'}}:

\setlist[itemize,1]{leftmargin=58pt}
\begin{itemize}


\item[{\bf Condition 1.}]
The measure needs to be left-inverse to \ensuremath{\Varid{pure}}:
\footnote{The symbol \ensuremath{\doteq} denotes \emph{extensional} equality,
  see Appendix~\ref{appendix:monadLaws}}
\begin{hscode}\SaveRestoreHook
\column{B}{@{}>{\hspre}l<{\hspost}@{}}%
\column{3}{@{}>{\hspre}l<{\hspost}@{}}%
\column{20}{@{}>{\hspre}c<{\hspost}@{}}%
\column{20E}{@{}l@{}}%
\column{23}{@{}>{\hspre}l<{\hspost}@{}}%
\column{E}{@{}>{\hspre}l<{\hspost}@{}}%
\>[3]{}\Varid{measPureSpec}{}\<[20]%
\>[20]{} \mathop{:} {}\<[20E]%
\>[23]{}\Varid{meas}\mathbin{\circ}\Varid{pure}\doteq\Varid{id}{}\<[E]%
\ColumnHook
\end{hscode}\resethooks

\begin{center}
  \includegraphics{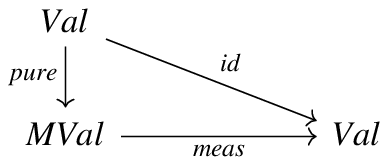}
\end{center}


\item[{\bf Condition 2.}]
  Applying the measure after \ensuremath{\Varid{join}} needs to be extensionally
equal to applying it after \ensuremath{\Varid{map}\;\Varid{meas}}:
\begin{hscode}\SaveRestoreHook
\column{B}{@{}>{\hspre}l<{\hspost}@{}}%
\column{3}{@{}>{\hspre}l<{\hspost}@{}}%
\column{17}{@{}>{\hspre}l<{\hspost}@{}}%
\column{E}{@{}>{\hspre}l<{\hspost}@{}}%
\>[3]{}\Varid{measJoinSpec}{}\<[17]%
\>[17]{} \mathop{:} \Varid{meas}\mathbin{\circ}\Varid{join}\doteq\Varid{meas}\mathbin{\circ}\Varid{map}\;\Varid{meas}{}\<[E]%
\ColumnHook
\end{hscode}\resethooks
\begin{center}
  \includegraphics{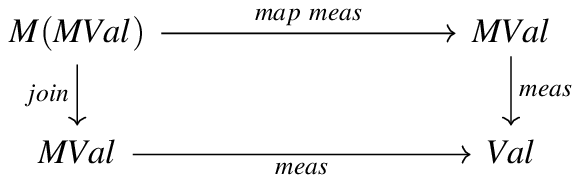}
\end{center}


\item[\bf{Condition 3.}]
For arbitrary \ensuremath{\Varid{v} \mathop{:} \Conid{Val}} and non-empty \ensuremath{\Varid{mv} \mathop{:} \Conid{M}\;\Conid{Val}} applying
the measure after mapping \ensuremath{(\Varid{v} \mathbin{\oplus} )} onto \ensuremath{\Varid{mv}} needs to be equal to
applying \ensuremath{(\Varid{v} \mathbin{\oplus} )} after the measure:
\begin{hscode}\SaveRestoreHook
\column{B}{@{}>{\hspre}l<{\hspost}@{}}%
\column{3}{@{}>{\hspre}l<{\hspost}@{}}%
\column{17}{@{}>{\hspre}c<{\hspost}@{}}%
\column{17E}{@{}l@{}}%
\column{20}{@{}>{\hspre}l<{\hspost}@{}}%
\column{E}{@{}>{\hspre}l<{\hspost}@{}}%
\>[3]{}\Varid{measPlusSpec}{}\<[17]%
\>[17]{} \mathop{:} {}\<[17E]%
\>[20]{}(\Varid{v} \mathop{:} \Conid{Val}) \to (\Varid{mv} \mathop{:} \Conid{M}\;\Conid{Val}) \to (\Conid{NotEmpty}\;\Varid{mv}) \to {}\<[E]%
\\
\>[20]{}(\Varid{meas}\mathbin{\circ}\Varid{map}\;(\Varid{v} \mathbin{\oplus} ))\;\Varid{mv}\mathrel{=}((\Varid{v} \mathbin{\oplus} )\mathbin{\circ}\Varid{meas})\;\Varid{mv}{}\<[E]%
\ColumnHook
\end{hscode}\resethooks
\begin{center}
  \includegraphics{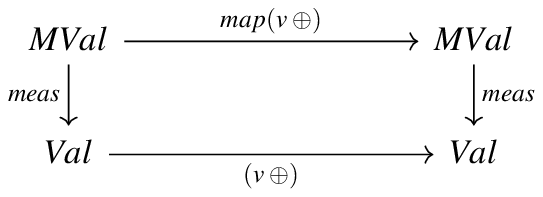}
%
\end{center}

\end{itemize}
\setlist[itemize,1]{leftmargin=10pt}

Essentially, these conditions assure that the measure is well-behaved
relative to the monad structure and the \ensuremath{ \mathbin{\oplus} }-operation.
They arise by, again, generalising from the
standard example of stochastic SDPs with a probability monad,
the \emph{expected value} as measure and ordinary addition as \ensuremath{ \mathbin{\oplus} }.
The first two conditions are lifting properties that allow to do
computations either in the monad or the underlying structure
with the same result.
The third condition is a distributivity law. For the computation of the
measured total reward it means that instead of adding the current reward to
the outcome of each trajectory and then measuring, one may as well first measure the
outcomes and then add the current reward.

\vspace{0.2cm}
To illustrate the conditions, let us consider a simple representation of
discrete probability distributions like in \citep{DBLP:journals/jfp/ErwigK06}.
\begin{hscode}\SaveRestoreHook
\column{B}{@{}>{\hspre}l<{\hspost}@{}}%
\column{3}{@{}>{\hspre}l<{\hspost}@{}}%
\column{11}{@{}>{\hspre}l<{\hspost}@{}}%
\column{15}{@{}>{\hspre}c<{\hspost}@{}}%
\column{15E}{@{}l@{}}%
\column{18}{@{}>{\hspre}l<{\hspost}@{}}%
\column{E}{@{}>{\hspre}l<{\hspost}@{}}%
\>[3]{}\Conid{Dist} \mathop{:} {}\<[11]%
\>[11]{}\Conid{Type} \to \Conid{Type}{}\<[E]%
\\
\>[3]{}\Conid{Dist}\;\Conid{Omega}{}\<[15]%
\>[15]{}\mathrel{=}{}\<[15E]%
\>[18]{}\Conid{List}\;(\Conid{Omega},\Conid{Double}){}\<[E]%
\ColumnHook
\end{hscode}\resethooks
with
\begin{hscode}\SaveRestoreHook
\column{B}{@{}>{\hspre}l<{\hspost}@{}}%
\column{3}{@{}>{\hspre}l<{\hspost}@{}}%
\column{12}{@{}>{\hspre}c<{\hspost}@{}}%
\column{12E}{@{}l@{}}%
\column{15}{@{}>{\hspre}l<{\hspost}@{}}%
\column{18}{@{}>{\hspre}c<{\hspost}@{}}%
\column{18E}{@{}l@{}}%
\column{21}{@{}>{\hspre}l<{\hspost}@{}}%
\column{E}{@{}>{\hspre}l<{\hspost}@{}}%
\>[3]{}\Varid{distMap}{}\<[12]%
\>[12]{} \mathop{:} {}\<[12E]%
\>[15]{}\{\mskip1.5mu \Conid{A},\Conid{B} \mathop{:} \Conid{Type}\mskip1.5mu\} \to (\Varid{f} \mathop{:} \Conid{A} \to \Conid{B}) \to \Conid{Dist}\;\Conid{A} \to \Conid{Dist}\;\Conid{B}{}\<[E]%
\\
\>[3]{}\Varid{distMap}\;\Varid{f}\;\Varid{aps}{}\<[18]%
\>[18]{}\mathrel{=}{}\<[18E]%
\>[21]{}[\mskip1.5mu (\Varid{f}\;(\Varid{fst}\;\Varid{ap}),\Varid{snd}\;\Varid{ap})\mid \Varid{ap}\leftarrow \Varid{aps}\mskip1.5mu]{}\<[E]%
\ColumnHook
\end{hscode}\resethooks
\begin{hscode}\SaveRestoreHook
\column{B}{@{}>{\hspre}l<{\hspost}@{}}%
\column{3}{@{}>{\hspre}l<{\hspost}@{}}%
\column{15}{@{}>{\hspre}c<{\hspost}@{}}%
\column{15E}{@{}l@{}}%
\column{18}{@{}>{\hspre}l<{\hspost}@{}}%
\column{E}{@{}>{\hspre}l<{\hspost}@{}}%
\>[3]{}\Varid{distPure}{}\<[15]%
\>[15]{} \mathop{:} {}\<[15E]%
\>[18]{}\{\mskip1.5mu \Conid{A} \mathop{:} \Conid{Type}\mskip1.5mu\} \to \Conid{A} \to \Conid{Dist}\;\Conid{A}{}\<[E]%
\\
\>[3]{}\Varid{distPure}\;\Varid{a}{}\<[15]%
\>[15]{}\mathrel{=}{}\<[15E]%
\>[18]{}[\mskip1.5mu (\Varid{a},\mathrm{1.0})\mskip1.5mu]{}\<[E]%
\ColumnHook
\end{hscode}\resethooks
\begin{hscode}\SaveRestoreHook
\column{B}{@{}>{\hspre}l<{\hspost}@{}}%
\column{3}{@{}>{\hspre}l<{\hspost}@{}}%
\column{E}{@{}>{\hspre}l<{\hspost}@{}}%
\>[3]{}\Varid{distJoin} \mathop{:} \{\mskip1.5mu \Conid{A} \mathop{:} \Conid{Type}\mskip1.5mu\} \to (\Conid{Dist}\;(\Conid{Dist}\;\Conid{A})) \to \Conid{Dist}\;\Conid{A}{}\<[E]%
\\
\>[3]{}\Varid{distJoin}\;\Varid{apsps}\mathrel{=}\Varid{concat}\;[\mskip1.5mu [\mskip1.5mu (\Varid{fst}\;\Varid{ap},\Varid{snd}\;\Varid{ap}\mathbin{*}\Varid{snd}\;\Varid{aps})\mid \Varid{ap}\leftarrow \Varid{fst}\;\Varid{aps}\mskip1.5mu]\mid \Varid{aps}\leftarrow \Varid{apsps}\mskip1.5mu]{}\<[E]%
\ColumnHook
\end{hscode}\resethooks
\ensuremath{\Conid{Val}\mathrel{=}\Conid{Double}} and as measure the expected value
\begin{hscode}\SaveRestoreHook
\column{B}{@{}>{\hspre}l<{\hspost}@{}}%
\column{3}{@{}>{\hspre}l<{\hspost}@{}}%
\column{E}{@{}>{\hspre}l<{\hspost}@{}}%
\>[3]{}\Varid{expected} \mathop{:} \Conid{Dist}\;\Conid{Double} \to \Conid{Double}{}\<[E]%
\\
\>[3]{}\Varid{expected}\;\Varid{dps}\mathrel{=}\Varid{sum}\;[\mskip1.5mu \Varid{fst}\;\Varid{dp}\mathbin{*}\Varid{snd}\;\Varid{dp}\mid \Varid{dp}\leftarrow \Varid{dps}\mskip1.5mu]{}\<[E]%
\ColumnHook
\end{hscode}\resethooks
With \ensuremath{\Varid{meas}\mathrel{=}\Varid{expected}} and \ensuremath{ \mathbin{\oplus} \mathrel{=}\mathbin{+}}, we can now consider the three conditions from above.

\paragraph*{Condition 1.} \hspace{0.1cm}
The first condition \ensuremath{\Varid{measPureSpec}} holds since \ensuremath{\mathrm{1.0}}
is neutral for \ensuremath{\mathbin{*}}.
\begin{hscode}\SaveRestoreHook
\column{B}{@{}>{\hspre}l<{\hspost}@{}}%
\column{3}{@{}>{\hspre}l<{\hspost}@{}}%
\column{E}{@{}>{\hspre}l<{\hspost}@{}}%
\>[3]{}\Varid{expected}\mathbin{\circ}\Varid{distPure}\;\Varid{a}\mathrel{=}\Varid{a}\mathbin{*}\mathrm{1.0}\mathrel{=}\Varid{a}{}\<[E]%
\ColumnHook
\end{hscode}\resethooks
The second and the third condition require some arithmetic reasoning, so
let us just consider them for two examples. 
Let \ensuremath{\Varid{a},\Varid{b},\Varid{c},\Varid{d}} be variables of type \ensuremath{\Conid{Double}} and say we have distributions

\begin{hscode}\SaveRestoreHook
\column{B}{@{}>{\hspre}l<{\hspost}@{}}%
\column{3}{@{}>{\hspre}l<{\hspost}@{}}%
\column{9}{@{}>{\hspre}c<{\hspost}@{}}%
\column{9E}{@{}l@{}}%
\column{12}{@{}>{\hspre}l<{\hspost}@{}}%
\column{E}{@{}>{\hspre}l<{\hspost}@{}}%
\>[3]{}\Varid{dps1}{}\<[9]%
\>[9]{} \mathop{:} {}\<[9E]%
\>[12]{}\Conid{Dist}\;\Conid{Double}{}\<[E]%
\\
\>[3]{}\Varid{dps1}{}\<[9]%
\>[9]{}\mathrel{=}{}\<[9E]%
\>[12]{}[\mskip1.5mu (\Varid{a},\mathrm{0.5}),(\Varid{b},\mathrm{0.3}),(\Varid{c},\mathrm{0.2})\mskip1.5mu]{}\<[E]%
\ColumnHook
\end{hscode}\resethooks
\begin{hscode}\SaveRestoreHook
\column{B}{@{}>{\hspre}l<{\hspost}@{}}%
\column{3}{@{}>{\hspre}l<{\hspost}@{}}%
\column{9}{@{}>{\hspre}c<{\hspost}@{}}%
\column{9E}{@{}l@{}}%
\column{12}{@{}>{\hspre}l<{\hspost}@{}}%
\column{E}{@{}>{\hspre}l<{\hspost}@{}}%
\>[3]{}\Varid{dps2}{}\<[9]%
\>[9]{} \mathop{:} {}\<[9E]%
\>[12]{}\Conid{Dist}\;\Conid{Double}{}\<[E]%
\\
\>[3]{}\Varid{dps2}{}\<[9]%
\>[9]{}\mathrel{=}{}\<[9E]%
\>[12]{}[\mskip1.5mu (\Varid{a},\mathrm{0.4}),(\Varid{d},\mathrm{0.6})\mskip1.5mu]{}\<[E]%
\ColumnHook
\end{hscode}\resethooks
\begin{hscode}\SaveRestoreHook
\column{B}{@{}>{\hspre}l<{\hspost}@{}}%
\column{3}{@{}>{\hspre}l<{\hspost}@{}}%
\column{10}{@{}>{\hspre}l<{\hspost}@{}}%
\column{13}{@{}>{\hspre}l<{\hspost}@{}}%
\column{E}{@{}>{\hspre}l<{\hspost}@{}}%
\>[3]{}\Varid{dpdps}{}\<[10]%
\>[10]{} \mathop{:} {}\<[13]%
\>[13]{}\Conid{Dist}\;(\Conid{Dist}\;\Conid{Double}){}\<[E]%
\\
\>[3]{}\Varid{dpdps}{}\<[10]%
\>[10]{}\mathrel{=}[\mskip1.5mu (\Varid{dps1},\mathrm{0.1}),(\Varid{dps2},\mathrm{0.9})\mskip1.5mu]{}\<[E]%
\ColumnHook
\end{hscode}\resethooks

\paragraph*{Condition 2.} \hspace{0.1cm}
Then the second condition \ensuremath{\Varid{measJoinSpec}} instantiates to
\begin{hscode}\SaveRestoreHook
\column{B}{@{}>{\hspre}l<{\hspost}@{}}%
\column{3}{@{}>{\hspre}l<{\hspost}@{}}%
\column{E}{@{}>{\hspre}l<{\hspost}@{}}%
\>[3]{}(\Varid{expected}\mathbin{\circ}\Varid{distJoin})\;\Varid{dpdps}\mathrel{=}(\Varid{expected}\mathbin{\circ}\Varid{distMap}\;\Varid{expected})\;\Varid{dpdps}{}\<[E]%
\ColumnHook
\end{hscode}\resethooks
This equality holds because of the standard properties of
addition and multiplication:
\begin{hscode}\SaveRestoreHook
\column{B}{@{}>{\hspre}l<{\hspost}@{}}%
\column{3}{@{}>{\hspre}l<{\hspost}@{}}%
\column{84}{@{}>{\hspre}c<{\hspost}@{}}%
\column{84E}{@{}l@{}}%
\column{94}{@{}>{\hspre}c<{\hspost}@{}}%
\column{94E}{@{}l@{}}%
\column{E}{@{}>{\hspre}l<{\hspost}@{}}%
\>[3]{}(\Varid{expected}\mathbin{\circ}\Varid{distJoin})\;\Varid{dpdps}{}\<[94]%
\>[94]{}\mathrel{=}{}\<[94E]%
\\[\blanklineskip]%
\>[3]{}\Varid{expected}\;[\mskip1.5mu (\Varid{a},\mathrm{0.5}\mathbin{*}\mathrm{0.1}),(\Varid{b},\mathrm{0.3}\mathbin{*}\mathrm{0.1}),(\Varid{c},\mathrm{0.2}\mathbin{*}\mathrm{0.1}),(\Varid{a},\mathrm{0.4}\mathbin{*}\mathrm{0.9}),(\Varid{d},\mathrm{0.6}\mathbin{*}\mathrm{0.9})\mskip1.5mu]{}\<[94]%
\>[94]{}\mathrel{=}{}\<[94E]%
\\[\blanklineskip]%
\>[3]{}(\Varid{a}\mathbin{*}\mathrm{0.5}\mathbin{*}\mathrm{0.1})\mathbin{+}(\Varid{b}\mathbin{*}\mathrm{0.3}\mathbin{*}\mathrm{0.1})\mathbin{+}(\Varid{c}\mathbin{*}\mathrm{0.2}\mathbin{*}\mathrm{0.1})\mathbin{+}(\Varid{a}\mathbin{*}\mathrm{0.4}\mathbin{*}\mathrm{0.9})\mathbin{+}(\Varid{d}\mathbin{*}\mathrm{0.6}\mathbin{*}\mathrm{0.9}){}\<[94]%
\>[94]{}\mathrel{=}{}\<[94E]%
\\[\blanklineskip]%
\>[3]{}((\Varid{a}\mathbin{*}\mathrm{0.5}\mathbin{+}\Varid{b}\mathbin{*}\mathrm{0.3}\mathbin{+}\Varid{c}\mathbin{*}\mathrm{0.2})\mathbin{*}\mathrm{0.1}\mathbin{+}(\Varid{a}\mathbin{*}\mathrm{0.4}\mathbin{+}\Varid{d}\mathbin{*}\mathrm{0.6})\mathbin{*}\mathrm{0.9}{}\<[84]%
\>[84]{}\mathrel{=}{}\<[84E]%
\\[\blanklineskip]%
\>[3]{}\Varid{expected}\;[\mskip1.5mu (\Varid{a}\mathbin{*}\mathrm{0.5}\mathbin{+}\Varid{b}\mathbin{*}\mathrm{0.3}\mathbin{+}\Varid{c}\mathbin{*}\mathrm{0.2},\mathrm{0.1}),(\Varid{a}\mathbin{*}\mathrm{0.4}\mathbin{+}\Varid{d}\mathbin{*}\mathrm{0.6},\mathrm{0.9})\mskip1.5mu]{}\<[84]%
\>[84]{}\mathrel{=}{}\<[84E]%
\\[\blanklineskip]%
\>[3]{}(\Varid{expected}\mathbin{\circ}\Varid{distMap}\;\Varid{expected})\;\Varid{dpdps}{}\<[E]%
\ColumnHook
\end{hscode}\resethooks

\paragraph*{Condition 3.} \hspace{0.1cm}
For the third condition \ensuremath{\Varid{measPlusSpec}}, consider for some \ensuremath{\Varid{v} \mathop{:} \Conid{Double}}
the equation
\begin{hscode}\SaveRestoreHook
\column{B}{@{}>{\hspre}l<{\hspost}@{}}%
\column{3}{@{}>{\hspre}l<{\hspost}@{}}%
\column{E}{@{}>{\hspre}l<{\hspost}@{}}%
\>[3]{}(\Varid{expected}\mathbin{\circ}\Varid{distMap}\;(\Varid{v}\mathbin{+}))\;\Varid{dps1}\mathrel{=}((\Varid{v}\mathbin{+})\mathbin{\circ}\Varid{expected})\;\Varid{dps1}{}\<[E]%
\ColumnHook
\end{hscode}\resethooks
Again using the usual arithmetic laws for \ensuremath{\mathbin{+}} and \ensuremath{\mathbin{*}}, we can calculate
\begin{hscode}\SaveRestoreHook
\column{B}{@{}>{\hspre}l<{\hspost}@{}}%
\column{3}{@{}>{\hspre}l<{\hspost}@{}}%
\column{66}{@{}>{\hspre}c<{\hspost}@{}}%
\column{66E}{@{}l@{}}%
\column{70}{@{}>{\hspre}c<{\hspost}@{}}%
\column{70E}{@{}l@{}}%
\column{E}{@{}>{\hspre}l<{\hspost}@{}}%
\>[3]{}\Varid{expected}\;(\Varid{distMap}\;(\Varid{v}\mathbin{+})\;[\mskip1.5mu (\Varid{a},\mathrm{0.5}),(\Varid{b},\mathrm{0.3}),(\Varid{c},\mathrm{0.2})\mskip1.5mu]){}\<[70]%
\>[70]{}\mathrel{=}{}\<[70E]%
\\[\blanklineskip]%
\>[3]{}(\Varid{v}\mathbin{+}\Varid{a})\mathbin{*}\mathrm{0.5}\mathbin{+}(\Varid{v}\mathbin{+}\Varid{b})\mathbin{*}\mathrm{0.3}\mathbin{+}(\Varid{v}\mathbin{+}\Varid{c})\mathbin{*}\mathrm{0.2}{}\<[70]%
\>[70]{}\mathrel{=}{}\<[70E]%
\\[\blanklineskip]%
\>[3]{}(\Varid{v}\mathbin{*}\mathrm{0.5}\mathbin{+}\Varid{a}\mathbin{*}\mathrm{0.5})\mathbin{+}(\Varid{v}\mathbin{*}\mathrm{0.3}\mathbin{+}\Varid{b}\mathbin{*}\mathrm{0.3})\mathbin{+}(\Varid{v}\mathbin{*}\mathrm{0.2}\mathbin{+}\Varid{c}\mathbin{*}\mathrm{0.2}){}\<[70]%
\>[70]{}\mathrel{=}{}\<[70E]%
\\[\blanklineskip]%
\>[3]{}(\Varid{v}\mathbin{*}\mathrm{0.5}\mathbin{+}\Varid{v}\mathbin{*}\mathrm{0.3}\mathbin{+}\Varid{v}\mathbin{*}\mathrm{0.2})\mathbin{+}(\Varid{a}\mathbin{*}\mathrm{0.5}\mathbin{+}\Varid{b}\mathbin{*}\mathrm{0.3}\mathbin{+}\Varid{c}\mathbin{*}\mathrm{0.2}){}\<[66]%
\>[66]{}\mathrel{=}{}\<[66E]%
\\[\blanklineskip]%
\>[3]{}\Varid{v}\mathbin{+}\Varid{expected}\;[\mskip1.5mu (\Varid{a},\mathrm{0.5}),(\Varid{b},\mathrm{0.3}),(\Varid{c},\mathrm{0.2})\mskip1.5mu]{}\<[E]%
\ColumnHook
\end{hscode}\resethooks
As we can see, an essential ingredient for the equality to hold is that
the mapped occurrences of
\ensuremath{(\Varid{v}\mathbin{+})} are weighted by the probabilities which add up to 1.

\vspace{0.2cm}
Note that in this example, we have glossed over problems that might arise
from the use of \ensuremath{\Conid{Dist}} to represent probability distributions.
\footnote{For the sake of simplicity,
  we do not address (important)
  conceptional questions concerning the representation
  of probability distributions or the problems caused by the use of
  floating point arithmetic in this example.
  Note however, that the chosen type \ensuremath{\Conid{Prob}} does e.g. neither
  enforce that the probabilities lie in the interval \ensuremath{[\mskip1.5mu \mathrm{0},\mathrm{1}\mskip1.5mu]} nor
  that they add up to \ensuremath{\mathrm{1}}. These properties would however
  be crucial for actual proofs.}
We will briefly address probability monads and the expected value
from a more abstract perspective in Subsection~\ref{subsection:impactMeas}.

\subsection{Examples and counter-examples}
\label{subsection:exAndCounterEx}

Besides the motivating example above, let us now consider some more
functions that have the correct type to serve as a measure, 
and that do or do not fulfil the three conditions.

Simple examples of admissible measures are the minimum (\ensuremath{\Varid{minList}} as
defined in Fig.~\ref{fig:example1Formal}) or maximum (\ensuremath{\Varid{maxList}\mathrel{=}\Varid{foldr}\mathbin{`\Varid{maximum}`}\mathrm{0}}) of a list for \ensuremath{\Conid{M}\mathrel{=}\Conid{List}} with \ensuremath{\mathbb{N}} as type of values
and ordinary addition as \ensuremath{ \mathbin{\oplus} }. It is straightforward to prove that
the conditions hold for these two measures and the proofs for
\ensuremath{\Varid{maxList}} are included in the supplementary material.

The function \ensuremath{\Varid{length}} is a very simple counter-example:
It has the right type for a list measure but fails all three
of the conditions.
As to other counter-examples, let us revisit the conditions one by one.
Throughout, we use \ensuremath{\Conid{M}\mathrel{=}\Conid{List}} with \ensuremath{\Varid{map}\mathrel{=}\Varid{listMap}}, \ensuremath{\Varid{join}\mathrel{=}\Varid{concat}}
and \ensuremath{ \mathbin{\oplus} \mathrel{=}\mathbin{+}}
(the canonical addition for the respective type of \ensuremath{\Conid{Val}}). 

\paragraph*{Condition~1.} \hspace{0.1cm}
We remain in the setting of
Ex.~\ref{subsection:example1SDPs} with \ensuremath{\Conid{Val}\mathrel{=}\mathbb{N}},
and just vary the measure. Using a somewhat contrived
variation of \ensuremath{\Varid{maxList}}
\begin{hscode}\SaveRestoreHook
\column{B}{@{}>{\hspre}l<{\hspost}@{}}%
\column{3}{@{}>{\hspre}l<{\hspost}@{}}%
\column{15}{@{}>{\hspre}c<{\hspost}@{}}%
\column{15E}{@{}l@{}}%
\column{18}{@{}>{\hspre}l<{\hspost}@{}}%
\column{E}{@{}>{\hspre}l<{\hspost}@{}}%
\>[3]{}\Varid{maxListVar}{}\<[15]%
\>[15]{} \mathop{:} {}\<[15E]%
\>[18]{}\Conid{List}\;\mathbb{N} \to \mathbb{N}{}\<[E]%
\\
\>[3]{}\Varid{maxListVar}{}\<[15]%
\>[15]{}\mathrel{=}{}\<[15E]%
\>[18]{}\Varid{foldr}\;(\lambda \Varid{x},\Varid{v}\Rightarrow (\Varid{x}\mathbin{+}\mathrm{1}\mathbin{`\Varid{maximum}`}\Varid{v}))\;\mathrm{0}{}\<[E]%
\ColumnHook
\end{hscode}\resethooks
with \ensuremath{\Varid{meas}\mathrel{=}\Varid{maxListVar}} it suffices to consider
that for an arbitrary \ensuremath{\Varid{n} \mathop{:} \mathbb{N}}
\begin{hscode}\SaveRestoreHook
\column{B}{@{}>{\hspre}l<{\hspost}@{}}%
\column{3}{@{}>{\hspre}l<{\hspost}@{}}%
\column{E}{@{}>{\hspre}l<{\hspost}@{}}%
\>[3]{}(\Varid{maxListVar}\mathbin{\circ}\Varid{pure})\;\Varid{n}\mathrel{=}\Varid{maxListVar}\;[\mskip1.5mu \Varid{n}\mskip1.5mu]\mathrel{=}(\Varid{n}\mathbin{+}\mathrm{1})\mathbin{`\Varid{maximum}`}\mathrm{0}\mathrel{=}\Varid{n}\mathbin{+}\mathrm{1}\neq\Varid{n}\mathrel{=}\Varid{id}\;\Varid{n}{}\<[E]%
\ColumnHook
\end{hscode}\resethooks
to see that now the condition \ensuremath{\Varid{measPureSpec}} fails.

\paragraph*{Condition~2.} \hspace{0.1cm}
To exhibit a measure that fails the condition \ensuremath{\Varid{measJoinSpec}}, we
switch to \ensuremath{\Conid{Val}\mathrel{=}\Conid{Double}} and use the arithmetic average
\begin{hscode}\SaveRestoreHook
\column{B}{@{}>{\hspre}l<{\hspost}@{}}%
\column{3}{@{}>{\hspre}l<{\hspost}@{}}%
\column{8}{@{}>{\hspre}c<{\hspost}@{}}%
\column{8E}{@{}l@{}}%
\column{11}{@{}>{\hspre}l<{\hspost}@{}}%
\column{14}{@{}>{\hspre}l<{\hspost}@{}}%
\column{E}{@{}>{\hspre}l<{\hspost}@{}}%
\>[3]{}\Varid{avg}{}\<[8]%
\>[8]{} \mathop{:} {}\<[8E]%
\>[11]{}\Conid{List}\;\Conid{Double} \to \Conid{Double}{}\<[E]%
\\
\>[3]{}\Varid{avg}\;[\mskip1.5mu \mskip1.5mu]{}\<[11]%
\>[11]{}\mathrel{=}{}\<[14]%
\>[14]{}\mathrm{0.0}{}\<[E]%
\\
\>[3]{}\Varid{avg}\;\Varid{ds}{}\<[11]%
\>[11]{}\mathrel{=}\Varid{sum}\;\Varid{ds}\mathbin{/}\Varid{cast}\;(\Varid{length}\;\Varid{ds}){}\<[E]%
\ColumnHook
\end{hscode}\resethooks
as measure \ensuremath{\Varid{meas}\mathrel{=}\Varid{avg}}. Taking a list of lists of different lengths
like [[1], [2, 3]] we have
\begin{hscode}\SaveRestoreHook
\column{B}{@{}>{\hspre}l<{\hspost}@{}}%
\column{3}{@{}>{\hspre}l<{\hspost}@{}}%
\column{39}{@{}>{\hspre}c<{\hspost}@{}}%
\column{39E}{@{}l@{}}%
\column{43}{@{}>{\hspre}l<{\hspost}@{}}%
\column{58}{@{}>{\hspre}l<{\hspost}@{}}%
\column{E}{@{}>{\hspre}l<{\hspost}@{}}%
\>[3]{}\Varid{avg}\;(\Varid{concat}\;[\mskip1.5mu [\mskip1.5mu \mathrm{1}\mskip1.5mu],[\mskip1.5mu \mathrm{2},\mathrm{3}\mskip1.5mu]\mskip1.5mu]){}\<[39]%
\>[39]{}\mathrel{=}{}\<[39E]%
\>[43]{}\Varid{avg}\;[\mskip1.5mu \mathrm{1},\mathrm{2},\mathrm{3}\mskip1.5mu]{}\<[58]%
\>[58]{}\mathrel{=}\mathrm{2}{}\<[E]%
\\
\>[39]{}\neq{}\<[39E]%
\\
\>[3]{}\Varid{avg}\;(\Varid{listMap}\;\Varid{avg}\;[\mskip1.5mu [\mskip1.5mu \mathrm{1}\mskip1.5mu],[\mskip1.5mu \mathrm{2},\mathrm{3}\mskip1.5mu]\mskip1.5mu]){}\<[39]%
\>[39]{}\mathrel{=}{}\<[39E]%
\>[43]{}\Varid{avg}\;[\mskip1.5mu \mathrm{1},\mathrm{2.5}\mskip1.5mu]{}\<[58]%
\>[58]{}\mathrel{=}\mathrm{1.75}{}\<[E]%
\ColumnHook
\end{hscode}\resethooks
\paragraph*{Condition~3.} \hspace{0.1cm}
Let again \ensuremath{\Conid{Val}\mathrel{=}\mathbb{N}} to take another look at our counter-example
from the last section with \ensuremath{\Varid{meas}\mathrel{=}\Varid{sum}}, the arithmetic sum of a list.
It does fulfil \ensuremath{\Varid{measPureSpec}} and \ensuremath{\Varid{measJoinSpec}}, the first by definition,
the second by structural induction using the associativity of \ensuremath{\mathbin{+}}.
But it fails to fulfil
\ensuremath{\Varid{measPlusSpec}}. If the list has the form \ensuremath{\Varid{a}\mathbin{::}\Varid{as}}, we would have to
show the following equality for \ensuremath{\Varid{measPlusSpec}} to hold:
\begin{hscode}\SaveRestoreHook
\column{B}{@{}>{\hspre}l<{\hspost}@{}}%
\column{3}{@{}>{\hspre}l<{\hspost}@{}}%
\column{E}{@{}>{\hspre}l<{\hspost}@{}}%
\>[3]{}(\Varid{sum}\mathbin{\circ}\Varid{listMap}\;(\Varid{v}\mathbin{+}))\;(\Varid{a}\mathbin{::}\Varid{as})\mathrel{=}((\Varid{v}\mathbin{+})\mathbin{\circ}\Varid{sum})\;(\Varid{a}\mathbin{::}\Varid{as}){}\<[E]%
\ColumnHook
\end{hscode}\resethooks
Clearly, if \ensuremath{\Varid{v}\neq\mathrm{0}} and \ensuremath{\Varid{as}\neq[\mskip1.5mu \mskip1.5mu]} this equality cannot hold.
This is why in the last section the equality of \ensuremath{\Varid{val}} and \ensuremath{\Varid{val'}}
failed for \ensuremath{\Varid{meas}\mathrel{=}\Varid{sum}}.\\
A similar failure would arise if we chose \ensuremath{\Varid{meas}\mathrel{=}\Varid{foldr}\;(\mathbin{*})\;\mathrm{1}} instead,
as \ensuremath{\mathbin{+}} does not distribute over \ensuremath{\mathbin{*}}. But if we turned the situation
around by setting \ensuremath{ \mathbin{\oplus} \mathrel{=}\mathbin{*}} and \ensuremath{\Varid{meas}\mathrel{=}\Varid{sum}}, the condition
\ensuremath{\Varid{measPlusSpec}} would hold thanks to the usual arithmetic
distributivity law for \ensuremath{\mathbin{*}} over \ensuremath{\mathbin{+}}.

\vspace{0.2cm}
All of the measures considered in this subsection do fulfil the
\ensuremath{\Varid{measMonSpec}} condition imposed by the BJI-theory. This raises the
question how previously admissible measures are impacted by adding the
three new conditions to the framework.

\subsection{Impact on previously admissible measures}
\label{subsection:impactMeas}
As we have seen in Sec.~\ref{subsection:solution_components}, the
BJI-framework already requires measures to fulfil the monotonicity
condition
\begin{hscode}\SaveRestoreHook
\column{B}{@{}>{\hspre}l<{\hspost}@{}}%
\column{3}{@{}>{\hspre}l<{\hspost}@{}}%
\column{16}{@{}>{\hspre}c<{\hspost}@{}}%
\column{16E}{@{}l@{}}%
\column{19}{@{}>{\hspre}l<{\hspost}@{}}%
\column{E}{@{}>{\hspre}l<{\hspost}@{}}%
\>[3]{}\Varid{measMonSpec}{}\<[16]%
\>[16]{} \mathop{:} {}\<[16E]%
\>[19]{}\{\mskip1.5mu \Conid{A} \mathop{:} \Conid{Type}\mskip1.5mu\} \to (\Varid{f},\Varid{g} \mathop{:} \Conid{A} \to \Conid{Val}) \to ((\Varid{a} \mathop{:} \Conid{A}) \to (\Varid{f}\;\Varid{a}) \,\sqsubseteq\, (\Varid{g}\;\Varid{a})) \to {}\<[E]%
\\
\>[19]{}(\Varid{ma} \mathop{:} \Conid{M}\;\Conid{A}) \to \Varid{meas}\;(\Varid{map}\;\Varid{f}\;\Varid{ma}) \,\sqsubseteq\, \Varid{meas}\;(\Varid{map}\;\Varid{g}\;\Varid{ma}){}\<[E]%
\ColumnHook
\end{hscode}\resethooks
\bottaetal show that the arithmetic average (for \ensuremath{\Conid{M}\mathrel{=}\Conid{List}}), the
worst-case measure (for \ensuremath{\Conid{M}\mathrel{=}\Conid{List}} and for a probability monad \ensuremath{\Conid{M}\mathrel{=}\Conid{Prob}})
and the expected value measure (for \ensuremath{\Conid{M}\mathrel{=}\Conid{Prob}}) all fulfil \ensuremath{\Varid{measMonSpec}}.
Thus, a natural question is whether these measures also fulfil the three
additional requirements.

\paragraph*{Expected value for probability distributions.} \hspace{0.1cm}
As already discussed, most applications of backward induction concern
stochastic SDPs where possible rewards are aggregated using the
expected value measure from probability theory, commonly denoted as \ensuremath{\Conid{E}}.

Essentially, for a numerical type \ensuremath{\Conid{Q}}, the
expected value of a probability distribution on \ensuremath{\Conid{Q}} is
\begin{hscode}\SaveRestoreHook
\column{B}{@{}>{\hspre}l<{\hspost}@{}}%
\column{3}{@{}>{\hspre}l<{\hspost}@{}}%
\column{6}{@{}>{\hspre}c<{\hspost}@{}}%
\column{6E}{@{}l@{}}%
\column{9}{@{}>{\hspre}l<{\hspost}@{}}%
\column{E}{@{}>{\hspre}l<{\hspost}@{}}%
\>[3]{}\Conid{E}{}\<[6]%
\>[6]{} \mathop{:} {}\<[6E]%
\>[9]{}\Conid{Num}\;\Conid{Q}\Rightarrow \Conid{Prob}\;\Conid{Q} \to \Conid{Q}{}\<[E]%
\\
\>[3]{}\Conid{E}\;\Varid{pq}\mathrel{=}\Varid{sum}\;[\mskip1.5mu \Varid{q}\mathbin{*}\Varid{prob}\;\Varid{pq}\;\Varid{q}\mid \Varid{q}\leftarrow \Varid{supp}\;\Varid{pq}\mskip1.5mu]{}\<[E]%
\ColumnHook
\end{hscode}\resethooks
where \ensuremath{\Varid{prob}} and \ensuremath{\Varid{supp}} are generic functions that encode the notions of
\emph{probability} and of \emph{support} associated with a finite
probability distribution:
\begin{hscode}\SaveRestoreHook
\column{B}{@{}>{\hspre}l<{\hspost}@{}}%
\column{3}{@{}>{\hspre}l<{\hspost}@{}}%
\column{E}{@{}>{\hspre}l<{\hspost}@{}}%
\>[3]{}\Varid{prob} \mathop{:} \{\mskip1.5mu \Conid{A} \mathop{:} \Conid{Type}\mskip1.5mu\} \to \Conid{Prob}\;\Conid{A} \to \Conid{A} \to \Conid{Q}{}\<[E]%
\\[\blanklineskip]%
\>[3]{}\Varid{supp} \mathop{:} \{\mskip1.5mu \Conid{A} \mathop{:} \Conid{Type}\mskip1.5mu\} \to \Conid{Prob}\;\Conid{A} \to \Conid{List}\;\Conid{A}{}\<[E]%
\ColumnHook
\end{hscode}\resethooks
For \ensuremath{\Varid{pa}} and \ensuremath{\Varid{a}} of suitable types, \ensuremath{\Varid{prob}\;\Varid{pa}\;\Varid{a}} represents the
probability of \ensuremath{\Varid{a}} according to \ensuremath{\Varid{pa}}. Similarly, \ensuremath{\Varid{supp}\;\Varid{pa}} returns a
list of those values whose probability is not zero in \ensuremath{\Varid{pa}}.
The probability function \ensuremath{\Varid{prob}} has to fulfil the axioms of
probability theory. In particular,
\begin{hscode}\SaveRestoreHook
\column{B}{@{}>{\hspre}l<{\hspost}@{}}%
\column{3}{@{}>{\hspre}l<{\hspost}@{}}%
\column{E}{@{}>{\hspre}l<{\hspost}@{}}%
\>[3]{}\Varid{sum}\;[\mskip1.5mu \Varid{prob}\;\Varid{pa}\;\Varid{a}\mid \Varid{a}\leftarrow \Varid{supp}\;\Varid{pa}\mskip1.5mu]\mathrel{=}\mathrm{1}{}\<[E]%
\ColumnHook
\end{hscode}\resethooks
This condition implies that probability distributions cannot be empty, a
precondition of \ensuremath{\Varid{measPlusSpec}}. Putting forward minimal specifications
for \ensuremath{\Varid{prob}} and \ensuremath{\Varid{supp}} is not completely trivial but if the \ensuremath{\mathbin{+}}-operation
associated with \ensuremath{\Conid{Q}} is commutative and associative, if \ensuremath{\mathbin{*}}
distributes over \ensuremath{\mathbin{+}} and if the \ensuremath{\Varid{map}} and \ensuremath{\Varid{join}} associated with
\ensuremath{\Conid{Prob}} -- for \ensuremath{\Varid{f}}, \ensuremath{\Varid{a}}, \ensuremath{\Varid{b}}, \ensuremath{\Varid{pa}} and \ensuremath{\Varid{ppa}} of suitable types --
fulfil the conservation law
\begin{hscode}\SaveRestoreHook
\column{B}{@{}>{\hspre}l<{\hspost}@{}}%
\column{3}{@{}>{\hspre}l<{\hspost}@{}}%
\column{E}{@{}>{\hspre}l<{\hspost}@{}}%
\>[3]{}\Varid{prob}\;(\Varid{map}\;\Varid{f}\;\Varid{pa})\;\Varid{b}\mathrel{=}\Varid{sum}\;[\mskip1.5mu \Varid{prob}\;\Varid{pa}\;\Varid{a}\mid \Varid{a}\leftarrow \Varid{supp}\;\Varid{pa},\Varid{f}\;\Varid{a}\doubleequals\Varid{b}\mskip1.5mu]{}\<[E]%
\ColumnHook
\end{hscode}\resethooks
and the total probability law
\begin{hscode}\SaveRestoreHook
\column{B}{@{}>{\hspre}l<{\hspost}@{}}%
\column{3}{@{}>{\hspre}l<{\hspost}@{}}%
\column{E}{@{}>{\hspre}l<{\hspost}@{}}%
\>[3]{}\Varid{prob}\;(\Varid{join}\;\Varid{ppa})\;\Varid{a}\mathrel{=}\Varid{sum}\;[\mskip1.5mu \Varid{prob}\;\Varid{pa}\;\Varid{a}\mathbin{*}\Varid{prob}\;\Varid{ppa}\;\Varid{pa}\mid \Varid{pa}\leftarrow \Varid{supp}\;\Varid{ppa}\mskip1.5mu]{}\<[E]%
\ColumnHook
\end{hscode}\resethooks
 then the expected value fulfils \ensuremath{\Varid{measPureSpec}}, \ensuremath{\Varid{measJoinSpec}} and
 \ensuremath{\Varid{measPlusSpec}}.
 This is not surprising since -- as stated above -- this has been the
 guiding example for the generalisation to monadic SDPs and the formulation
 of the three conditions.
 
\paragraph*{Average and arithmetic sum.} \hspace{0.1cm}
As can already be concluded
from the corresponding counter-examples in the previous subsection,
neither the plain arithmetic average nor the arithmetic sum are
suited as measure when using the standard monad structure on
\ensuremath{\Conid{List}} to represent non-deterministic
uncertainty. We think this is an important observation, as the
average seems innocent enough to come to mind as a simple way
to represent uniformly distributed outcomes:
\emph{``The probability of each element can simply be inferred from the length of the list 
-- so why bother to explicitly deal with probabilities?''} 
Although our counter-example shows that this idea is flawed, the intuition
behind it can be employed to define an alternative, but less general monad
structure on lists by incorporating the averaging operation into the joining
of lists (i.e. by choosing \ensuremath{\Varid{join}\mathrel{=}\Varid{map}\;\Varid{avg}}).
However, this only makes sense for types that are instances of the \ensuremath{\Conid{Num}}
and \ensuremath{\Conid{Fractional}} type classes, and naturality only holds for a restricted class of
functions (namely additive functions). As a consequence, this alternative structure
does not seem particularly useful for our current purpose either.

\paragraph*{Worst-case measures.} \hspace{0.1cm}
In many important applications in
climate impact research but also in portfolio management and sports,
decisions are taken as to minimise the consequences of worst case
outcomes. Depending on how ``worse'' is defined, the corresponding
measures might pick the maximum or
minimum from an \ensuremath{\Conid{M}}-structure of values. In the previous subsection we
considered an example in which the monad was \ensuremath{\Conid{List}}, the operation
\ensuremath{ \mathbin{\oplus} } plain addition together with either \ensuremath{\Varid{maxList}} or \ensuremath{\Varid{minList}} as
measure. And indeed we can prove that for both measures the three
requirements hold (the proofs for \ensuremath{\Varid{maxList}} can be found in the
supplementary material). This gives us a useful notion of worst-case
measure that is admissible for monadic backward induction.

\vspace{0.2cm}
We can thus conclude that the new requirements hold for certain
familiar measures, but that they also rule out
  certain instances that were considered admissible in the BJI-framework.
Given the three conditions \ensuremath{\Varid{measPureSpec}}, \ensuremath{\Varid{measJoinSpec}},
\ensuremath{\Varid{measPlusSpec}} hold, we can prove the
extensional equality of the functions \ensuremath{\Varid{val}} and \ensuremath{\Varid{val'}} generically.
This is what we will do in the next section.



\section{Correctness Proofs}
\label{section:valval}

In this section we show that \ensuremath{\Varid{val}}
(Sec.~\ref{subsection:solution_components}) and
\ensuremath{\Varid{val'}} (Sec.~\ref{section:preparation}) are extensionally equal
\begin{hscode}\SaveRestoreHook
\column{B}{@{}>{\hspre}l<{\hspost}@{}}%
\column{3}{@{}>{\hspre}l<{\hspost}@{}}%
\column{23}{@{}>{\hspre}c<{\hspost}@{}}%
\column{23E}{@{}l@{}}%
\column{26}{@{}>{\hspre}l<{\hspost}@{}}%
\column{E}{@{}>{\hspre}l<{\hspost}@{}}%
\>[3]{}\Varid{valMeasTotalReward}{}\<[23]%
\>[23]{} \mathop{:} {}\<[23E]%
\>[26]{}\{\mskip1.5mu \Varid{t},\Varid{n} \mathop{:} \mathbb{N}\mskip1.5mu\} \to (\Varid{ps} \mathop{:} \Conid{PolicySeq}\;\Varid{t}\;\Varid{n}) \to (\Varid{x} \mathop{:} \Conid{X}\;\Varid{t}) \to \Varid{val'}\;\Varid{ps}\;\Varid{x}\mathrel{=}\Varid{val}\;\Varid{ps}\;\Varid{x}{}\<[E]%
\ColumnHook
\end{hscode}\resethooks
given the three conditions from the previous section hold. As a
corollary we then obtain our correctness result for monadic backward
induction.

We can understand the proof of \ensuremath{\Varid{valMeasTotalReward}} as an optimising
program transformation from the less efficient but ``obviously
correct'' implementation \ensuremath{\Varid{val'}} to the more efficient implementation
\ensuremath{\Varid{val}}. Therefore the equational reasoning proofs in this section will
proceed from \ensuremath{\Varid{val'}} to \ensuremath{\Varid{val}}. In Sec.~\ref{section:conditions} we
have stated sufficient conditions for this transformation to be
possible: \ensuremath{\Varid{measPureSpec}}, \ensuremath{\Varid{measJoinSpec}}, \ensuremath{\Varid{measPlusSpec}}. We also have
seen the different computational patterns that the two implementations
exhibit: While \ensuremath{\Varid{val'}} first computes all possible trajectories for the
given policy sequence and initial state, then computes their
individual sum of rewards and finally applies the measure once, \ensuremath{\Varid{val}}
computes its final result by adding the current reward to an
intermediate outcome and applying the measure locally at each decision
step. This suggests that a transformation from \ensuremath{\Varid{val'}} to \ensuremath{\Varid{val}} will
essentially have to push the application of the measure into the
recursive computation of the sum of rewards. The proof will be carried
out by induction on the structure of policy sequences.

\subsection{Deterministic Case}
\label{subsection:detCase}
To get a first intuition, let's have a look at what
the induction step looks like in the deterministic case
(i.e. if we fix monad and measure to be identities):

{\linespread{1.5}\begin{hscode}\SaveRestoreHook
\column{B}{@{}>{\hspre}l<{\hspost}@{}}%
\column{3}{@{}>{\hspre}l<{\hspost}@{}}%
\column{7}{@{}>{\hspre}l<{\hspost}@{}}%
\column{48}{@{}>{\hspre}l<{\hspost}@{}}%
\column{E}{@{}>{\hspre}l<{\hspost}@{}}%
\>[3]{}\Varid{valMeasTotalReward}\;(\Varid{p}\mathbin{::}\Varid{ps})\;\Varid{x}\mathrel{=}{}\<[E]%
\\
\>[3]{}\hsindent{4}{}\<[7]%
\>[7]{}(\Varid{val'}\;(\Varid{p}\mathbin{::}\Varid{ps})\;\Varid{x}){}\<[48]%
\>[48]{}=\hspace{-3pt}\{\; \text{by definition of }\;\Varid{val'}\;\}\hspace{-3pt}={}\<[E]%
\\
\>[3]{}\hsindent{4}{}\<[7]%
\>[7]{}(\Varid{sumR}\;((\Varid{x}  \mathbin{*\!*} \Varid{y})  \mathbin{\#\!\#} \Varid{trj}\;\Varid{ps}\;\Varid{x'})){}\<[48]%
\>[48]{}=\hspace{-3pt}\{\; \text{by definition of }\;\Varid{sumR}\;\}\hspace{-3pt}={}\<[E]%
\\
\>[3]{}\hsindent{4}{}\<[7]%
\>[7]{}(\Varid{r}\;(\Varid{head}\;(\Varid{trj}\;\Varid{ps}\;\Varid{x'})) \mathbin{\oplus} \Varid{val'}\;\Varid{ps}\;\Varid{x'}){}\<[48]%
\>[48]{}=\hspace{-3pt}\{\; \text{by }\;\Varid{headLemma}\;\}\hspace{-3pt}={}\<[E]%
\\
\>[3]{}\hsindent{4}{}\<[7]%
\>[7]{}(\Varid{r}\;\Varid{x'} \mathbin{\oplus} \Varid{val'}\;\Varid{ps}\;\Varid{x'}){}\<[48]%
\>[48]{}=\hspace{-3pt}\{\; \text{by }\;\text{induction hypothesis}\;\}\hspace{-3pt}={}\<[E]%
\\
\>[3]{}\hsindent{4}{}\<[7]%
\>[7]{}(\Varid{r}\;\Varid{x'} \mathbin{\oplus} \Varid{val}\;\Varid{ps}\;\Varid{x'}){}\<[48]%
\>[48]{}=\hspace{-3pt}\{\; \text{by definition of }\;\Varid{val}\;\}\hspace{-3pt}={}\<[E]%
\\
\>[3]{}\hsindent{4}{}\<[7]%
\>[7]{}(\Varid{val}\;(\Varid{p}\mathbin{::}\Varid{ps})\;\Varid{x})\;{}\<[48]%
\>[48]{}\hfill\Box{}\<[E]%
\ColumnHook
\end{hscode}\resethooks
}

\noindent where \ensuremath{\Varid{y}\mathrel{=}\Varid{p}\;\Varid{x}}, \ensuremath{\Varid{x'}\mathrel{=}\Varid{next}\;\Varid{t}\;\Varid{x}\;\Varid{y}} and \ensuremath{\Varid{r}\mathrel{=}\Varid{reward}\;\Varid{t}\;\Varid{x}\;\Varid{y}}. In the
proof sketch, we have first applied the definitions of \ensuremath{\Varid{val'}} and
\ensuremath{\Varid{sumR}}. Using the fact that in the deterministic case \ensuremath{\Varid{trj}} returns
exactly one state-control sequence and that the \ensuremath{\Varid{head}} of any
trajectory starting in \ensuremath{\Varid{x'}} is just \ensuremath{\Varid{x'}} (let us call the latter
\ensuremath{\Varid{headLemma}}), the left-hand side of the sum simplifies to \ensuremath{\Varid{r}\;\Varid{x'}}. Its
right-hand side amounts to \ensuremath{\Varid{val'}\;\Varid{ps}\;\Varid{x'}} so that we can apply the
induction hypothesis. The rest of the proof
only relies on definitional equalities. Thus in the deterministic case
\ensuremath{\Varid{val}} and \ensuremath{\Varid{val'}} are unconditionally extensionally equal -- or rather,
the conditions of Sec.~\ref{section:conditions} are trivially fulfilled.

\subsection{Lemmas}
\label{subsection:lemmas}
To prove the general, monadic case, we proceed similarly.
This time, however, the situation is complicated by the presence of
the abstract monad \ensuremath{\Conid{M}}. Instead of being able to use the type structure
of some concrete monad, we need to leverage on the properties of \ensuremath{\Conid{M}},
\ensuremath{\Varid{meas}} and \ensuremath{ \mathbin{\oplus} } postulated in Sec.~\ref{section:conditions}. To
facilitate the main proof, we first prove three lemmas about the
interaction of the measure with the monad structure and the
\ensuremath{ \mathbin{\oplus} }-operator on \ensuremath{\Conid{Val}}.
Machine-checked proofs are given in the
Appendices~\ref{appendix:theorem},~\ref{appendix:biCorrectness} and
\ref{appendix:lemmas}. The monad laws we use are stated in
Appendix~\ref{appendix:monadLaws}. In the remainder of this section,
we discuss semi-formal versions of the proofs.\\

\paragraph*{Monad algebras.} \hspace{0.1cm} The first lemma allows us
to lift and eliminate an application of the monad's \ensuremath{\Varid{join}} operation:
\begin{hscode}\SaveRestoreHook
\column{B}{@{}>{\hspre}l<{\hspost}@{}}%
\column{3}{@{}>{\hspre}l<{\hspost}@{}}%
\column{17}{@{}>{\hspre}c<{\hspost}@{}}%
\column{17E}{@{}l@{}}%
\column{20}{@{}>{\hspre}l<{\hspost}@{}}%
\column{E}{@{}>{\hspre}l<{\hspost}@{}}%
\>[3]{}\Varid{measAlgLemma}{}\<[17]%
\>[17]{} \mathop{:} {}\<[17E]%
\>[20]{}\{\mskip1.5mu \Conid{A},\Conid{B} \mathop{:} \Conid{Type}\mskip1.5mu\} \to (\Varid{f} \mathop{:} \Conid{B} \to \Conid{Val}) \to (\Varid{g} \mathop{:} \Conid{A} \to \Conid{M}\;\Conid{B}) \to {}\<[E]%
\\
\>[20]{}(\Varid{meas}\mathbin{\circ}\Varid{map}\;(\Varid{meas}\mathbin{\circ}\Varid{map}\;\Varid{f}\mathbin{\circ}\Varid{g}))\doteq(\Varid{meas}\mathbin{\circ}\Varid{map}\;\Varid{f}\mathbin{\circ}\Varid{join}\mathbin{\circ}\Varid{map}\;\Varid{g}){}\<[E]%
\ColumnHook
\end{hscode}\resethooks
The proof of this lemma hinges on the condition \ensuremath{\Varid{measJoinSpec}}. It
allows to trade the application of \ensuremath{\Varid{join}} against an application of
\ensuremath{\Varid{map}\;\Varid{meas}}. The rest is just standard reasoning with monad and functor
laws, i.e. we use that the functorial map for \ensuremath{\Conid{M}} preserves
composition and that \ensuremath{\Varid{join}} is a natural transformation:
\begin{hscode}\SaveRestoreHook
\column{B}{@{}>{\hspre}l<{\hspost}@{}}%
\column{3}{@{}>{\hspre}l<{\hspost}@{}}%
\column{5}{@{}>{\hspre}l<{\hspost}@{}}%
\column{53}{@{}>{\hspre}l<{\hspost}@{}}%
\column{E}{@{}>{\hspre}l<{\hspost}@{}}%
\>[3]{}\Varid{measAlgLemma}\;\Varid{f}\;\Varid{g}\;\Varid{ma}\mathrel{=}{}\<[E]%
\\[\blanklineskip]%
\>[3]{}\hsindent{2}{}\<[5]%
\>[5]{}((\Varid{meas}\mathbin{\circ}\Varid{map}\;(\Varid{meas}\mathbin{\circ}\Varid{map}\;\Varid{f}\mathbin{\circ}\Varid{g}))\;\Varid{ma}){}\<[53]%
\>[53]{}=\hspace{-3pt}\{\; map \text{ preserves composition}\;\}\hspace{-3pt}={}\<[E]%
\\[\blanklineskip]%
\>[3]{}\hsindent{2}{}\<[5]%
\>[5]{}((\Varid{meas}\mathbin{\circ}\Varid{map}\;(\Varid{meas}\mathbin{\circ}\Varid{map}\;\Varid{f})\mathbin{\circ}\Varid{map}\;\Varid{g})\;\Varid{ma}){}\<[53]%
\>[53]{}=\hspace{-3pt}\{\; map \text{ preserves composition}\;\}\hspace{-3pt}={}\<[E]%
\\[\blanklineskip]%
\>[3]{}\hsindent{2}{}\<[5]%
\>[5]{}((\Varid{meas}\mathbin{\circ}\Varid{map}\;\Varid{meas}\mathbin{\circ}\Varid{map}\;(\Varid{map}\;\Varid{f})\mathbin{\circ}\Varid{map}\;\Varid{g})\;\Varid{ma}){}\<[53]%
\>[53]{}=\hspace{-3pt}\{\; \text{by }\;\Varid{measJoinSpec}\;\}\hspace{-3pt}={}\<[E]%
\\[\blanklineskip]%
\>[3]{}\hsindent{2}{}\<[5]%
\>[5]{}((\Varid{meas}\mathbin{\circ}\Varid{join}\mathbin{\circ}\Varid{map}\;(\Varid{map}\;\Varid{f})\mathbin{\circ}\Varid{map}\;\Varid{g})\;\Varid{ma}){}\<[53]%
\>[53]{}=\hspace{-3pt}\{\; join \text{ is a natural transformation}\;\}\hspace{-3pt}={}\<[E]%
\\[\blanklineskip]%
\>[3]{}\hsindent{2}{}\<[5]%
\>[5]{}((\Varid{meas}\mathbin{\circ}\Varid{map}\;\Varid{f}\mathbin{\circ}\Varid{join}\mathbin{\circ}\Varid{map}\;\Varid{g})\;\Varid{ma})\;{}\<[53]%
\>[53]{}\hfill\Box{}\<[E]%
\ColumnHook
\end{hscode}\resethooks

This lemma is generic in the sense that it holds for arbitrary
Eilenberg-Moore algebras of a monad. Here we prove it for the
framework's measure \ensuremath{\Varid{meas}}, but note that in the appendix we prove a
generic version that is then appropriately instantiated.

\paragraph*{Head/trajectory interaction.} \hspace{0.1cm}
The second lemma amounts to a lifted version of \ensuremath{\Varid{headLemma}} in the
deterministic case. Mapping \ensuremath{\Varid{head}} onto an \ensuremath{\Conid{M}}-structure of
trajectories computed with \ensuremath{\Varid{trj}} results in an \ensuremath{\Conid{M}}-structure filled
with the initial states of these trajectories; similarly, mapping \ensuremath{(\Varid{r}\mathbin{\circ}\Varid{head} \mathbin{\medoplus} \Varid{s})} onto \ensuremath{\Varid{trj}\;\Varid{ps}\;\Varid{x}} for functions \ensuremath{\Varid{r}} and \ensuremath{\Varid{s}} of
appropriate type is the same as mapping \ensuremath{(\Varid{const}\;(\Varid{r}\;\Varid{x}) \mathbin{\medoplus} \Varid{s})} onto
\ensuremath{\Varid{trj}\;\Varid{ps}\;\Varid{x}} (where \ensuremath{\Varid{const}} is the constant function). We can prove
\begin{hscode}\SaveRestoreHook
\column{B}{@{}>{\hspre}l<{\hspost}@{}}%
\column{3}{@{}>{\hspre}l<{\hspost}@{}}%
\column{17}{@{}>{\hspre}c<{\hspost}@{}}%
\column{17E}{@{}l@{}}%
\column{20}{@{}>{\hspre}l<{\hspost}@{}}%
\column{E}{@{}>{\hspre}l<{\hspost}@{}}%
\>[3]{}\Varid{headTrjLemma}{}\<[17]%
\>[17]{} \mathop{:} {}\<[17E]%
\>[20]{}\{\mskip1.5mu \Varid{t},\Varid{n} \mathop{:} \mathbb{N}\mskip1.5mu\} \to (\Varid{ps} \mathop{:} \Conid{PolicySeq}\;\Varid{t}\;\Varid{n}) \to (\Varid{r} \mathop{:} \Conid{X}\;\Varid{t} \to \Conid{Val}) \to {}\<[E]%
\\
\>[20]{}(\Varid{s} \mathop{:} \Conid{StateCtrlSeq}\;\Varid{t}\;(\Conid{S}\;\Varid{n}) \to \Conid{Val}) \to (\Varid{x} \mathop{:} \Conid{X}\;\Varid{t}) \to {}\<[E]%
\\
\>[20]{}(\Varid{map}\;(\Varid{r}\mathbin{\circ}\Varid{head} \mathbin{\medoplus} \Varid{s})\mathbin{\circ}\Varid{trj}\;\Varid{ps})\;\Varid{x}\mathrel{=}{}\<[E]%
\\
\>[20]{}(\Varid{map}\;(\Varid{const}\;(\Varid{r}\;\Varid{x}) \mathbin{\medoplus} \Varid{s})\mathbin{\circ}\Varid{trj}\;\Varid{ps})\;\Varid{x}{}\<[E]%
\ColumnHook
\end{hscode}\resethooks
by doing a case split on \ensuremath{\Varid{ps}}. In case \ensuremath{\Varid{ps}\mathrel{=}\Conid{Nil}}, the equality holds
because the monad's \ensuremath{\Varid{pure}} is a natural transformation and in case \ensuremath{\Varid{ps}\mathrel{=}\Varid{p}\mathbin{::}\Varid{ps'}} because \ensuremath{\Conid{M}}'s functorial \ensuremath{\Varid{map}} preserves composition.

\paragraph*{Measure/sum interaction.} \hspace{0.1cm}
 The third lemma allows us to both commute the measure into the right
 summand of an \ensuremath{ \mathbin{\medoplus} }-sum and to perform the head/trajectory
 simplification. It lies at the core of the relationship between \ensuremath{\Varid{val}} and \ensuremath{\Varid{val'}}.
\begin{hscode}\SaveRestoreHook
\column{B}{@{}>{\hspre}l<{\hspost}@{}}%
\column{3}{@{}>{\hspre}l<{\hspost}@{}}%
\column{17}{@{}>{\hspre}c<{\hspost}@{}}%
\column{17E}{@{}l@{}}%
\column{20}{@{}>{\hspre}l<{\hspost}@{}}%
\column{E}{@{}>{\hspre}l<{\hspost}@{}}%
\>[3]{}\Varid{measSumLemma}{}\<[17]%
\>[17]{} \mathop{:} {}\<[17E]%
\>[20]{}\{\mskip1.5mu \Varid{t},\Varid{n} \mathop{:} \mathbb{N}\mskip1.5mu\} \to (\Varid{ps} \mathop{:} \Conid{PolicySeq}\;\Varid{t}\;\Varid{n}) \to {}\<[E]%
\\
\>[20]{}(\Varid{r} \mathop{:} \Conid{X}\;\Varid{t} \to \Conid{Val}) \to {}\<[E]%
\\
\>[20]{}(\Varid{s} \mathop{:} \Conid{StateCtrlSeq}\;\Varid{t}\;(\Conid{S}\;\Varid{n}) \to \Conid{Val}) \to {}\<[E]%
\\
\>[20]{}(\Varid{meas}\mathbin{\circ}\Varid{map}\;(\Varid{r}\mathbin{\circ}\Varid{head} \mathbin{\medoplus} \Varid{s})\mathbin{\circ}\Varid{trj}\;\Varid{ps})\doteq{}\<[E]%
\\
\>[20]{}(\Varid{r} \mathbin{\medoplus} \Varid{meas}\mathbin{\circ}\Varid{map}\;\Varid{s}\mathbin{\circ}\Varid{trj}\;\Varid{ps}){}\<[E]%
\ColumnHook
\end{hscode}\resethooks
Recall that our third condition from
Sec.~\ref{section:conditions}, \ensuremath{\Varid{measPlusSpec}}, plays the role of a
distributive law and allows us to ``factor out'' a partially applied
sum \ensuremath{(\Varid{v} \mathbin{\oplus} )} for arbitrary \ensuremath{\Varid{v} \mathop{:} \Conid{Val}}.
Given that \ensuremath{\Varid{measPlusSpec}} holds, the lemma is provable by simple
equational reasoning using the above head-trajectory lemma and the fact
that map preserves composition:
\begin{hscode}\SaveRestoreHook
\column{B}{@{}>{\hspre}l<{\hspost}@{}}%
\column{3}{@{}>{\hspre}l<{\hspost}@{}}%
\column{5}{@{}>{\hspre}l<{\hspost}@{}}%
\column{66}{@{}>{\hspre}l<{\hspost}@{}}%
\column{E}{@{}>{\hspre}l<{\hspost}@{}}%
\>[3]{}\Varid{measSumLemma}\;\Varid{ps}\;\Varid{r}\;\Varid{s}\;\Varid{x'}\mathrel{=}{}\<[E]%
\\[\blanklineskip]%
\>[3]{}\hsindent{2}{}\<[5]%
\>[5]{}((\Varid{meas}\mathbin{\circ}\Varid{map}\;(\Varid{r}\mathbin{\circ}\Varid{head} \mathbin{\medoplus} \Varid{s})\mathbin{\circ}\Varid{trj}\;\Varid{ps})\;\Varid{x'}){}\<[66]%
\>[66]{}=\hspace{-3pt}\{\; \text{by }\;\Varid{headTrjLemma}\;\}\hspace{-3pt}={}\<[E]%
\\[\blanklineskip]%
\>[3]{}\hsindent{2}{}\<[5]%
\>[5]{}((\Varid{meas}\mathbin{\circ}\Varid{map}\;(\Varid{const}\;(\Varid{r}\;\Varid{x'}) \mathbin{\medoplus} \Varid{s})\mathbin{\circ}\Varid{trj}\;\Varid{ps})\;\Varid{x'}){}\<[66]%
\>[66]{}=\hspace{-3pt}\{\; \text{by definition of } \mathbin{\medoplus} ,\mathbin{\circ}\;\}\hspace{-3pt}={}\<[E]%
\\[\blanklineskip]%
\>[3]{}\hsindent{2}{}\<[5]%
\>[5]{}((\Varid{meas}\mathbin{\circ}\Varid{map}\;((\Varid{const}\;(\Varid{r}\;\Varid{x'}) \mathbin{\medoplus} \Varid{id})\mathbin{\circ}\Varid{s})\mathbin{\circ}\Varid{trj}\;\Varid{ps})\;\Varid{x'}){}\<[66]%
\>[66]{}=\hspace{-3pt}\{\; map \text{ preserves composition}\;\}\hspace{-3pt}={}\<[E]%
\\[\blanklineskip]%
\>[3]{}\hsindent{2}{}\<[5]%
\>[5]{}((\Varid{meas}\mathbin{\circ}\Varid{map}\;(\Varid{const}\;(\Varid{r}\;\Varid{x'}) \mathbin{\medoplus} \Varid{id})\mathbin{\circ}\Varid{map}\;\Varid{s}\mathbin{\circ}\Varid{trj}\;\Varid{ps})\;\Varid{x'}){}\<[66]%
\>[66]{}=\hspace{-3pt}\{\; \text{by definition of } \mathbin{\medoplus} \;\}\hspace{-3pt}={}\<[E]%
\\[\blanklineskip]%
\>[3]{}\hsindent{2}{}\<[5]%
\>[5]{}((\Varid{meas}\mathbin{\circ}\Varid{map}\;((\Varid{r}\;\Varid{x'}) \mathbin{\oplus} )\mathbin{\circ}\Varid{map}\;\Varid{s}\mathbin{\circ}\Varid{trj}\;\Varid{ps})\;\Varid{x'}){}\<[66]%
\>[66]{}=\hspace{-3pt}\{\; \text{by }\;\Varid{measPlusSpec}\;\}\hspace{-3pt}={}\<[E]%
\\[\blanklineskip]%
\>[3]{}\hsindent{2}{}\<[5]%
\>[5]{}((((\Varid{r}\;\Varid{x'}) \mathbin{\oplus} )\mathbin{\circ}\Varid{meas}\mathbin{\circ}\Varid{map}\;\Varid{s}\mathbin{\circ}\Varid{trj}\;\Varid{ps})\;\Varid{x'}){}\<[66]%
\>[66]{}=\hspace{-3pt}\{\; \text{by definition of } \mathbin{\medoplus} \;\}\hspace{-3pt}={}\<[E]%
\\[\blanklineskip]%
\>[3]{}\hsindent{2}{}\<[5]%
\>[5]{}((\Varid{r} \mathbin{\medoplus} \Varid{meas}\mathbin{\circ}\Varid{map}\;\Varid{s}\mathbin{\circ}\Varid{trj}\;\Varid{ps})\;\Varid{x'})\;{}\<[66]%
\>[66]{}\hfill\Box{}\<[E]%
\ColumnHook
\end{hscode}\resethooks
Notice how \ensuremath{\Varid{measPlusSpec}} is used to transform an application of
\ensuremath{\Varid{meas}\mathbin{\circ}\Varid{map}\;((\Varid{r}\;\Varid{x'}) \mathbin{\oplus} )} into an application of \ensuremath{((\Varid{r}\;\Varid{x'}) \mathbin{\oplus} )\mathbin{\circ}\Varid{meas}}.
This is essential to simplify the computation of the measured total reward:
instead of first adding the current reward to the intermediate outcome of each
individual trajectory and then measuring the outcomes, one can first measure the
intermediate outcomes of the trajectories and then add the current reward.

\subsection{Correctness of the BJI-value function}
\label{subsection:valvalTh}
With the above lemmas in place, we now prove that \ensuremath{\Varid{val}} is
extensionally equal to \ensuremath{\Varid{val'}}.

\vspace{0.15cm}
Let \ensuremath{\Varid{t},\Varid{n} \mathop{:} \mathbb{N}}, \ensuremath{\Varid{ps} \mathop{:} \Conid{PolicySeq}\;\Varid{t}\;\Varid{n}}. We prove \ensuremath{\Varid{valMeasTotalReward}}
by induction on \ensuremath{\Varid{ps}}.

\vspace{0.15cm}
\paragraph*{Base case.}\hspace{0.1cm}
We need to show that for all \ensuremath{\Varid{x} \mathop{:} \Conid{X}\;\Varid{t}}, \ensuremath{\Varid{val'}\;\Conid{Nil}\;\Varid{x}} = \ensuremath{\Varid{val}\;\Conid{Nil}\;\Varid{x}}. The right hand side of this equation
reduces to \ensuremath{\Varid{zero}} by definition.
The left hand side can be simplified to \ensuremath{\Varid{meas}\;(\Varid{pure}\;\Varid{zero})}
since \ensuremath{\Varid{pure}} is a natural transformation. At this point,
our first condition, \ensuremath{\Varid{measPureSpec}}, comes into play: Using that
\ensuremath{\Varid{meas}} is inverse to \ensuremath{\Varid{pure}} on the left, we can conclude
that the equality holds.\\

\noindent In equational reasoning style: For all \ensuremath{\Varid{x} \mathop{:} \Conid{X}\;\Varid{t}},

{\linespread{1.2}
\begin{hscode}\SaveRestoreHook
\column{B}{@{}>{\hspre}l<{\hspost}@{}}%
\column{3}{@{}>{\hspre}l<{\hspost}@{}}%
\column{5}{@{}>{\hspre}l<{\hspost}@{}}%
\column{44}{@{}>{\hspre}l<{\hspost}@{}}%
\column{E}{@{}>{\hspre}l<{\hspost}@{}}%
\>[3]{}\Varid{valMeasTotalReward}\;\Conid{Nil}\;\Varid{x}\mathrel{=}{}\<[E]%
\\[\blanklineskip]%
\>[3]{}\hsindent{2}{}\<[5]%
\>[5]{}(\Varid{val'}\;\Conid{Nil}\;\Varid{x}){}\<[44]%
\>[44]{}=\hspace{-3pt}\{\; \text{by definition of }\;\Varid{val'}\;\}\hspace{-3pt}={}\<[E]%
\\[\blanklineskip]%
\>[3]{}\hsindent{2}{}\<[5]%
\>[5]{}(\Varid{meas}\;(\Varid{map}\;\Varid{sumR}\;(\Varid{trj}\;\Conid{Nil}\;\Varid{x}))){}\<[44]%
\>[44]{}=\hspace{-3pt}\{\; \text{by definition of }\;\Varid{trj}\;\}\hspace{-3pt}={}\<[E]%
\\[\blanklineskip]%
\>[3]{}\hsindent{2}{}\<[5]%
\>[5]{}(\Varid{meas}\;(\Varid{map}\;\Varid{sumR}\;(\Varid{pure}\;(\Conid{Last}\;\Varid{x})))){}\<[44]%
\>[44]{}=\hspace{-3pt}\{\; pure \text{ is a natural transformation}\;\}\hspace{-3pt}={}\<[E]%
\\[\blanklineskip]%
\>[3]{}\hsindent{2}{}\<[5]%
\>[5]{}(\Varid{meas}\;(\Varid{pure}\;(\Varid{sumR}\;(\Conid{Last}\;\Varid{x})))){}\<[44]%
\>[44]{}=\hspace{-3pt}\{\; \text{by definition of }\;\Varid{sumR}\;\}\hspace{-3pt}={}\<[E]%
\\[\blanklineskip]%
\>[3]{}\hsindent{2}{}\<[5]%
\>[5]{}(\Varid{meas}\;(\Varid{pure}\;\Varid{zero})){}\<[44]%
\>[44]{}=\hspace{-3pt}\{\; \text{by }\;\Varid{measPureSpec}\;\}\hspace{-3pt}={}\<[E]%
\\[\blanklineskip]%
\>[3]{}\hsindent{2}{}\<[5]%
\>[5]{}(\Varid{zero}){}\<[44]%
\>[44]{}=\hspace{-3pt}\{\; \text{by definition of }\;\Varid{val}\;\}\hspace{-3pt}={}\<[E]%
\\[\blanklineskip]%
\>[3]{}\hsindent{2}{}\<[5]%
\>[5]{}(\Varid{val}\;\Conid{Nil}\;\Varid{x}){}\<[E]%
\ColumnHook
\end{hscode}\resethooks
}

\paragraph*{Step case.}\hspace{0.1cm}
The induction hypothesis (\ensuremath{\Conid{IH}}) is:
for all \ensuremath{\Varid{x} \mathop{:} \Conid{X}\;\Varid{t}}, \ensuremath{\Varid{val'}\;\Varid{ps}\;\Varid{x}\mathrel{=}\Varid{val}\;\Varid{ps}\;\Varid{x}}. We have to show that
\ensuremath{\Conid{IH}} implies that for all \ensuremath{\Varid{p} \mathop{:} \Conid{Policy}\;\Varid{t}} and \ensuremath{\Varid{x} \mathop{:} \Conid{X}\;\Varid{t}}, the equality
\ensuremath{\Varid{val'}\;(\Varid{p}\mathbin{::}\Varid{ps})\;\Varid{x}\mathrel{=}\Varid{val}\;(\Varid{p}\mathbin{::}\Varid{ps})\;\Varid{x}} holds.

For brevity (and to economise on brackets), let in the following
\ensuremath{\Varid{y}\mathrel{=}\Varid{p}\;\Varid{x}}, \ensuremath{\Varid{mx'}\mathrel{=}\Varid{next}\;\Varid{t}\;\Varid{x}\;\Varid{y}}, \ensuremath{\Varid{r}\mathrel{=}\Varid{reward}\;\Varid{t}\;\Varid{x}\;\Varid{y}}, \ensuremath{\Varid{trjps}\mathrel{=}\Varid{trj}\;\Varid{ps}}, and \ensuremath{\Varid{consxy}\mathrel{=}((\Varid{x}  \mathbin{*\!*} \Varid{y}) \mathbin{\#\!\#} )}.

As in the base case, all that has to be done on the \ensuremath{\Varid{val}}-side of the
equation only depends on definitional equality. However it is more
involved to bring the \ensuremath{\Varid{val'}}-side into a form in which the induction
hypothesis can be applied. This is where we leverage on the lemmas
proved above.

By definition and because \ensuremath{\Varid{map}} preserves
composition, we know that \ensuremath{\Varid{val'}\;(\Varid{p}\mathbin{::}\Varid{ps})\;\Varid{x}} is equal to
\ensuremath{(\Varid{meas}\mathbin{\circ}\Varid{map}\;((\Varid{r}\mathbin{\circ}\Varid{head}) \mathbin{\medoplus} \Varid{sumR}))\;(\Varid{mx'} \mathbin{>\!\!>\!\!=} \Varid{trjps})}.
We use the relation between the monad's \ensuremath{\Varid{bind}} and \ensuremath{\Varid{join}} to eliminate
the \ensuremath{\Varid{bind}}-operator from the term.
Now we can apply the first lemma from above, \ensuremath{\Varid{measAlgLemma}}, to lift
and eliminate the \ensuremath{\Varid{join}} operation.

To commute the measure under the \ensuremath{ \mathbin{\medoplus} } and get rid of the application
of \ensuremath{\Varid{head}}, we use our third lemma, \ensuremath{\Varid{measSumLemma}}.
At this point we can apply the induction hypothesis and the resulting
term is equal to \ensuremath{\Varid{val}\;\Varid{ps}\;\Varid{x}} by definition.\\

\noindent The more detailed equational reasoning proof:
\footnote{We are very grateful to the anonymous reviewer who suggested
  an alternative proof for the induction step. The proof presented
  here is based on his proof, and his suggestions have lead to
  significantly weaker conditions on the measure and thus a stronger
  result.}
\begin{hscode}\SaveRestoreHook
\column{B}{@{}>{\hspre}l<{\hspost}@{}}%
\column{3}{@{}>{\hspre}l<{\hspost}@{}}%
\column{4}{@{}>{\hspre}l<{\hspost}@{}}%
\column{66}{@{}>{\hspre}l<{\hspost}@{}}%
\column{E}{@{}>{\hspre}l<{\hspost}@{}}%
\>[3]{}\Varid{valMeasTotalReward}\;(\Varid{p}\mathbin{::}\Varid{ps})\;\Varid{x}\mathrel{=}{}\<[E]%
\\[\blanklineskip]%
\>[3]{}\hsindent{1}{}\<[4]%
\>[4]{}(\Varid{val'}\;(\Varid{p}\mathbin{::}\Varid{ps})\;\Varid{x}){}\<[66]%
\>[66]{}=\hspace{-3pt}\{\; \text{by definition of }\;\Varid{val'}\;\}\hspace{-3pt}={}\<[E]%
\\[\blanklineskip]%
\>[3]{}\hsindent{1}{}\<[4]%
\>[4]{}(\Varid{meas}\;(\Varid{map}\;\Varid{sumR}\;(\Varid{trj}\;(\Varid{p}\mathbin{::}\Varid{ps})\;\Varid{x}))){}\<[66]%
\>[66]{}=\hspace{-3pt}\{\; \text{by definition of }\;\Varid{trj}\;\}\hspace{-3pt}={}\<[E]%
\\[\blanklineskip]%
\>[3]{}\hsindent{1}{}\<[4]%
\>[4]{}(\Varid{meas}\;(\Varid{map}\;\Varid{sumR}\;(\Varid{map}\;\Varid{consxy}\;(\Varid{mx'} \mathbin{>\!\!>\!\!=} \Varid{trjps})))){}\<[66]%
\>[66]{}=\hspace{-3pt}\{\; map \text{ preserves composition}\;\}\hspace{-3pt}={}\<[E]%
\\[\blanklineskip]%
\>[3]{}\hsindent{1}{}\<[4]%
\>[4]{}(\Varid{meas}\;(\Varid{map}\;(\Varid{sumR}\mathbin{\circ}\Varid{consxy})\;(\Varid{mx'} \mathbin{>\!\!>\!\!=} \Varid{trjps}))){}\<[66]%
\>[66]{}=\hspace{-3pt}\{\; \text{by definition of }\;\Varid{sumR}\;\}\hspace{-3pt}={}\<[E]%
\\[\blanklineskip]%
\>[3]{}\hsindent{1}{}\<[4]%
\>[4]{}(\Varid{meas}\;(\Varid{map}\;((\Varid{r}\mathbin{\circ}\Varid{head}) \mathbin{\medoplus} \Varid{sumR})\;(\Varid{mx'} \mathbin{>\!\!>\!\!=} \Varid{trjps}))){}\<[66]%
\>[66]{}=\hspace{-3pt}\{\; \text{relation } bind/join \;\}\hspace{-3pt}={}\<[E]%
\\[\blanklineskip]%
\>[3]{}\hsindent{1}{}\<[4]%
\>[4]{}(\Varid{meas}\;(\Varid{map}\;((\Varid{r}\mathbin{\circ}\Varid{head}) \mathbin{\medoplus} \Varid{sumR})\;(\Varid{join}\;(\Varid{map}\;\Varid{trjps}\;\Varid{mx'})))){}\<[66]%
\>[66]{}=\hspace{-3pt}\{\; \text{by }\;\Varid{measAlgLemma}\;\}\hspace{-3pt}={}\<[E]%
\\[\blanklineskip]%
\>[3]{}\hsindent{1}{}\<[4]%
\>[4]{}(\Varid{meas}\;(\Varid{map}\;(\Varid{meas}\mathbin{\circ}\Varid{map}\;(\Varid{r}\mathbin{\circ}\Varid{head} \mathbin{\medoplus} \Varid{sumR})\mathbin{\circ}\Varid{trjps})\;\Varid{mx'})){}\<[66]%
\>[66]{}=\hspace{-3pt}\{\; \text{by }\;\Varid{measSumLemma}\;\}\hspace{-3pt}={}\<[E]%
\\[\blanklineskip]%
\>[3]{}\hsindent{1}{}\<[4]%
\>[4]{}(\Varid{meas}\;(\Varid{map}\;(\Varid{r} \mathbin{\medoplus} \Varid{meas}\mathbin{\circ}\Varid{map}\;\Varid{sumR}\mathbin{\circ}\Varid{trjps})\;\Varid{mx'})){}\<[66]%
\>[66]{}=\hspace{-3pt}\{\; \text{by definition of }\;\Varid{val'}\;\}\hspace{-3pt}={}\<[E]%
\\[\blanklineskip]%
\>[3]{}\hsindent{1}{}\<[4]%
\>[4]{}(\Varid{meas}\;(\Varid{map}\;(\Varid{r} \mathbin{\medoplus} \Varid{val'}\;\Varid{ps})\;\Varid{mx'})){}\<[66]%
\>[66]{}=\hspace{-3pt}\{\; \text{by }\;\text{induction hypothesis}\;\}\hspace{-3pt}={}\<[E]%
\\[\blanklineskip]%
\>[3]{}\hsindent{1}{}\<[4]%
\>[4]{}(\Varid{meas}\;(\Varid{map}\;(\Varid{r} \mathbin{\medoplus} \Varid{val}\;\Varid{ps})\;\Varid{mx'})){}\<[66]%
\>[66]{}=\hspace{-3pt}\{\; \text{by definition of }\;\Varid{val}\;\}\hspace{-3pt}={}\<[E]%
\\[\blanklineskip]%
\>[3]{}\hsindent{1}{}\<[4]%
\>[4]{}(\Varid{val}\;(\Varid{p}\mathbin{::}\Varid{ps})\;\Varid{x}){}\<[E]%
\ColumnHook
\end{hscode}\resethooks
\hfill $\Box$

\paragraph*{Technical remarks.} \hspace{0.1cm}
The above proof of \ensuremath{\Varid{valMeasTotalReward}} omits some technical details that
may be uninteresting for a pen and paper proof, but turn out to be
crucial in the setting of an intensional type theory -- like Idris --
where function extensionality does not hold in general.
In particular, we have to postulate that the functorial \ensuremath{\Varid{map}}
preserves extensional equality (see Appendix~\ref{appendix:monadLaws}
and \citep{botta2020extensional}) for Idris to accept the proof.
In fact, most of the reasoning proceeds by replacing functions that are mapped
onto monadic values by other functions that are only extensionally
equal. Using that \ensuremath{\Varid{map}} preserves extensional equality
allows to carry out such proofs generically without knowledge of
the concrete structure of the functor.

\subsection{Correctness of monadic backward induction}
\label{subsection:biCorrectness}

As corollary, we can now prove the correctness of monadic backward induction,
namely that the policy sequences computed by \ensuremath{\Varid{bi}} are optimal with
respect to the measured total reward computed by \ensuremath{\Varid{val'}}:  
\begin{hscode}\SaveRestoreHook
\column{B}{@{}>{\hspre}l<{\hspost}@{}}%
\column{3}{@{}>{\hspre}l<{\hspost}@{}}%
\column{6}{@{}>{\hspre}l<{\hspost}@{}}%
\column{18}{@{}>{\hspre}l<{\hspost}@{}}%
\column{25}{@{}>{\hspre}c<{\hspost}@{}}%
\column{25E}{@{}l@{}}%
\column{28}{@{}>{\hspre}l<{\hspost}@{}}%
\column{57}{@{}>{\hspre}l<{\hspost}@{}}%
\column{E}{@{}>{\hspre}l<{\hspost}@{}}%
\>[3]{}\Varid{biOptMeasTotalReward}{}\<[25]%
\>[25]{} \mathop{:} {}\<[25E]%
\>[28]{}(\Varid{t},\Varid{n} \mathop{:} \mathbb{N}) \to \Conid{GenOptPolicySeq}\;\Varid{val'}\;(\Varid{bi}\;\Varid{t}\;\Varid{n}){}\<[E]%
\\[\blanklineskip]%
\>[3]{}\Varid{biOptMeasTotalReward}\;\Varid{t}\;\Varid{n}\;\Varid{ps'}\;\Varid{x}\mathrel{=}{}\<[E]%
\\
\>[3]{}\hsindent{3}{}\<[6]%
\>[6]{}\mathbf{let}\;\Varid{vvEqL}{}\<[18]%
\>[18]{}\mathrel{=}\Varid{sym}\;(\Varid{valMeasTotalReward}\;\Varid{ps'}\;\Varid{x})\;{}\<[57]%
\>[57]{}\mathbf{in}{}\<[E]%
\\
\>[3]{}\hsindent{3}{}\<[6]%
\>[6]{}\mathbf{let}\;\Varid{vvEqR}{}\<[18]%
\>[18]{}\mathrel{=}\Varid{sym}\;(\Varid{valMeasTotalReward}\;(\Varid{bi}\;\Varid{t}\;\Varid{n})\;\Varid{x})\;{}\<[57]%
\>[57]{}\mathbf{in}{}\<[E]%
\\
\>[3]{}\hsindent{3}{}\<[6]%
\>[6]{}\mathbf{let}\;\Varid{biOpt}{}\<[18]%
\>[18]{}\mathrel{=}\Varid{biOptVal}\;\Varid{t}\;\Varid{n}\;\Varid{ps'}\;\Varid{x}\;{}\<[57]%
\>[57]{}\mathbf{in}{}\<[E]%
\\
\>[3]{}\hsindent{3}{}\<[6]%
\>[6]{}\Varid{replace}\;\Varid{vvEqR}\;(\Varid{replace}\;\Varid{vvEqL}\;\Varid{biOpt}){}\<[E]%
\ColumnHook
\end{hscode}\resethooks

\section{Discussion}
\label{section:discussion}
In the last two sections we have seen what the three conditions mean 
for concrete examples and how they are used in the correctness proof.
In this section we take a step back and consider them from a more abstract 
point of view.

\paragraph*{Category-theoretical perspective.}\hspace{0.1cm}
Readers familiar with the theory of monads might have recognised that the
first two conditions ensure that \ensuremath{\Varid{meas}} is the structure map of a
monad algebra for \ensuremath{\Conid{M}} on \ensuremath{\Conid{Val}} and thus the pair \ensuremath{(\Conid{Val},\Varid{meas})} is an
object of
the Eilenberg-Moore category associated with the monad \ensuremath{\Conid{M}}. The third
condition requires the map \ensuremath{(\Varid{v} \mathbin{\oplus} )} to be an \ensuremath{\Conid{M}}-algebra homomorphism
-- a structure preserving map -- for arbitrary values \ensuremath{\Varid{v}}.

This perspective allows us to use existing knowledge about monad
algebras as a first criterion for choosing measures. For example, the
Eilenberg-Moore-algebras of the list monad are monoids -- this
implicitly played a role in the examples we considered
above. \cite{DBLP:journals/tcs/Jacobs11} shows that the algebras of
the distribution monad for probability distributions with finite
support correspond to convex sets. Interestingly, convex sets play an
important role in the theory of optimal control
\citep{bertsekas2003convex}.

\paragraph*{Measures for the list monad.} \hspace{0.1cm}
The knowledge that monoids are \ensuremath{\Conid{List}}-algebras suggests a generic
description of admissible measures for \ensuremath{\Conid{M}\mathrel{=}\Conid{List}}:
Given a monoid \ensuremath{(\Conid{Val}, \mathbin{\odot} ,\Varid{b})}, we can prove that monoid
homomorphisms of the form \ensuremath{\Varid{foldr}\; \mathbin{\odot} \;\Varid{b}} fulfil the three conditions,
if \ensuremath{ \mathbin{\oplus} } distributes over $\odot$ on the left. I.e. for \ensuremath{\Varid{meas}\mathrel{=}\Varid{foldr}\; \mathbin{\odot} \;\Varid{b}} the three conditions can be proven from
\begin{hscode}\SaveRestoreHook
\column{B}{@{}>{\hspre}l<{\hspost}@{}}%
\column{3}{@{}>{\hspre}l<{\hspost}@{}}%
\column{23}{@{}>{\hspre}c<{\hspost}@{}}%
\column{23E}{@{}l@{}}%
\column{26}{@{}>{\hspre}l<{\hspost}@{}}%
\column{43}{@{}>{\hspre}c<{\hspost}@{}}%
\column{43E}{@{}l@{}}%
\column{47}{@{}>{\hspre}l<{\hspost}@{}}%
\column{70}{@{}>{\hspre}c<{\hspost}@{}}%
\column{70E}{@{}l@{}}%
\column{73}{@{}>{\hspre}l<{\hspost}@{}}%
\column{E}{@{}>{\hspre}l<{\hspost}@{}}%
\>[3]{}\Varid{odotNeutrRight}{}\<[23]%
\>[23]{} \mathop{:} {}\<[23E]%
\>[26]{}(\Varid{l} \mathop{:} \Conid{Val}){}\<[43]%
\>[43]{} \to {}\<[43E]%
\>[47]{}\Varid{l} \mathbin{\odot} \Varid{neutr}{}\<[70]%
\>[70]{}\mathrel{=}{}\<[70E]%
\>[73]{}\Varid{l}{}\<[E]%
\\
\>[3]{}\Varid{odotNeutrLeft}{}\<[23]%
\>[23]{} \mathop{:} {}\<[23E]%
\>[26]{}(\Varid{r} \mathop{:} \Conid{Val}){}\<[43]%
\>[43]{} \to {}\<[43E]%
\>[47]{}\Varid{neutr} \mathbin{\odot} \Varid{r}{}\<[70]%
\>[70]{}\mathrel{=}{}\<[70E]%
\>[73]{}\Varid{r}{}\<[E]%
\\
\>[3]{}\Varid{odotAssociative}{}\<[23]%
\>[23]{} \mathop{:} {}\<[23E]%
\>[26]{}(\Varid{l},\Varid{v},\Varid{r} \mathop{:} \Conid{Val}){}\<[43]%
\>[43]{} \to {}\<[43E]%
\>[47]{}\Varid{l} \mathbin{\odot} (\Varid{v} \mathbin{\odot} \Varid{r}){}\<[70]%
\>[70]{}\mathrel{=}{}\<[70E]%
\>[73]{}(\Varid{l} \mathbin{\odot} \Varid{v}) \mathbin{\odot} \Varid{r}{}\<[E]%
\\
\>[3]{}\Varid{oplusOdotDistrLeft}{}\<[23]%
\>[23]{} \mathop{:} {}\<[23E]%
\>[26]{}(\Varid{n},\Varid{l},\Varid{r} \mathop{:} \Conid{Val}){}\<[43]%
\>[43]{} \to {}\<[43E]%
\>[47]{}\Varid{n} \mathbin{\oplus} (\Varid{l} \mathbin{\odot} \Varid{r}){}\<[70]%
\>[70]{}\mathrel{=}{}\<[70E]%
\>[73]{}(\Varid{n} \mathbin{\oplus} \Varid{l}) \mathbin{\odot} (\Varid{n} \mathbin{\oplus} \Varid{r}){}\<[E]%
\ColumnHook
\end{hscode}\resethooks
Neutrality of \ensuremath{\Varid{b}} on the right is needed for \ensuremath{\Varid{measPureSpec}},
while \ensuremath{\Varid{measJoinSpec}} follows from neutrality on the left and
the associativity of \ensuremath{ \mathbin{\odot} }. The algebra morphism condition on \ensuremath{(\Varid{v} \mathbin{\oplus} )}
is provable from the distributivity of \ensuremath{ \mathbin{\oplus} } over \ensuremath{ \mathbin{\odot} } and again
neutrality of \ensuremath{\Varid{b}} on the right.
If moreover \ensuremath{ \mathbin{\odot} } is monotone with respect to \ensuremath{ \,\sqsubseteq\, }
\begin{hscode}\SaveRestoreHook
\column{B}{@{}>{\hspre}l<{\hspost}@{}}%
\column{3}{@{}>{\hspre}l<{\hspost}@{}}%
\column{13}{@{}>{\hspre}c<{\hspost}@{}}%
\column{13E}{@{}l@{}}%
\column{16}{@{}>{\hspre}l<{\hspost}@{}}%
\column{E}{@{}>{\hspre}l<{\hspost}@{}}%
\>[3]{}\Varid{odotMon}{}\<[13]%
\>[13]{} \mathop{:} {}\<[13E]%
\>[16]{}\{\mskip1.5mu \Varid{a},\Varid{b},\Varid{c},\Varid{d} \mathop{:} \Conid{Val}\mskip1.5mu\} \to \Varid{a} \,\sqsubseteq\, \Varid{b} \to \Varid{c} \,\sqsubseteq\, \Varid{d} \to (\Varid{a} \mathbin{\odot} \Varid{c}) \,\sqsubseteq\, (\Varid{b} \mathbin{\odot} \Varid{d}){}\<[E]%
\ColumnHook
\end{hscode}\resethooks
then we can also prove \ensuremath{\Varid{measMonSpec}} using the transitivity of \ensuremath{ \,\sqsubseteq\, }.
The proofs are simple and can be found in the
supplementary material to this paper.
This also illustrates how the three abstract conditions follow from
more familiar algebraic properties.

\paragraph*{Mutual independence.}\hspace{0.1cm}
Although it does not seem surprising, it should be noted that the
three conditions are mutually independent. This can be concluded from
the counter-examples in Sec.~\ref{subsection:exAndCounterEx}: The
sum, the modified list maximum and the arithmetic average each fail
exactly one of the three conditions. 

\paragraph*{Sufficient vs.\ necessary.}\hspace{0.1cm}
The three conditions are sufficient to prove the extensional equality
of the functions \ensuremath{\Varid{val}} and \ensuremath{\Varid{val'}}. They are justified by their level
of generality and the fact that they hold for standard \ensuremath{\Varid{measures}} used
in control theory. However, we leave open the interesting question
whether these conditions are also necessary for the correctness of
monadic backward induction.

\paragraph*{Non-emptiness requirement.}\hspace{0.1cm}
Note that \ensuremath{\Varid{mv}} in the premises of \ensuremath{\Varid{measPlusSpec}} is required to be
non-empty. 
This condition arises from a pragmatic consideration.
As an example, let us again use the list monad with \ensuremath{\Conid{Val}\mathrel{=}\mathbb{N}} and \ensuremath{ \mathbin{\oplus} \mathrel{=}\mathbin{+}}. It is not hard to see that  
for any natural number \ensuremath{\Varid{n}} greater than 0 the equality
\ensuremath{\Varid{meas}\;(\Varid{map}\;(\Varid{n}\mathbin{+})\;[\mskip1.5mu \mskip1.5mu])\mathrel{=}\Varid{n}\mathbin{+}\Varid{meas}\;[\mskip1.5mu \mskip1.5mu]} must fail.
So, if we wish to use the standard list data type instead of
defining a custom type of non-empty lists, the only way to prove
the base case of \ensuremath{\Varid{measPlusSpec}} is by contradiction with the
non-emptiness premise.

However, omitting the premise \ensuremath{\Varid{mv} \mathop{:} \Conid{NotEmpty}} would not prevent us
from generically proving the correctness result of
Section~\ref{section:valval} -- it would even simplify matters as it
would spare us reasoning about preservation of non-emptiness.
But it would implicitly restrict the class of monads that can be used
to instantiate \ensuremath{\Conid{M}}. For example, we have seen above, that
\ensuremath{\Varid{measPlusSpec}} is not provable for the empty list without the
non-emptiness premise and we would therefore need to resort to a custom 
type of non-empty lists instead. 

The price to pay for including the non-emptiness premise is
the additional condition \ensuremath{\Varid{nextNotEmpty}} on the transition function
\ensuremath{\Varid{next}} that was already stated in Sec.~\ref{subsection:wrap-up}.
Moreover, we have to postulate non-emptiness preservation laws for the
monad operations (Appendix~\ref{appendix:monadLaws}) and to prove an
additional lemma about the preservation of non-emptiness
(Appendix~\ref{appendix:lemmas}).
Conceptually, it might seem cleaner to omit the non-emptiness
condition: In this case, the remaining conditions would only concern
the interaction between the monad, the measure, the type of values and the
binary operation \ensuremath{ \mathbin{\oplus} }. However, the non-emptiness preservation laws seem
less restrictive with respect to the monad. In particular, for our
above example of ordinary lists they hold (the relevant proofs can be
found in the supplementary material).
Thus we have opted for explicitly restricting the \ensuremath{\Varid{next}} function
instead of implicitly restricting the class of monads for which the
result of Sec.~\ref{section:valval} holds.

\section{Conclusion}
\label{section:conclusion}

In this paper, we have proposed correctness criteria for monadic backward
induction and its underlying value function in the
framework for specifying and solving finite-horizon, monadic SDPs proposed
in \citep{2017_Botta_Jansson_Ionescu}.
After having shown that these criteria are not necessarily met for arbitrary
monadic SDPs, we have formulated three general compatibility conditions.
We have given a proof that monadic backward induction and its underlying value
function are correct if these conditions are fulfilled.

The main theorem has been proved via the extensional equality of
two functions: 1) the value function of Bellman's dynamic programming
\citep{bellman1957} and optimal control theory \citep{bertsekas1995,
  puterman2014markov} that is also at the core of the generic
backward induction algorithm of \citep{2017_Botta_Jansson_Ionescu} and
2) the measured total reward function that specifies the objective of
decision making in monadic SDPs: the maximisation of a measure of the
sum of the rewards along the trajectories rooted at the state
associated with the first decision.

Our contribution to verified optimal decision making is twofold: On the
one hand, we have implemented a machine-checked generalisation of the
semi-formal results for deterministic and stochastic SDPs
discussed in \citep[Prop.~1.3.1]{bertsekas1995} and
\citep[Theorem~4.5.1.c]{puterman2014markov}.
As a consequence, we now have a provably correct method for solving
deterministic and stochastic sequential decision problems with their
canonical measure functions.
On the other hand, we have identified three general conditions that are
sufficient for the equivalence between the two functions and thus the
correctness result to hold.
The first two conditions are natural compatibility conditions
between the measure of uncertainty \ensuremath{\Varid{meas}} and the monadic operations
associated with the uncertainty monad \ensuremath{\Conid{M}}. The third condition is a
distributivity principle concerning the relationship between \ensuremath{\Varid{meas}},
the functorial map associated with \ensuremath{\Conid{M}} and
the rule for adding rewards \ensuremath{ \mathbin{\oplus} }. All three conditions have a
straightforward category-theoretical interpretation in terms of
Eilenberg-Moore algebras \citep[ch.~VI.2]{maclane}.
As discussed in Sec.~\ref{section:discussion}, the three
conditions are independent and have non-trivial implications for the
measure and the addition function that cannot be derived from the
monotonicity condition on \ensuremath{\Varid{meas}} already imposed in
\citep{ionescu2009, 2017_Botta_Jansson_Ionescu}.

A consequence of this contribution is more flexibility:
We can now compute verified solutions of stochastic sequential
decision problems in which the
measure of uncertainty is different from the expected value
measure. This is important for applications in which the goal of
decision making is, for example, of maximising the value of
worst-case outcomes.
To the best of our knowledge, the formulation of the compatibility
condition and the proof of the equivalence between the two value
functions are novel results.

The latter can be employed in a wider context than the one that has
motivated our study: in many practical problems in science and
engineering, the computation of optimal policies via backward induction
(let apart brute-force or gradient methods) is simply not feasible.
In these problems one often still needs to generate, evaluate and
compare different policies and our result shows under which conditions
such evaluation can safely be done via the ``fast'' value function \ensuremath{\Varid{val}}
of standard control theory.

Finally, our contribution is an application of verified, literal
programming to optimal decision making: the sources of this document
have been written in literal Idris and are available at
\citep{IdrisLibsValVal}, where the reader can also find the bare code
and some examples. Although the development has been carried out in
Idris, it should be readily reproducible in other implementations of
type theory like Agda or Coq. 


\section*{Acknowledgements} \label{section:acknowledgements}

The work presented in this paper was motivated by a remark of Marina
Mart{\'i}nez Montero who raised the question of the equivalence between
\ensuremath{\Varid{val}} and \ensuremath{\Varid{val'}} (and, thus, of the correctness of the
\bottaetal framework) during an introduction to verified decision making
that the authors gave at UCL (Université catholique de Louvain) in
2019. We are especially thankful to Marina for that question!

We are grateful to Jeremy Gibbons, Christoph Kreitz, Patrik Jansson, Tim
Richter and to the JFP editors and reviewers, whose comments and
recommendations have lead to significant improvements of the original
manuscript.

A very special thanks goes to the anonymous reviewer who has suggested
both a more straightforward proof of the \ensuremath{\Varid{val}}-\ensuremath{\Varid{val'}} equality and,
crucially, weaker conditions on the measure function for the
result to hold. This warrants the applicability of the
\bottaetal framework for verified decision making to a wider class of
problems than our original conditions.

The work presented in this paper heavily relies on free software, among
others on Coq, Idris, Agda, GHC, git, vi, Emacs, \LaTeX\ and on the
FreeBSD and Debian GNU/Linux operating systems.  It is our pleasure to
thank all developers of these excellent products.
This is TiPES contribution No 37. This project has received funding from
the European Union’s Horizon 2020 research and innovation programme
under grant agreement No 820970 (TiPES
--Tipping Points in the Earth System, \citeyear{TiPES::Website}).

\subsection*{Conflicts of Interest}
None.

\bibliographystyle{jfplike}
\bibliography{references}






\appendix
\label{section:appendix}

\addcontentsline{toc}{section}{Appendices}
\section*{Appendices}

\section{Preliminaries}
\label{appendix:prelim}

\subsection{General remarks concerning the Idris formalisation}
\label{appendix:idrisRemarks}

\begin{itemize}

\item Idris' type checker often struggles with dependencies in implicit
   arguments. We sometimes use abbreviations to avoid
   cluttering the proofs with implicit arguments.

\item As a standard, we write \ensuremath{(\Varid{f}\mathbin{\circ}\Varid{g}\mathbin{\circ}\Varid{h})\;\Varid{x}} instead of \ensuremath{\Varid{f}\;(\Varid{g}\;(\Varid{h}\;\Varid{x}))}.\\ When this is a problem for the type checker, we use the second
  notation.

\item Functions that are not defined explicitly are from the Idris
  standard library.\\ Examples are \ensuremath{\Varid{cong}}, \ensuremath{\Varid{void}} and \ensuremath{\Varid{concat}}.

\item Proofs in Idris can be implemented by preorder reasoning.\\
I.e.\ equational reasoning steps of the form
\begin{hscode}\SaveRestoreHook
\column{B}{@{}>{\hspre}l<{\hspost}@{}}%
\column{3}{@{}>{\hspre}l<{\hspost}@{}}%
\column{E}{@{}>{\hspre}l<{\hspost}@{}}%
\>[3]{}(\Varid{t}_{1})=\hspace{-3pt}\{\; \Varid{step}\;\}\hspace{-3pt}={}\<[E]%
\\
\>[3]{}(\Varid{t}_{2})\;\hfill\Box{}\<[E]%
\ColumnHook
\end{hscode}\resethooks
as displayed in this appendix are actual type-checkable implementations
of proofs.

\end{itemize}

\subsection{Monad Laws}
\label{appendix:monadLaws}

In the BJI-framework, \ensuremath{\Conid{M}} is required to be a \emph{container monad} but none
of the standard monad laws \citep{bird2014thinking} is required for
the verification result to hold. By contrast, to prove our extended correctness
result, we need \ensuremath{\Conid{M}} to be a full-fledged
monad. More specifically, we require of the monad \ensuremath{\Conid{M}} that

\begin{itemize}
\item it is equipped with functor and monad operations:
\begin{hscode}\SaveRestoreHook
\column{B}{@{}>{\hspre}l<{\hspost}@{}}%
\column{3}{@{}>{\hspre}l<{\hspost}@{}}%
\column{12}{@{}>{\hspre}c<{\hspost}@{}}%
\column{12E}{@{}l@{}}%
\column{15}{@{}>{\hspre}l<{\hspost}@{}}%
\column{E}{@{}>{\hspre}l<{\hspost}@{}}%
\>[3]{}\Varid{map}{}\<[12]%
\>[12]{} \mathop{:} {}\<[12E]%
\>[15]{}\{\mskip1.5mu \Conid{A},\Conid{B} \mathop{:} \Conid{Type}\mskip1.5mu\} \to (\Conid{A} \to \Conid{B}) \to \Conid{M}\;\Conid{A} \to \Conid{M}\;\Conid{B}{}\<[E]%
\\[\blanklineskip]%
\>[3]{}\Varid{pure}{}\<[12]%
\>[12]{} \mathop{:} {}\<[12E]%
\>[15]{}\{\mskip1.5mu \Conid{A} \mathop{:} \Conid{Type}\mskip1.5mu\} \to \Conid{A} \to \Conid{M}\;\Conid{A}{}\<[E]%
\\[\blanklineskip]%
\>[3]{}( \mathbin{>\!\!>\!\!=} ){}\<[12]%
\>[12]{} \mathop{:} {}\<[12E]%
\>[15]{}\{\mskip1.5mu \Conid{A},\Conid{B} \mathop{:} \Conid{Type}\mskip1.5mu\} \to \Conid{M}\;\Conid{A} \to (\Conid{A} \to \Conid{M}\;\Conid{B}) \to \Conid{M}\;\Conid{B}{}\<[E]%
\\[\blanklineskip]%
\>[3]{}\Varid{join}{}\<[12]%
\>[12]{} \mathop{:} {}\<[12E]%
\>[15]{}\{\mskip1.5mu \Conid{A} \mathop{:} \Conid{Type}\mskip1.5mu\} \to \Conid{M}\;(\Conid{M}\;\Conid{A}) \to \Conid{M}\;\Conid{A}{}\<[E]%
\ColumnHook
\end{hscode}\resethooks
\item it preserves extensional equality \citep{botta2020extensional},
  identity and composition of arrows:
\begin{hscode}\SaveRestoreHook
\column{B}{@{}>{\hspre}l<{\hspost}@{}}%
\column{3}{@{}>{\hspre}l<{\hspost}@{}}%
\column{17}{@{}>{\hspre}c<{\hspost}@{}}%
\column{17E}{@{}l@{}}%
\column{20}{@{}>{\hspre}l<{\hspost}@{}}%
\column{E}{@{}>{\hspre}l<{\hspost}@{}}%
\>[3]{} mapPresEE {}\<[17]%
\>[17]{} \mathop{:} {}\<[17E]%
\>[20]{}\{\mskip1.5mu \Conid{A},\Conid{B} \mathop{:} \Conid{Type}\mskip1.5mu\} \to (\Varid{f},\Varid{g} \mathop{:} \Conid{A} \to \Conid{B}) \to \Varid{f}\doteq\Varid{g} \to \Varid{map}\;\Varid{f}\doteq\Varid{map}\;\Varid{g}{}\<[E]%
\\[\blanklineskip]%
\>[3]{}\Varid{mapPresId}{}\<[17]%
\>[17]{} \mathop{:} {}\<[17E]%
\>[20]{}\{\mskip1.5mu \Conid{A} \mathop{:} \Conid{Type}\mskip1.5mu\} \to \Varid{map}\;\Varid{id}\doteq\Varid{id}{}\<[E]%
\\[\blanklineskip]%
\>[3]{}\Varid{mapPresComp}{}\<[17]%
\>[17]{} \mathop{:} {}\<[17E]%
\>[20]{}\{\mskip1.5mu \Conid{A},\Conid{B},\Conid{C} \mathop{:} \Conid{Type}\mskip1.5mu\} \to (\Varid{f} \mathop{:} \Conid{A} \to \Conid{B}) \to (\Varid{g} \mathop{:} \Conid{B} \to \Conid{C}) \to {}\<[E]%
\\
\>[20]{}\Varid{map}\;(\Varid{g}\mathbin{\circ}\Varid{f})\doteq\Varid{map}\;\Varid{g}\mathbin{\circ}\Varid{map}\;\Varid{f}{}\<[E]%
\ColumnHook
\end{hscode}\resethooks
\item Its \ensuremath{\Varid{pure}} and \ensuremath{\Varid{join}} operations are natural transformations \citep[see][I.4]{maclane}:
\begin{hscode}\SaveRestoreHook
\column{B}{@{}>{\hspre}l<{\hspost}@{}}%
\column{3}{@{}>{\hspre}l<{\hspost}@{}}%
\column{17}{@{}>{\hspre}c<{\hspost}@{}}%
\column{17E}{@{}l@{}}%
\column{20}{@{}>{\hspre}l<{\hspost}@{}}%
\column{E}{@{}>{\hspre}l<{\hspost}@{}}%
\>[3]{}\Varid{pureNatTrans}{}\<[17]%
\>[17]{} \mathop{:} {}\<[17E]%
\>[20]{}\{\mskip1.5mu \Conid{A},\Conid{B} \mathop{:} \Conid{Type}\mskip1.5mu\} \to (\Varid{f} \mathop{:} \Conid{A} \to \Conid{B}) \to \Varid{map}\;\Varid{f}\mathbin{\circ}\Varid{pure}\doteq\Varid{pure}\mathbin{\circ}\Varid{f}{}\<[E]%
\\[\blanklineskip]%
\>[3]{}\Varid{joinNatTrans}{}\<[17]%
\>[17]{} \mathop{:} {}\<[17E]%
\>[20]{}\{\mskip1.5mu \Conid{A},\Conid{B} \mathop{:} \Conid{Type}\mskip1.5mu\} \to (\Varid{f} \mathop{:} \Conid{A} \to \Conid{B}) \to \Varid{map}\;\Varid{f}\mathbin{\circ}\Varid{join}\doteq\Varid{join}\mathbin{\circ}\Varid{map}\;(\Varid{map}\;\Varid{f}){}\<[E]%
\ColumnHook
\end{hscode}\resethooks
and fulfil the neutrality and associativity axioms:
\begin{hscode}\SaveRestoreHook
\column{B}{@{}>{\hspre}l<{\hspost}@{}}%
\column{3}{@{}>{\hspre}l<{\hspost}@{}}%
\column{21}{@{}>{\hspre}c<{\hspost}@{}}%
\column{21E}{@{}l@{}}%
\column{24}{@{}>{\hspre}l<{\hspost}@{}}%
\column{39}{@{}>{\hspre}l<{\hspost}@{}}%
\column{E}{@{}>{\hspre}l<{\hspost}@{}}%
\>[3]{}\Varid{pureNeutralLeft}{}\<[21]%
\>[21]{} \mathop{:} {}\<[21E]%
\>[24]{}\{\mskip1.5mu \Conid{A} \mathop{:} \Conid{Type}\mskip1.5mu\} \to {}\<[39]%
\>[39]{}\Varid{join}\mathbin{\circ}\Varid{pure}\doteq\Varid{id}{}\<[E]%
\\[\blanklineskip]%
\>[3]{}\Varid{pureNeutralRight}{}\<[21]%
\>[21]{} \mathop{:} {}\<[21E]%
\>[24]{}\{\mskip1.5mu \Conid{A} \mathop{:} \Conid{Type}\mskip1.5mu\} \to {}\<[39]%
\>[39]{}\Varid{join}\mathbin{\circ}\Varid{map}\;\Varid{pure}\doteq\Varid{id}{}\<[E]%
\\[\blanklineskip]%
\>[3]{}\Varid{joinAssoc}{}\<[21]%
\>[21]{} \mathop{:} {}\<[21E]%
\>[24]{}\{\mskip1.5mu \Conid{A} \mathop{:} \Conid{Type}\mskip1.5mu\} \to {}\<[39]%
\>[39]{}\Varid{join}\mathbin{\circ}\Varid{map}\;\Varid{join}\doteq\Varid{join}\mathbin{\circ}\Varid{join}{}\<[E]%
\ColumnHook
\end{hscode}\resethooks
\end{itemize}
Notice that the above specification of the monad operations is not
minimal: \ensuremath{( \mathbin{>\!\!>\!\!=} )} is not assumed to be implemented in terms of \ensuremath{\Varid{join}}
and \ensuremath{\Varid{map}} (or \ensuremath{\Varid{map}} and \ensuremath{\Varid{join}} via \ensuremath{( \mathbin{>\!\!>\!\!=} )} and \ensuremath{\Varid{pure}}).
Therefore \ensuremath{( \mathbin{>\!\!>\!\!=} )} (pronounced ``\ensuremath{\Varid{bind}}''), \ensuremath{\Varid{join}} and
\ensuremath{\Varid{map}} have to fulfil:
\begin{hscode}\SaveRestoreHook
\column{B}{@{}>{\hspre}l<{\hspost}@{}}%
\column{3}{@{}>{\hspre}l<{\hspost}@{}}%
\column{17}{@{}>{\hspre}c<{\hspost}@{}}%
\column{17E}{@{}l@{}}%
\column{20}{@{}>{\hspre}l<{\hspost}@{}}%
\column{E}{@{}>{\hspre}l<{\hspost}@{}}%
\>[3]{}\Varid{bindJoinSpec}{}\<[17]%
\>[17]{} \mathop{:} {}\<[17E]%
\>[20]{}\{\mskip1.5mu \Conid{A},\Conid{B} \mathop{:} \Conid{Type}\mskip1.5mu\} \to (\Varid{ma} \mathop{:} \Conid{M}\;\Conid{A}) \to (\Varid{f} \mathop{:} \Conid{A} \to \Conid{M}\;\Conid{B}) \to (\Varid{ma} \mathbin{>\!\!>\!\!=} \Varid{f})\mathrel{=}\Varid{join}\;(\Varid{map}\;\Varid{f}\;\Varid{ma}){}\<[E]%
\ColumnHook
\end{hscode}\resethooks
The equality in the axioms is extensional equality, not the type
theory's definitional equality:
\begin{hscode}\SaveRestoreHook
\column{B}{@{}>{\hspre}l<{\hspost}@{}}%
\column{3}{@{}>{\hspre}l<{\hspost}@{}}%
\column{14}{@{}>{\hspre}c<{\hspost}@{}}%
\column{14E}{@{}l@{}}%
\column{17}{@{}>{\hspre}l<{\hspost}@{}}%
\column{E}{@{}>{\hspre}l<{\hspost}@{}}%
\>[3]{}(\doteq){}\<[14]%
\>[14]{} \mathop{:} {}\<[14E]%
\>[17]{}\{\mskip1.5mu \Conid{A},\Conid{B} \mathop{:} \Conid{Type}\mskip1.5mu\} \to (\Varid{f},\Varid{g} \mathop{:} \Conid{A} \to \Conid{B}) \to \Conid{Type}{}\<[E]%
\\
\>[3]{}(\doteq)\;{}\<[17]%
\>[17]{}\{\mskip1.5mu \Conid{A}\mskip1.5mu\}\;\Varid{f}\;\Varid{g}\mathrel{=}(\Varid{a} \mathop{:} \Conid{A}) \to \Varid{f}\;\Varid{a}\mathrel{=}\Varid{g}\;\Varid{a}{}\<[E]%
\ColumnHook
\end{hscode}\resethooks
As Idris does not have function extensionality, not postulating
definitional equalities makes the axioms strictly weaker.

The BJI-framework also requires \ensuremath{\Conid{M}} to be equipped with type-level membership,
with a universal quantifier and with a type-valued predicate
\begin{hscode}\SaveRestoreHook
\column{B}{@{}>{\hspre}l<{\hspost}@{}}%
\column{3}{@{}>{\hspre}l<{\hspost}@{}}%
\column{13}{@{}>{\hspre}c<{\hspost}@{}}%
\column{13E}{@{}l@{}}%
\column{16}{@{}>{\hspre}l<{\hspost}@{}}%
\column{E}{@{}>{\hspre}l<{\hspost}@{}}%
\>[3]{}\Conid{NotEmpty}{}\<[13]%
\>[13]{} \mathop{:} {}\<[13E]%
\>[16]{}\{\mskip1.5mu \Conid{A} \mathop{:} \Conid{Type}\mskip1.5mu\} \to \Conid{M}\;\Conid{A} \to \Conid{Type}{}\<[E]%
\ColumnHook
\end{hscode}\resethooks
For our purposes, the monad operations are moreover required to fulfil the
following preservation laws:

\begin{itemize}

  \item The \ensuremath{\Conid{M}}-structure obtained from lifting an element into the
    monad is not empty:
\begin{hscode}\SaveRestoreHook
\column{B}{@{}>{\hspre}l<{\hspost}@{}}%
\column{3}{@{}>{\hspre}l<{\hspost}@{}}%
\column{17}{@{}>{\hspre}c<{\hspost}@{}}%
\column{17E}{@{}l@{}}%
\column{20}{@{}>{\hspre}l<{\hspost}@{}}%
\column{E}{@{}>{\hspre}l<{\hspost}@{}}%
\>[3]{}\Varid{pureNotEmpty}{}\<[17]%
\>[17]{} \mathop{:} {}\<[17E]%
\>[20]{}\{\mskip1.5mu \Conid{A} \mathop{:} \Conid{Type}\mskip1.5mu\} \to (\Varid{a} \mathop{:} \Conid{A}) \to \Conid{NotEmpty}\;(\Varid{pure}\;\Varid{a}){}\<[E]%
\ColumnHook
\end{hscode}\resethooks
  \item The monad's \ensuremath{\Varid{map}} and \ensuremath{\Varid{bind}} preserve non-emptiness:
\begin{hscode}\SaveRestoreHook
\column{B}{@{}>{\hspre}l<{\hspost}@{}}%
\column{3}{@{}>{\hspre}l<{\hspost}@{}}%
\column{21}{@{}>{\hspre}c<{\hspost}@{}}%
\column{21E}{@{}l@{}}%
\column{24}{@{}>{\hspre}l<{\hspost}@{}}%
\column{E}{@{}>{\hspre}l<{\hspost}@{}}%
\>[3]{}\Varid{mapPresNotEmpty}{}\<[21]%
\>[21]{} \mathop{:} {}\<[21E]%
\>[24]{}\{\mskip1.5mu \Conid{A},\Conid{B} \mathop{:} \Conid{Type}\mskip1.5mu\} \to (\Varid{f} \mathop{:} \Conid{A} \to \Conid{B}) \to (\Varid{ma} \mathop{:} \Conid{M}\;\Conid{A}) \to {}\<[E]%
\\
\>[24]{}\Conid{NotEmpty}\;\Varid{ma} \to \Conid{NotEmpty}\;(\Varid{map}\;\Varid{f}\;\Varid{ma}){}\<[E]%
\ColumnHook
\end{hscode}\resethooks
\begin{hscode}\SaveRestoreHook
\column{B}{@{}>{\hspre}l<{\hspost}@{}}%
\column{3}{@{}>{\hspre}l<{\hspost}@{}}%
\column{21}{@{}>{\hspre}c<{\hspost}@{}}%
\column{21E}{@{}l@{}}%
\column{24}{@{}>{\hspre}l<{\hspost}@{}}%
\column{E}{@{}>{\hspre}l<{\hspost}@{}}%
\>[3]{}\Varid{bindPresNotEmpty}{}\<[21]%
\>[21]{} \mathop{:} {}\<[21E]%
\>[24]{}\{\mskip1.5mu \Conid{A},\Conid{B} \mathop{:} \Conid{Type}\mskip1.5mu\} \to (\Varid{f} \mathop{:} \Conid{A} \to \Conid{M}\;\Conid{B}) \to (\Varid{ma} \mathop{:} \Conid{M}\;\Conid{A}) \to {}\<[E]%
\\
\>[24]{}\Conid{NotEmpty}\;\Varid{ma} \to ((\Varid{a} \mathop{:} \Conid{A}) \to \Conid{NotEmpty}\;(\Varid{f}\;\Varid{a})) \to \Conid{NotEmpty}\;(\Varid{ma} \mathbin{>\!\!>\!\!=} \Varid{f}){}\<[E]%
\ColumnHook
\end{hscode}\resethooks
\end{itemize}

As discussed in Sec.~\ref{section:discussion}, we could have
omitted these non-emptiness preservation laws, but instead would have
implicitly restricted the class of monads for which the correctness
result holds.

\subsection{Preservation of extensional equality}
\label{appendix:presEE}

We have stated above that for our correctness proof the functor \ensuremath{\Conid{M}}
has to satisfy the monad laws and moreover its functorial \ensuremath{\Varid{map}} has to
preserve extensional equality.

This e.g.\ is the case for \ensuremath{\Conid{M}\mathrel{=}\Conid{Identity}} (no uncertainty), \ensuremath{\Conid{M}\mathrel{=}\Conid{List}}
(non-deterministic uncertainty) and for the finite probability
distributions (stochastic uncertainty, \ensuremath{\Conid{M}\mathrel{=}\Conid{Prob}}) discussed in
\citep{2017_Botta_Jansson_Ionescu}. For \ensuremath{\Conid{M}\mathrel{=}\Conid{List}} the
proof of \ensuremath{ mapPresEE } amounts to:
\begin{hscode}\SaveRestoreHook
\column{B}{@{}>{\hspre}l<{\hspost}@{}}%
\column{3}{@{}>{\hspre}l<{\hspost}@{}}%
\column{5}{@{}>{\hspre}l<{\hspost}@{}}%
\column{19}{@{}>{\hspre}l<{\hspost}@{}}%
\column{23}{@{}>{\hspre}c<{\hspost}@{}}%
\column{23E}{@{}l@{}}%
\column{26}{@{}>{\hspre}l<{\hspost}@{}}%
\column{34}{@{}>{\hspre}l<{\hspost}@{}}%
\column{36}{@{}>{\hspre}c<{\hspost}@{}}%
\column{36E}{@{}l@{}}%
\column{39}{@{}>{\hspre}l<{\hspost}@{}}%
\column{77}{@{}>{\hspre}l<{\hspost}@{}}%
\column{E}{@{}>{\hspre}l<{\hspost}@{}}%
\>[3]{} mapPresEE  \mathop{:} {}\<[19]%
\>[19]{}\{\mskip1.5mu \Conid{A},\Conid{B} \mathop{:} \Conid{Type}\mskip1.5mu\} \to (\Varid{f},\Varid{g} \mathop{:} \Conid{A} \to \Conid{B}) \to \Varid{f}\doteq\Varid{g} \to \Varid{map}\;\Varid{f}{}\<[77]%
\>[77]{}\doteq\Varid{map}\;\Varid{g}{}\<[E]%
\\[\blanklineskip]%
\>[3]{} mapPresEE \;\Varid{f}\;\Varid{g}\;\Varid{p}\;{}\<[26]%
\>[26]{}\Conid{Nil}{}\<[36]%
\>[36]{}\mathrel{=}{}\<[36E]%
\>[39]{}\Conid{Refl}{}\<[E]%
\\
\>[3]{} mapPresEE \;\Varid{f}\;\Varid{g}\;\Varid{p}\;{}\<[23]%
\>[23]{}({}\<[23E]%
\>[26]{}\Varid{a}\mathbin{::}\Varid{as}){}\<[36]%
\>[36]{}\mathrel{=}{}\<[36E]%
\\
\>[3]{}\hsindent{2}{}\<[5]%
\>[5]{}(\Varid{map}\;\Varid{f}\;(\Varid{a}\mathbin{::}\Varid{as})){}\<[34]%
\>[34]{}=\hspace{-3pt}\{\; \Conid{Refl}\;\}\hspace{-3pt}={}\<[E]%
\\
\>[3]{}\hsindent{2}{}\<[5]%
\>[5]{}(\Varid{f}\;\Varid{a}\mathbin{::}\Varid{map}\;\Varid{f}\;\Varid{as}){}\<[34]%
\>[34]{}=\hspace{-3pt}\{\; \Varid{cong}\;\{\mskip1.5mu \Varid{f}\mathrel{=}\lambda \Varid{x}\Rightarrow \Varid{x}\mathbin{::}\Varid{map}\;\Varid{f}\;\Varid{as}\mskip1.5mu\}\;(\Varid{p}\;\Varid{a})\;\}\hspace{-3pt}={}\<[E]%
\\
\>[3]{}\hsindent{2}{}\<[5]%
\>[5]{}(\Varid{g}\;\Varid{a}\mathbin{::}\Varid{map}\;\Varid{f}\;\Varid{as}){}\<[34]%
\>[34]{}=\hspace{-3pt}\{\; \Varid{cong}\;( mapPresEE \;\Varid{f}\;\Varid{g}\;\Varid{p}\;\Varid{as})\;\}\hspace{-3pt}={}\<[E]%
\\
\>[3]{}\hsindent{2}{}\<[5]%
\>[5]{}(\Varid{g}\;\Varid{a}\mathbin{::}\Varid{map}\;\Varid{g}\;\Varid{as}){}\<[34]%
\>[34]{}=\hspace{-3pt}\{\; \Conid{Refl}\;\}\hspace{-3pt}={}\<[E]%
\\
\>[3]{}\hsindent{2}{}\<[5]%
\>[5]{}(\Varid{map}\;\Varid{g}\;(\Varid{a}\mathbin{::}\Varid{as}))\;{}\<[34]%
\>[34]{}\hfill\Box{}\<[E]%
\ColumnHook
\end{hscode}\resethooks
The principle of extensional equality preservation is discussed in
more detail in \citep{botta2020extensional}.


\section{Correctness of the value function}
\label{appendix:theorem}


\begin{hscode}\SaveRestoreHook
\column{B}{@{}>{\hspre}l<{\hspost}@{}}%
\column{3}{@{}>{\hspre}l<{\hspost}@{}}%
\column{23}{@{}>{\hspre}c<{\hspost}@{}}%
\column{23E}{@{}l@{}}%
\column{26}{@{}>{\hspre}l<{\hspost}@{}}%
\column{E}{@{}>{\hspre}l<{\hspost}@{}}%
\>[3]{}\Varid{valMeasTotalReward}{}\<[23]%
\>[23]{} \mathop{:} {}\<[23E]%
\>[26]{}\{\mskip1.5mu \Varid{t},\Varid{n} \mathop{:} \mathbb{N}\mskip1.5mu\} \to (\Varid{ps} \mathop{:} \Conid{PolicySeq}\;\Varid{t}\;\Varid{n}) \to (\Varid{x} \mathop{:} \Conid{X}\;\Varid{t}) \to {}\<[E]%
\\
\>[26]{}\Varid{val'}\;\Varid{ps}\;\Varid{x}\mathrel{=}\Varid{val}\;\Varid{ps}\;\Varid{x}{}\<[E]%
\ColumnHook
\end{hscode}\resethooks
\begin{hscode}\SaveRestoreHook
\column{B}{@{}>{\hspre}l<{\hspost}@{}}%
\column{3}{@{}>{\hspre}l<{\hspost}@{}}%
\column{5}{@{}>{\hspre}l<{\hspost}@{}}%
\column{44}{@{}>{\hspre}l<{\hspost}@{}}%
\column{E}{@{}>{\hspre}l<{\hspost}@{}}%
\>[3]{}\Varid{valMeasTotalReward}\;\Conid{Nil}\;\Varid{x}\mathrel{=}{}\<[E]%
\\[\blanklineskip]%
\>[3]{}\hsindent{2}{}\<[5]%
\>[5]{}(\Varid{val'}\;\Conid{Nil}\;\Varid{x}){}\<[44]%
\>[44]{}=\hspace{-3pt}\{\; \Conid{Refl}\;\}\hspace{-3pt}={}\<[E]%
\\[\blanklineskip]%
\>[3]{}\hsindent{2}{}\<[5]%
\>[5]{}(\Varid{meas}\;(\Varid{map}\;\Varid{sumR}\;(\Varid{trj}\;\Conid{Nil}\;\Varid{x}))){}\<[44]%
\>[44]{}=\hspace{-3pt}\{\; \Conid{Refl}\;\}\hspace{-3pt}={}\<[E]%
\\[\blanklineskip]%
\>[3]{}\hsindent{2}{}\<[5]%
\>[5]{}(\Varid{meas}\;(\Varid{map}\;\Varid{sumR}\;(\Varid{pure}\;(\Conid{Last}\;\Varid{x})))){}\<[44]%
\>[44]{}=\hspace{-3pt}\{\; \Varid{cong}\;(\Varid{pureNatTrans}\;\Varid{sumR}\;(\Conid{Last}\;\Varid{x}))\;\}\hspace{-3pt}={}\<[E]%
\\[\blanklineskip]%
\>[3]{}\hsindent{2}{}\<[5]%
\>[5]{}(\Varid{meas}\;(\Varid{pure}\;(\Varid{sumR}\;(\Conid{Last}\;\Varid{x})))){}\<[44]%
\>[44]{}=\hspace{-3pt}\{\; \Conid{Refl}\;\}\hspace{-3pt}={}\<[E]%
\\[\blanklineskip]%
\>[3]{}\hsindent{2}{}\<[5]%
\>[5]{}(\Varid{meas}\;(\Varid{pure}\;\Varid{zero})){}\<[44]%
\>[44]{}=\hspace{-3pt}\{\; \Varid{measPureSpec}\;\Varid{zero}\;\}\hspace{-3pt}={}\<[E]%
\\[\blanklineskip]%
\>[3]{}\hsindent{2}{}\<[5]%
\>[5]{}(\Varid{zero}){}\<[44]%
\>[44]{}=\hspace{-3pt}\{\; \Conid{Refl}\;\}\hspace{-3pt}={}\<[E]%
\\[\blanklineskip]%
\>[3]{}\hsindent{2}{}\<[5]%
\>[5]{}(\Varid{val}\;\Conid{Nil}\;\Varid{x})\;{}\<[44]%
\>[44]{}\hfill\Box{}\<[E]%
\ColumnHook
\end{hscode}\resethooks
\begin{hscode}\SaveRestoreHook
\column{B}{@{}>{\hspre}l<{\hspost}@{}}%
\column{3}{@{}>{\hspre}l<{\hspost}@{}}%
\column{4}{@{}>{\hspre}l<{\hspost}@{}}%
\column{18}{@{}>{\hspre}l<{\hspost}@{}}%
\column{29}{@{}>{\hspre}l<{\hspost}@{}}%
\column{32}{@{}>{\hspre}l<{\hspost}@{}}%
\column{33}{@{}>{\hspre}l<{\hspost}@{}}%
\column{35}{@{}>{\hspre}l<{\hspost}@{}}%
\column{40}{@{}>{\hspre}l<{\hspost}@{}}%
\column{64}{@{}>{\hspre}l<{\hspost}@{}}%
\column{79}{@{}>{\hspre}l<{\hspost}@{}}%
\column{E}{@{}>{\hspre}l<{\hspost}@{}}%
\>[3]{}\Varid{valMeasTotalReward}\;\{\mskip1.5mu \Varid{t}\mskip1.5mu\}\;\{\mskip1.5mu \Varid{n}\mathrel{=}\Conid{S}\;\Varid{m}\mskip1.5mu\}\;(\Varid{p}\mathbin{::}\Varid{ps})\;\Varid{x}\mathrel{=}{}\<[E]%
\\[\blanklineskip]%
\>[3]{}\hsindent{1}{}\<[4]%
\>[4]{}\mbox{\onelinecomment  type abbreviations}{}\<[E]%
\\
\>[3]{}\hsindent{1}{}\<[4]%
\>[4]{}\mathbf{let}\;\Conid{SCS}{}\<[18]%
\>[18]{} \mathop{:} \Conid{Type}{}\<[40]%
\>[40]{}\mathrel{=}\Conid{StateCtrlSeq}\;(\Conid{S}\;\Varid{t})\;(\Conid{S}\;\Varid{m})\;{}\<[79]%
\>[79]{}\mathbf{in}{}\<[E]%
\\
\>[3]{}\hsindent{1}{}\<[4]%
\>[4]{}\mathbf{let}\;\Conid{SCS'}{}\<[18]%
\>[18]{} \mathop{:} \Conid{Type}{}\<[40]%
\>[40]{}\mathrel{=}\Conid{StateCtrlSeq}\;\Varid{t}\;(\Conid{S}\;(\Conid{S}\;\Varid{m}))\;{}\<[79]%
\>[79]{}\mathbf{in}{}\<[E]%
\\
\>[3]{}\hsindent{1}{}\<[4]%
\>[4]{}\mbox{\onelinecomment  element and function abbreviations}{}\<[E]%
\\
\>[3]{}\hsindent{1}{}\<[4]%
\>[4]{}\mathbf{let}\;\Varid{y}{}\<[18]%
\>[18]{} \mathop{:} \Conid{Y}\;\Varid{t}\;\Varid{x}{}\<[40]%
\>[40]{}\mathrel{=}\Varid{p}\;\Varid{x}\;{}\<[79]%
\>[79]{}\mathbf{in}{}\<[E]%
\\
\>[3]{}\hsindent{1}{}\<[4]%
\>[4]{}\mathbf{let}\;\Varid{mx'}{}\<[18]%
\>[18]{} \mathop{:} \Conid{M}\;(\Conid{X}\;(\Conid{S}\;\Varid{t})){}\<[40]%
\>[40]{}\mathrel{=}\Varid{next}\;\Varid{t}\;\Varid{x}\;\Varid{y}\;{}\<[79]%
\>[79]{}\mathbf{in}{}\<[E]%
\\
\>[3]{}\hsindent{1}{}\<[4]%
\>[4]{}\mathbf{let}\;\Varid{r}{}\<[18]%
\>[18]{} \mathop{:} (\Conid{X}\;(\Conid{S}\;\Varid{t}) \to \Conid{Val}){}\<[40]%
\>[40]{}\mathrel{=}\Varid{reward}\;\Varid{t}\;\Varid{x}\;\Varid{y}\;{}\<[79]%
\>[79]{}\mathbf{in}{}\<[E]%
\\
\>[3]{}\hsindent{1}{}\<[4]%
\>[4]{}\mathbf{let}\;\Varid{trjps}{}\<[18]%
\>[18]{} \mathop{:} (\Conid{X}\;(\Conid{S}\;\Varid{t}) \to \Conid{M}\;\Conid{SCS}){}\<[40]%
\>[40]{}\mathrel{=}\Varid{trj}\;\Varid{ps}\;{}\<[79]%
\>[79]{}\mathbf{in}{}\<[E]%
\\
\>[3]{}\hsindent{1}{}\<[4]%
\>[4]{}\mathbf{let}\;\Varid{consxy}{}\<[18]%
\>[18]{} \mathop{:} (\Conid{SCS} \to \Conid{SCS'}){}\<[40]%
\>[40]{}\mathrel{=}((\Varid{x}  \mathbin{*\!*} \Varid{y}) \mathbin{\#\!\#} )\;{}\<[79]%
\>[79]{}\mathbf{in}{}\<[E]%
\\
\>[3]{}\hsindent{1}{}\<[4]%
\>[4]{}\mathbf{let}\;\Varid{mx'trjps}{}\<[18]%
\>[18]{} \mathop{:} \Conid{M}\;\Conid{SCS}{}\<[40]%
\>[40]{}\mathrel{=}(\Varid{mx'} \mathbin{>\!\!>\!\!=} \Varid{trjps})\;{}\<[79]%
\>[79]{}\mathbf{in}{}\<[E]%
\\
\>[3]{}\hsindent{1}{}\<[4]%
\>[4]{}\mathbf{let}\;\Varid{sR}{}\<[18]%
\>[18]{} \mathop{:} (\Conid{SCS} \to \Conid{Val}){}\<[40]%
\>[40]{}\mathrel{=}\Varid{sumR}\;\{\mskip1.5mu \Varid{t}\mathrel{=}\Conid{S}\;\Varid{t}\mskip1.5mu\}\;\{\mskip1.5mu \Varid{n}\mathrel{=}\Conid{S}\;\Varid{m}\mskip1.5mu\}\;{}\<[79]%
\>[79]{}\mathbf{in}{}\<[E]%
\\
\>[3]{}\hsindent{1}{}\<[4]%
\>[4]{}\mathbf{let}\;\Varid{hd}{}\<[18]%
\>[18]{} \mathop{:} (\Conid{SCS} \to \Conid{X}\;(\Conid{S}\;\Varid{t})){}\<[40]%
\>[40]{}\mathrel{=}\Varid{head}\;\{\mskip1.5mu \Varid{t}\mathrel{=}\Conid{S}\;\Varid{t}\mskip1.5mu\}\;\{\mskip1.5mu \Varid{n}\mathrel{=}\Varid{m}\mskip1.5mu\}\;{}\<[79]%
\>[79]{}\mathbf{in}{}\<[E]%
\\
\>[3]{}\hsindent{1}{}\<[4]%
\>[4]{}\mbox{\onelinecomment  proof steps:}{}\<[E]%
\\
\>[3]{}\hsindent{1}{}\<[4]%
\>[4]{}\mathbf{let}\;\Varid{useMapPresComp}{}\<[29]%
\>[29]{}\mathrel{=}\Varid{mapPresComp}\;\Varid{consxy}\;\Varid{sumR}\;\Varid{mx'trjps}\;{}\<[79]%
\>[79]{}\mathbf{in}{}\<[E]%
\\
\>[3]{}\hsindent{1}{}\<[4]%
\>[4]{}\mathbf{let}\;\Varid{useBindJoinSpec}{}\<[29]%
\>[29]{}\mathrel{=}\Varid{bindJoinSpec}\;\{\mskip1.5mu \Conid{B}\mathrel{=}\Conid{SCS}\mskip1.5mu\}\;\Varid{trjps}\;\Varid{mx'}\;{}\<[79]%
\>[79]{}\mathbf{in}{}\<[E]%
\\
\>[3]{}\hsindent{1}{}\<[4]%
\>[4]{}\mathbf{let}\;\Varid{useAlgLemma}{}\<[29]%
\>[29]{}\mathrel{=}\Varid{algLemma}\;\{\mskip1.5mu \Conid{B}\mathrel{=}\Conid{SCS}\mskip1.5mu\}\;{}\<[E]%
\\
\>[29]{}\hsindent{3}{}\<[32]%
\>[32]{}\Varid{meas}\;\Varid{measJoinSpec}\;{}\<[E]%
\\
\>[29]{}\hsindent{3}{}\<[32]%
\>[32]{}((\Varid{r}\mathbin{\circ}\Varid{hd}) \mathbin{\medoplus} \Varid{sumR})\;\Varid{trjps}\;\Varid{mx'}\;{}\<[79]%
\>[79]{}\mathbf{in}{}\<[E]%
\\
\>[3]{}\hsindent{1}{}\<[4]%
\>[4]{}\mathbf{let}\;\Varid{useMeasSumLemma}{}\<[29]%
\>[29]{}\mathrel{=} mapPresEE \;{}\<[E]%
\\
\>[29]{}\hsindent{3}{}\<[32]%
\>[32]{}(\Varid{meas}\mathbin{\circ}\Varid{map}\;(\Varid{r}\mathbin{\circ}\Varid{hd} \mathbin{\medoplus} \Varid{sR})\mathbin{\circ}\Varid{trjps})\;{}\<[E]%
\\
\>[29]{}\hsindent{3}{}\<[32]%
\>[32]{}(\Varid{r} \mathbin{\medoplus} \Varid{meas}\mathbin{\circ}\Varid{map}\;\Varid{sR}\mathbin{\circ}\Varid{trjps})\;{}\<[E]%
\\
\>[29]{}\hsindent{3}{}\<[32]%
\>[32]{}(\Varid{measSumLemma}\;\Varid{ps}\;\Varid{r}\;\Varid{sR})\;{}\<[E]%
\\
\>[29]{}\hsindent{3}{}\<[32]%
\>[32]{}\Varid{mx'}\;{}\<[79]%
\>[79]{}\mathbf{in}{}\<[E]%
\\
\>[3]{}\hsindent{1}{}\<[4]%
\>[4]{}\mathbf{let}\;\Varid{useIH}{}\<[29]%
\>[29]{}\mathrel{=} mapPresEE \;{}\<[E]%
\\
\>[29]{}\hsindent{3}{}\<[32]%
\>[32]{}(\Varid{r} \mathbin{\medoplus} \Varid{val'}\;\Varid{ps})\;{}\<[E]%
\\
\>[29]{}\hsindent{3}{}\<[32]%
\>[32]{}(\Varid{r} \mathbin{\medoplus} \Varid{val}\;\Varid{ps})\;{}\<[E]%
\\
\>[29]{}\hsindent{3}{}\<[32]%
\>[32]{}(\Varid{oplusLiftEERight}\;(\Varid{val'}\;\Varid{ps})\;{}\<[E]%
\\
\>[32]{}\hsindent{3}{}\<[35]%
\>[35]{}(\Varid{val}\;\Varid{ps})\;\Varid{r}\;(\Varid{valMeasTotalReward}\;\Varid{ps}))\;{}\<[E]%
\\
\>[32]{}\hsindent{1}{}\<[33]%
\>[33]{}\Varid{mx'}\;{}\<[79]%
\>[79]{}\mathbf{in}{}\<[E]%
\\
\>[3]{}\hsindent{1}{}\<[4]%
\>[4]{}\mathbf{let}\;\Varid{ctx}{}\<[29]%
\>[29]{}\mathrel{=}\lambda \Varid{a}\Rightarrow \Varid{meas}\;(\Varid{map}\;((\Varid{r}\mathbin{\circ}\Varid{hd}) \mathbin{\medoplus} \Varid{sumR})\;\Varid{a})\;{}\<[79]%
\>[79]{}\mathbf{in}{}\<[E]%
\\[\blanklineskip]%
\>[3]{}\hsindent{1}{}\<[4]%
\>[4]{}(\Varid{val'}\;(\Varid{p}\mathbin{::}\Varid{ps})\;\Varid{x}){}\<[64]%
\>[64]{}=\hspace{-3pt}\{\; \Conid{Refl}\;\}\hspace{-3pt}={}\<[E]%
\\[\blanklineskip]%
\>[3]{}\hsindent{1}{}\<[4]%
\>[4]{}(\Varid{meas}\;(\Varid{map}\;\Varid{sumR}\;(\Varid{trj}\;(\Varid{p}\mathbin{::}\Varid{ps})\;\Varid{x}))){}\<[64]%
\>[64]{}=\hspace{-3pt}\{\; \Conid{Refl}\;\}\hspace{-3pt}={}\<[E]%
\\[\blanklineskip]%
\>[3]{}\hsindent{1}{}\<[4]%
\>[4]{}(\Varid{meas}\;(\Varid{map}\;\Varid{sumR}\;(\Varid{map}\;\Varid{consxy}\;\Varid{mx'trjps}))){}\<[64]%
\>[64]{}=\hspace{-3pt}\{\; \Varid{cong}\;(\Varid{sym}\;\Varid{useMapPresComp})\;\}\hspace{-3pt}={}\<[E]%
\\[\blanklineskip]%
\>[3]{}\hsindent{1}{}\<[4]%
\>[4]{}(\Varid{meas}\;(\Varid{map}\;(\Varid{sumR}\mathbin{\circ}\Varid{consxy})\;\Varid{mx'trjps})){}\<[64]%
\>[64]{}=\hspace{-3pt}\{\; \Conid{Refl}\;\}\hspace{-3pt}={}\<[E]%
\\[\blanklineskip]%
\>[3]{}\hsindent{1}{}\<[4]%
\>[4]{}(\Varid{meas}\;(\Varid{map}\;((\Varid{r}\mathbin{\circ}\Varid{hd}) \mathbin{\medoplus} \Varid{sR})\;\Varid{mx'trjps})){}\<[64]%
\>[64]{}=\hspace{-3pt}\{\; \Varid{cong}\;\{\mskip1.5mu \Varid{f}\mathrel{=}\Varid{ctx}\mskip1.5mu\}\;\Varid{useBindJoinSpec}\;\}\hspace{-3pt}={}\<[E]%
\\[\blanklineskip]%
\>[3]{}\hsindent{1}{}\<[4]%
\>[4]{}(\Varid{meas}\;(\Varid{map}\;((\Varid{r}\mathbin{\circ}\Varid{hd}) \mathbin{\medoplus} \Varid{sR})\;(\Varid{join}\;(\Varid{map}\;\Varid{trjps}\;\Varid{mx'})))){}\<[64]%
\>[64]{}=\hspace{-3pt}\{\; \Varid{sym}\;\Varid{useAlgLemma}\;\}\hspace{-3pt}={}\<[E]%
\\[\blanklineskip]%
\>[3]{}\hsindent{1}{}\<[4]%
\>[4]{}(\Varid{meas}\;(\Varid{map}\;(\Varid{meas}\mathbin{\circ}\Varid{map}\;(\Varid{r}\mathbin{\circ}\Varid{hd} \mathbin{\medoplus} \Varid{sR})\mathbin{\circ}\Varid{trjps})\;\Varid{mx'})){}\<[64]%
\>[64]{}=\hspace{-3pt}\{\; \Varid{cong}\;\Varid{useMeasSumLemma}\;\}\hspace{-3pt}={}\<[E]%
\\[\blanklineskip]%
\>[3]{}\hsindent{1}{}\<[4]%
\>[4]{}(\Varid{meas}\;(\Varid{map}\;(\Varid{r} \mathbin{\medoplus} \Varid{meas}\mathbin{\circ}\Varid{map}\;\Varid{sR}\mathbin{\circ}\Varid{trjps})\;\Varid{mx'})){}\<[64]%
\>[64]{}=\hspace{-3pt}\{\; \Conid{Refl}\;\}\hspace{-3pt}={}\<[E]%
\\[\blanklineskip]%
\>[3]{}\hsindent{1}{}\<[4]%
\>[4]{}(\Varid{meas}\;(\Varid{map}\;(\Varid{r} \mathbin{\medoplus} \Varid{val'}\;\Varid{ps})\;\Varid{mx'})){}\<[64]%
\>[64]{}=\hspace{-3pt}\{\; \Varid{cong}\;\Varid{useIH}\;\}\hspace{-3pt}={}\<[E]%
\\[\blanklineskip]%
\>[3]{}\hsindent{1}{}\<[4]%
\>[4]{}(\Varid{meas}\;(\Varid{map}\;(\Varid{r} \mathbin{\medoplus} \Varid{val}\;\Varid{ps})\;\Varid{mx'})){}\<[64]%
\>[64]{}=\hspace{-3pt}\{\; \Conid{Refl}\;\}\hspace{-3pt}={}\<[E]%
\\[\blanklineskip]%
\>[3]{}\hsindent{1}{}\<[4]%
\>[4]{}(\Varid{val}\;(\Varid{p}\mathbin{::}\Varid{ps})\;\Varid{x})\;{}\<[64]%
\>[64]{}\hfill\Box{}\<[E]%
\ColumnHook
\end{hscode}\resethooks


\section{Correctness of monadic backward induction}
\label{appendix:biCorrectness}


Together with the result of \bottaetal (\ensuremath{\Varid{biOptVal}}, see
Appendix~\ref{appendix:bilemma} below) we can prove the correctness
of monadic backward induction as corollary, using a
generalised optimality of policy sequences predicate:
\begin{hscode}\SaveRestoreHook
\column{B}{@{}>{\hspre}l<{\hspost}@{}}%
\column{3}{@{}>{\hspre}l<{\hspost}@{}}%
\column{20}{@{}>{\hspre}c<{\hspost}@{}}%
\column{20E}{@{}l@{}}%
\column{23}{@{}>{\hspre}l<{\hspost}@{}}%
\column{33}{@{}>{\hspre}c<{\hspost}@{}}%
\column{33E}{@{}l@{}}%
\column{36}{@{}>{\hspre}l<{\hspost}@{}}%
\column{E}{@{}>{\hspre}l<{\hspost}@{}}%
\>[3]{}\Conid{GenOptPolicySeq}{}\<[20]%
\>[20]{} \mathop{:} {}\<[20E]%
\>[23]{}\{\mskip1.5mu \Varid{t},\Varid{n} \mathop{:} \mathbb{N}\mskip1.5mu\} \to (\Varid{f} \mathop{:} \Conid{PolicySeq}\;\Varid{t}\;\Varid{n} \to \Conid{X}\;\Varid{t} \to \Conid{Val}) \to \Conid{PolicySeq}\;\Varid{t}\;\Varid{n} \to \Conid{Type}{}\<[E]%
\\[\blanklineskip]%
\>[3]{}\Conid{GenOptPolicySeq}\;\{\mskip1.5mu \Varid{t}\mskip1.5mu\}\;\{\mskip1.5mu \Varid{n}\mskip1.5mu\}\;\Varid{f}\;\Varid{ps}{}\<[33]%
\>[33]{}\mathrel{=}{}\<[33E]%
\>[36]{}(\Varid{ps'} \mathop{:} \Conid{PolicySeq}\;\Varid{t}\;\Varid{n}) \to (\Varid{x} \mathop{:} \Conid{X}\;\Varid{t}) \to \Varid{f}\;\Varid{ps'}\;\Varid{x} \,\sqsubseteq\, \Varid{f}\;\Varid{ps}\;\Varid{x}{}\<[E]%
\ColumnHook
\end{hscode}\resethooks
\begin{hscode}\SaveRestoreHook
\column{B}{@{}>{\hspre}l<{\hspost}@{}}%
\column{3}{@{}>{\hspre}l<{\hspost}@{}}%
\column{6}{@{}>{\hspre}l<{\hspost}@{}}%
\column{18}{@{}>{\hspre}l<{\hspost}@{}}%
\column{25}{@{}>{\hspre}c<{\hspost}@{}}%
\column{25E}{@{}l@{}}%
\column{28}{@{}>{\hspre}l<{\hspost}@{}}%
\column{57}{@{}>{\hspre}l<{\hspost}@{}}%
\column{61}{@{}>{\hspre}l<{\hspost}@{}}%
\column{E}{@{}>{\hspre}l<{\hspost}@{}}%
\>[3]{}\Varid{biOptMeasTotalReward}{}\<[25]%
\>[25]{} \mathop{:} {}\<[25E]%
\>[28]{}(\Varid{t} \mathop{:} \mathbb{N}) \to (\Varid{n} \mathop{:} \mathbb{N}) \to \Conid{GenOptPolicySeq}\;\Varid{val'}\;(\Varid{bi}\;\Varid{t}\;\Varid{n}){}\<[E]%
\\[\blanklineskip]%
\>[3]{}\Varid{biOptMeasTotalReward}\;\Varid{t}\;\Varid{n}\;\Varid{ps'}\;\Varid{x}\mathrel{=}{}\<[E]%
\\[\blanklineskip]%
\>[3]{}\hsindent{3}{}\<[6]%
\>[6]{}\mathbf{let}\;\Varid{vvEqL}{}\<[18]%
\>[18]{}\mathrel{=}\Varid{sym}\;(\Varid{valMeasTotalReward}\;\Varid{ps'}\;\Varid{x})\;{}\<[57]%
\>[57]{}\mathbf{in}\;{}\<[61]%
\>[61]{}\mbox{\onelinecomment  \ensuremath{\Varid{val'}\;\Varid{ps'}\;\Varid{x}\mathrel{=}\Varid{val}\;\Varid{ps'}\;\Varid{x}}}{}\<[E]%
\\
\>[3]{}\hsindent{3}{}\<[6]%
\>[6]{}\mathbf{let}\;\Varid{vvEqR}{}\<[18]%
\>[18]{}\mathrel{=}\Varid{sym}\;(\Varid{valMeasTotalReward}\;(\Varid{bi}\;\Varid{t}\;\Varid{n})\;\Varid{x})\;{}\<[57]%
\>[57]{}\mathbf{in}\;{}\<[61]%
\>[61]{}\mbox{\onelinecomment  \ensuremath{\Varid{val'}\;(\Varid{bi}\;\Varid{t}\;\Varid{n})\;\Varid{x}\mathrel{=}\Varid{val}\;(\Varid{bi}\;\Varid{t}\;\Varid{n})\;\Varid{x}}}{}\<[E]%
\\
\>[3]{}\hsindent{3}{}\<[6]%
\>[6]{}\mathbf{let}\;\Varid{biOpt}{}\<[18]%
\>[18]{}\mathrel{=}\Varid{biOptVal}\;\Varid{t}\;\Varid{n}\;\Varid{ps'}\;\Varid{x}\;{}\<[57]%
\>[57]{}\mathbf{in}\;{}\<[61]%
\>[61]{}\mbox{\onelinecomment  \ensuremath{\Varid{val}\;\Varid{ps'}\;\Varid{x} \,\sqsubseteq\, \Varid{val}\;(\Varid{bi}\;\Varid{t}\;\Varid{n})\;\Varid{x}}}{}\<[E]%
\\
\>[3]{}\hsindent{3}{}\<[6]%
\>[6]{}\mathbf{let}\;\Varid{lP}{}\<[18]%
\>[18]{}\mathrel{=}\lambda \Varid{v}\Rightarrow \Varid{v} \,\sqsubseteq\, \Varid{val}\;(\Varid{bi}\;\Varid{t}\;\Varid{n})\;\Varid{x}\;{}\<[57]%
\>[57]{}\mathbf{in}{}\<[E]%
\\
\>[3]{}\hsindent{3}{}\<[6]%
\>[6]{}\mathbf{let}\;\Varid{rP}{}\<[18]%
\>[18]{}\mathrel{=}\lambda \Varid{v}\Rightarrow \Varid{val'}\;\Varid{ps'}\;\Varid{x} \,\sqsubseteq\, \Varid{v}\;{}\<[57]%
\>[57]{}\mathbf{in}{}\<[E]%
\\[\blanklineskip]%
\>[3]{}\hsindent{3}{}\<[6]%
\>[6]{}\Varid{replace}\;\{\mskip1.5mu \Conid{P}\mathrel{=}\Varid{rP}\mskip1.5mu\}\;\Varid{vvEqR}\;(\Varid{replace}\;\{\mskip1.5mu \Conid{P}\mathrel{=}\Varid{lP}\mskip1.5mu\}\;\Varid{vvEqL}\;\Varid{biOpt}){}\<[E]%
\ColumnHook
\end{hscode}\resethooks


\section{Lemmas}
\label{appendix:lemmas}


The proof of \ensuremath{\Varid{valMeasTotalReward}} relies
on a few auxiliary results which we prove here.\\

\noindent Lemma about the interaction of \ensuremath{\Varid{head}} and \ensuremath{\Varid{trj}} interleaved with \ensuremath{\Varid{map}}:
\begin{hscode}\SaveRestoreHook
\column{B}{@{}>{\hspre}l<{\hspost}@{}}%
\column{3}{@{}>{\hspre}l<{\hspost}@{}}%
\column{17}{@{}>{\hspre}c<{\hspost}@{}}%
\column{17E}{@{}l@{}}%
\column{20}{@{}>{\hspre}l<{\hspost}@{}}%
\column{57}{@{}>{\hspre}c<{\hspost}@{}}%
\column{57E}{@{}l@{}}%
\column{60}{@{}>{\hspre}l<{\hspost}@{}}%
\column{E}{@{}>{\hspre}l<{\hspost}@{}}%
\>[3]{}\Varid{headTrjLemma}{}\<[17]%
\>[17]{} \mathop{:} {}\<[17E]%
\>[20]{}\{\mskip1.5mu \Varid{t},\Varid{n} \mathop{:} \mathbb{N}\mskip1.5mu\} \to (\Varid{ps} \mathop{:} \Conid{PolicySeq}\;\Varid{t}\;\Varid{n}) \to {}\<[E]%
\\
\>[20]{}(\Varid{r} \mathop{:} \Conid{X}\;\Varid{t} \to \Conid{Val}) \to (\Varid{s} \mathop{:} \Conid{StateCtrlSeq}\;\Varid{t}\;(\Conid{S}\;\Varid{n}) \to \Conid{Val}) \to (\Varid{x'} \mathop{:} \Conid{X}\;\Varid{t}) \to {}\<[E]%
\\
\>[20]{}(\Varid{map}\;(\Varid{r}\mathbin{\circ}\Varid{head} \mathbin{\medoplus} \Varid{s})\mathbin{\circ}\Varid{trj}\;\Varid{ps})\;\Varid{x'}{}\<[57]%
\>[57]{}\mathrel{=}{}\<[57E]%
\>[60]{}(\Varid{map}\;(\Varid{const}\;(\Varid{r}\;\Varid{x'}) \mathbin{\medoplus} \Varid{s})\mathbin{\circ}\Varid{trj}\;\Varid{ps})\;\Varid{x'}{}\<[E]%
\ColumnHook
\end{hscode}\resethooks
\begin{hscode}\SaveRestoreHook
\column{B}{@{}>{\hspre}l<{\hspost}@{}}%
\column{3}{@{}>{\hspre}l<{\hspost}@{}}%
\column{11}{@{}>{\hspre}l<{\hspost}@{}}%
\column{13}{@{}>{\hspre}l<{\hspost}@{}}%
\column{21}{@{}>{\hspre}l<{\hspost}@{}}%
\column{23}{@{}>{\hspre}l<{\hspost}@{}}%
\column{26}{@{}>{\hspre}l<{\hspost}@{}}%
\column{34}{@{}>{\hspre}l<{\hspost}@{}}%
\column{35}{@{}>{\hspre}l<{\hspost}@{}}%
\column{58}{@{}>{\hspre}l<{\hspost}@{}}%
\column{69}{@{}>{\hspre}l<{\hspost}@{}}%
\column{83}{@{}>{\hspre}l<{\hspost}@{}}%
\column{92}{@{}>{\hspre}l<{\hspost}@{}}%
\column{E}{@{}>{\hspre}l<{\hspost}@{}}%
\>[3]{}\Varid{headTrjLemma}\;\{\mskip1.5mu \Varid{t}\mskip1.5mu\}\;\{\mskip1.5mu \Varid{n}\mathrel{=}\Conid{Z}\mskip1.5mu\}\;\Conid{Nil}\;\Varid{r}\;\Varid{s}\;\Varid{x'}\mathrel{=}{}\<[E]%
\\[\blanklineskip]%
\>[3]{}\hsindent{8}{}\<[11]%
\>[11]{}\mathbf{let}\;\Varid{hd}{}\<[23]%
\>[23]{}\mathrel{=}\Varid{head}\;\{\mskip1.5mu \Varid{t}\mskip1.5mu\}\;{}\<[83]%
\>[83]{}\mathbf{in}{}\<[E]%
\\
\>[3]{}\hsindent{8}{}\<[11]%
\>[11]{}\mathbf{let}\;\Varid{lastx'}{}\<[23]%
\>[23]{}\mathrel{=}\Conid{Last}\;\{\mskip1.5mu \Varid{t}\mskip1.5mu\}\;\Varid{x'}\;{}\<[83]%
\>[83]{}\mathbf{in}{}\<[E]%
\\
\>[3]{}\hsindent{8}{}\<[11]%
\>[11]{}\mathbf{let}\;\Varid{nil}{}\<[23]%
\>[23]{}\mathrel{=}\Conid{Nil}\;\{\mskip1.5mu \Varid{t}\mskip1.5mu\}\;{}\<[83]%
\>[83]{}\mathbf{in}{}\<[E]%
\\
\>[3]{}\hsindent{8}{}\<[11]%
\>[11]{}\mathbf{let}\;\Varid{usePureNatTrans}{}\<[35]%
\>[35]{}\mathrel{=}\Varid{pureNatTrans}\;(\Varid{r}\mathbin{\circ}\Varid{hd} \mathbin{\medoplus} \Varid{s})\;\Varid{lastx'}\;{}\<[83]%
\>[83]{}\mathbf{in}{}\<[E]%
\\
\>[3]{}\hsindent{8}{}\<[11]%
\>[11]{}\mathbf{let}\;\Varid{usePureNatTransSym}{}\<[35]%
\>[35]{}\mathrel{=}\Varid{sym}\;(\Varid{pureNatTrans}\;(( \mathbin{\oplus} )\;(\Varid{r}\;\Varid{x'})\mathbin{\circ}\Varid{s})\;\Varid{lastx'})\;{}\<[83]%
\>[83]{}\mathbf{in}{}\<[E]%
\\[\blanklineskip]%
\>[11]{}\hsindent{2}{}\<[13]%
\>[13]{}((\Varid{map}\;(\Varid{r}\mathbin{\circ}\Varid{hd} \mathbin{\medoplus} \Varid{s}))\;(\Varid{trj}\;\Varid{nil}\;\Varid{x'})){}\<[58]%
\>[58]{}=\hspace{-3pt}\{\; \Conid{Refl}\;\}\hspace{-3pt}={}\<[E]%
\\[\blanklineskip]%
\>[11]{}\hsindent{2}{}\<[13]%
\>[13]{}((\Varid{map}\;(\Varid{r}\mathbin{\circ}\Varid{hd} \mathbin{\medoplus} \Varid{s}))\;(\Varid{pure}\;\Varid{lastx'})){}\<[58]%
\>[58]{}=\hspace{-3pt}\{\; \Varid{usePureNatTrans}\;\}\hspace{-3pt}={}\<[E]%
\\[\blanklineskip]%
\>[11]{}\hsindent{2}{}\<[13]%
\>[13]{}((\Varid{pure}\mathbin{\circ}(\Varid{r}\mathbin{\circ}\Varid{hd} \mathbin{\medoplus} \Varid{s}))\;\Varid{lastx'}){}\<[58]%
\>[58]{}=\hspace{-3pt}\{\; \Conid{Refl}\;\}\hspace{-3pt}={}\<[E]%
\\[\blanklineskip]%
\>[11]{}\hsindent{2}{}\<[13]%
\>[13]{}((\Varid{pure}\mathbin{\circ}(\Varid{const}\;(\Varid{r}\;\Varid{x'}) \mathbin{\medoplus} \Varid{s}))\;\Varid{lastx'}){}\<[58]%
\>[58]{}=\hspace{-3pt}\{\; \Varid{usePureNatTransSym}\;\}\hspace{-3pt}={}\<[E]%
\\[\blanklineskip]%
\>[11]{}\hsindent{2}{}\<[13]%
\>[13]{}(\Varid{map}\;(\Varid{const}\;(\Varid{r}\;\Varid{x'}) \mathbin{\medoplus} \Varid{s})\;(\Varid{pure}\;\Varid{lastx'})){}\<[58]%
\>[58]{}=\hspace{-3pt}\{\; \Conid{Refl}\;\}\hspace{-3pt}={}\<[E]%
\\[\blanklineskip]%
\>[11]{}\hsindent{2}{}\<[13]%
\>[13]{}(\Varid{map}\;(\Varid{const}\;(\Varid{r}\;\Varid{x'}) \mathbin{\medoplus} \Varid{s})\;(\Varid{trj}\;\Varid{nil}\;\Varid{x'}))\;{}\<[58]%
\>[58]{}\hfill\Box{}\<[E]%
\\[\blanklineskip]%
\>[3]{}\Varid{headTrjLemma}\;\{\mskip1.5mu \Varid{t}\mathrel{=}\Varid{t}\mskip1.5mu\}\;\{\mskip1.5mu \Varid{n}\mathrel{=}\Conid{S}\;\Varid{n}\mskip1.5mu\}\;(\Varid{p}\mathbin{::}\Varid{ps})\;\Varid{r}\;\Varid{s}\;\Varid{x'}\mathrel{=}{}\<[E]%
\\[\blanklineskip]%
\>[3]{}\hsindent{8}{}\<[11]%
\>[11]{}\mathbf{let}\;\Varid{y'}{}\<[21]%
\>[21]{}\mathrel{=}\Varid{p}\;\Varid{x'}\;{}\<[92]%
\>[92]{}\mathbf{in}{}\<[E]%
\\
\>[3]{}\hsindent{8}{}\<[11]%
\>[11]{}\mathbf{let}\;\Varid{mx''}{}\<[21]%
\>[21]{}\mathrel{=}\Varid{next}\;\Varid{t}\;\Varid{x'}\;\Varid{y'}\;{}\<[92]%
\>[92]{}\mathbf{in}{}\<[E]%
\\
\>[3]{}\hsindent{8}{}\<[11]%
\>[11]{}\mathbf{let}\;\Conid{SCS}{}\<[21]%
\>[21]{}\mathrel{=}(\Conid{StateCtrlSeq}\;(\Conid{S}\;\Varid{t})\;(\Conid{S}\;\Varid{n}))\;{}\<[92]%
\>[92]{}\mathbf{in}{}\<[E]%
\\
\>[3]{}\hsindent{8}{}\<[11]%
\>[11]{}\mathbf{let}\;\Varid{consx'y'}{}\<[26]%
\>[26]{}\mathrel{=}( \mathbin{\#\!\#} )\;\{\mskip1.5mu \Varid{t}\mskip1.5mu\}\;\{\mskip1.5mu \Varid{n}\mskip1.5mu\}\;(\Varid{x'}  \mathbin{*\!*} \Varid{y'})\;{}\<[92]%
\>[92]{}\mathbf{in}{}\<[E]%
\\
\>[3]{}\hsindent{8}{}\<[11]%
\>[11]{}\mathbf{let}\;\Varid{mx''trjps}{}\<[26]%
\>[26]{}\mathrel{=}( \mathbin{>\!\!>\!\!=} )\;\{\mskip1.5mu \Conid{B}\mathrel{=}\Conid{SCS}\mskip1.5mu\}\;\Varid{mx''}\;(\Varid{trj}\;\Varid{ps})\;{}\<[92]%
\>[92]{}\mathbf{in}{}\<[E]%
\\
\>[3]{}\hsindent{8}{}\<[11]%
\>[11]{}\mathbf{let}\;\Varid{traj}{}\<[26]%
\>[26]{}\mathrel{=}\Varid{trj}\;\{\mskip1.5mu \Varid{t}\mskip1.5mu\}\;{}\<[92]%
\>[92]{}\mathbf{in}{}\<[E]%
\\
\>[3]{}\hsindent{8}{}\<[11]%
\>[11]{}\mathbf{let}\;\Varid{useMapPresComp}{}\<[34]%
\>[34]{}\mathrel{=}\Varid{mapPresComp}\;\Varid{consx'y'}\;(\Varid{const}\;(\Varid{r}\;\Varid{x'}) \mathbin{\medoplus} \Varid{s})\;\Varid{mx''trjps}\;{}\<[92]%
\>[92]{}\mathbf{in}{}\<[E]%
\\
\>[3]{}\hsindent{8}{}\<[11]%
\>[11]{}\mathbf{let}\;\Varid{useMapPresCompSym}{}\<[34]%
\>[34]{}\mathrel{=}\Varid{sym}\;(\Varid{mapPresComp}\;\Varid{consx'y'}\;(\Varid{r}\mathbin{\circ}\Varid{head} \mathbin{\medoplus} \Varid{s})\;\Varid{mx''trjps})\;{}\<[92]%
\>[92]{}\mathbf{in}{}\<[E]%
\\[\blanklineskip]%
\>[11]{}\hsindent{2}{}\<[13]%
\>[13]{}((\Varid{map}\;(\Varid{r}\mathbin{\circ}\Varid{head} \mathbin{\medoplus} \Varid{s}))\;(\Varid{traj}\;(\Varid{p}\mathbin{::}\Varid{ps})\;\Varid{x'})){}\<[69]%
\>[69]{}=\hspace{-3pt}\{\; \Conid{Refl}\;\}\hspace{-3pt}={}\<[E]%
\\[\blanklineskip]%
\>[11]{}\hsindent{2}{}\<[13]%
\>[13]{}((\Varid{map}\;(\Varid{r}\mathbin{\circ}\Varid{head} \mathbin{\medoplus} \Varid{s})\mathbin{\circ}\Varid{map}\;\Varid{consx'y'})\;\Varid{mx''trjps}){}\<[69]%
\>[69]{}=\hspace{-3pt}\{\; \Varid{useMapPresCompSym}\;\}\hspace{-3pt}={}\<[E]%
\\[\blanklineskip]%
\>[11]{}\hsindent{2}{}\<[13]%
\>[13]{}((\Varid{map}\;((\Varid{r}\mathbin{\circ}\Varid{head} \mathbin{\medoplus} \Varid{s})\mathbin{\circ}\Varid{consx'y'}))\;\Varid{mx''trjps}){}\<[69]%
\>[69]{}=\hspace{-3pt}\{\; \Conid{Refl}\;\}\hspace{-3pt}={}\<[E]%
\\[\blanklineskip]%
\>[11]{}\hsindent{2}{}\<[13]%
\>[13]{}(\Varid{map}\;((\Varid{const}\;(\Varid{r}\;\Varid{x'}) \mathbin{\medoplus} \Varid{s})\mathbin{\circ}\Varid{consx'y'})\;\Varid{mx''trjps}){}\<[69]%
\>[69]{}=\hspace{-3pt}\{\; \Varid{useMapPresComp}\;\}\hspace{-3pt}={}\<[E]%
\\[\blanklineskip]%
\>[11]{}\hsindent{2}{}\<[13]%
\>[13]{}(\Varid{map}\;(\Varid{const}\;(\Varid{r}\;\Varid{x'}) \mathbin{\medoplus} \Varid{s})\;(\Varid{map}\;\Varid{consx'y'}\;\Varid{mx''trjps})){}\<[69]%
\>[69]{}=\hspace{-3pt}\{\; \Conid{Refl}\;\}\hspace{-3pt}={}\<[E]%
\\[\blanklineskip]%
\>[11]{}\hsindent{2}{}\<[13]%
\>[13]{}(\Varid{map}\;(\Varid{const}\;(\Varid{r}\;\Varid{x'}) \mathbin{\medoplus} \Varid{s})\;(\Varid{traj}\;(\Varid{p}\mathbin{::}\Varid{ps})\;\Varid{x'}))\;{}\<[69]%
\>[69]{}\hfill\Box{}\<[E]%
\ColumnHook
\end{hscode}\resethooks
Lemma about the commutation of \ensuremath{\Varid{meas}\mathbin{\circ}\Varid{map}} and \ensuremath{ \mathbin{\medoplus} }:
\begin{hscode}\SaveRestoreHook
\column{B}{@{}>{\hspre}l<{\hspost}@{}}%
\column{3}{@{}>{\hspre}l<{\hspost}@{}}%
\column{17}{@{}>{\hspre}c<{\hspost}@{}}%
\column{17E}{@{}l@{}}%
\column{20}{@{}>{\hspre}l<{\hspost}@{}}%
\column{E}{@{}>{\hspre}l<{\hspost}@{}}%
\>[3]{}\Varid{measSumLemma}{}\<[17]%
\>[17]{} \mathop{:} {}\<[17E]%
\>[20]{}\{\mskip1.5mu \Varid{t},\Varid{n} \mathop{:} \mathbb{N}\mskip1.5mu\} \to (\Varid{ps} \mathop{:} \Conid{PolicySeq}\;\Varid{t}\;\Varid{n}) \to {}\<[E]%
\\
\>[20]{}(\Varid{r} \mathop{:} \Conid{X}\;\Varid{t} \to \Conid{Val}) \to {}\<[E]%
\\
\>[20]{}(\Varid{s} \mathop{:} \Conid{StateCtrlSeq}\;\Varid{t}\;(\Conid{S}\;\Varid{n}) \to \Conid{Val}) \to {}\<[E]%
\\
\>[20]{}(\Varid{meas}\mathbin{\circ}\Varid{map}\;(\Varid{r}\mathbin{\circ}\Varid{head} \mathbin{\medoplus} \Varid{s})\mathbin{\circ}\Varid{trj}\;\Varid{ps})\doteq{}\<[E]%
\\
\>[20]{}(\Varid{r} \mathbin{\medoplus} \Varid{meas}\mathbin{\circ}\Varid{map}\;\Varid{s}\mathbin{\circ}\Varid{trj}\;\Varid{ps}){}\<[E]%
\ColumnHook
\end{hscode}\resethooks
\begin{hscode}\SaveRestoreHook
\column{B}{@{}>{\hspre}l<{\hspost}@{}}%
\column{3}{@{}>{\hspre}l<{\hspost}@{}}%
\column{5}{@{}>{\hspre}l<{\hspost}@{}}%
\column{11}{@{}>{\hspre}l<{\hspost}@{}}%
\column{32}{@{}>{\hspre}l<{\hspost}@{}}%
\column{38}{@{}>{\hspre}l<{\hspost}@{}}%
\column{66}{@{}>{\hspre}l<{\hspost}@{}}%
\column{75}{@{}>{\hspre}l<{\hspost}@{}}%
\column{83}{@{}>{\hspre}l<{\hspost}@{}}%
\column{E}{@{}>{\hspre}l<{\hspost}@{}}%
\>[3]{}\Varid{measSumLemma}\;\{\mskip1.5mu \Varid{t}\mskip1.5mu\}\;\{\mskip1.5mu \Varid{n}\mskip1.5mu\}\;\Varid{ps}\;\Varid{r}\;\Varid{s}\;\Varid{x'}\mathrel{=}{}\<[E]%
\\[\blanklineskip]%
\>[3]{}\hsindent{8}{}\<[11]%
\>[11]{}\mbox{\onelinecomment  non-emptiness proofs}{}\<[E]%
\\
\>[3]{}\hsindent{8}{}\<[11]%
\>[11]{}\mathbf{let}\;\Varid{trjNE}{}\<[32]%
\>[32]{}\mathrel{=}\Varid{trjNotEmptyLemma}\;\Varid{ps}\;\Varid{x'}\;{}\<[83]%
\>[83]{}\mathbf{in}{}\<[E]%
\\
\>[3]{}\hsindent{8}{}\<[11]%
\>[11]{}\mathbf{let}\;\Varid{sNE}{}\<[32]%
\>[32]{}\mathrel{=}\Varid{mapPresNotEmpty}\;\Varid{s}\;(\Varid{trj}\;\Varid{ps}\;\Varid{x'})\;\Varid{trjNE}\;{}\<[83]%
\>[83]{}\mathbf{in}{}\<[E]%
\\
\>[3]{}\hsindent{8}{}\<[11]%
\>[11]{}\mbox{\onelinecomment  proof steps}{}\<[E]%
\\
\>[3]{}\hsindent{8}{}\<[11]%
\>[11]{}\mathbf{let}\;\Varid{useMeasPlusSpec}{}\<[32]%
\>[32]{}\mathrel{=}\Varid{measPlusSpec}\;(\Varid{r}\;\Varid{x'})\;(\Varid{map}\;\Varid{s}\;(\Varid{trj}\;\Varid{ps}\;\Varid{x'}))\;\Varid{sNE}\;{}\<[83]%
\>[83]{}\mathbf{in}{}\<[E]%
\\
\>[3]{}\hsindent{8}{}\<[11]%
\>[11]{}\mathbf{let}\;\Varid{useMapPresComp}{}\<[32]%
\>[32]{}\mathrel{=}\Varid{cong}\;\{\mskip1.5mu \Varid{f}\mathbin{=\char92 }\Varid{prf}\Rightarrow \Varid{meas}\;\Varid{prf}\mskip1.5mu\}\;{}\<[E]%
\\
\>[32]{}\hsindent{6}{}\<[38]%
\>[38]{}(\Varid{mapPresComp}\;\Varid{s}\;((\Varid{r}\;\Varid{x'}) \mathbin{\oplus} )\;(\Varid{trj}\;\Varid{ps}\;\Varid{x'}))\;{}\<[83]%
\>[83]{}\mathbf{in}{}\<[E]%
\\
\>[3]{}\hsindent{8}{}\<[11]%
\>[11]{}\mathbf{let}\;\Varid{useHdTrjLemma}{}\<[32]%
\>[32]{}\mathrel{=}\Varid{cong}\;(\Varid{headTrjLemma}\;\Varid{ps}\;\Varid{r}\;\Varid{s}\;\Varid{x'})\;{}\<[83]%
\>[83]{}\mathbf{in}{}\<[E]%
\\[\blanklineskip]%
\>[3]{}\hsindent{2}{}\<[5]%
\>[5]{}((\Varid{meas}\mathbin{\circ}\Varid{map}\;(\Varid{r}\mathbin{\circ}\Varid{head} \mathbin{\medoplus} \Varid{s})\mathbin{\circ}\Varid{trj}\;\Varid{ps})\;\Varid{x'}){}\<[66]%
\>[66]{}=\hspace{-3pt}\{\; \Varid{useHdTrjLemma}\;\}\hspace{-3pt}={}\<[E]%
\\[\blanklineskip]%
\>[3]{}\hsindent{2}{}\<[5]%
\>[5]{}((\Varid{meas}\mathbin{\circ}\Varid{map}\;(\Varid{const}\;(\Varid{r}\;\Varid{x'}) \mathbin{\medoplus} \Varid{s})\mathbin{\circ}\Varid{trj}\;\Varid{ps})\;\Varid{x'}){}\<[66]%
\>[66]{}=\hspace{-3pt}\{\; \Conid{Refl}{}\<[75]%
\>[75]{}\;\}\hspace{-3pt}={}\<[E]%
\\[\blanklineskip]%
\>[3]{}\hsindent{2}{}\<[5]%
\>[5]{}((\Varid{meas}\mathbin{\circ}\Varid{map}\;((\Varid{const}\;(\Varid{r}\;\Varid{x'}) \mathbin{\medoplus} \Varid{id})\mathbin{\circ}\Varid{s})\mathbin{\circ}\Varid{trj}\;\Varid{ps})\;\Varid{x'}){}\<[66]%
\>[66]{}=\hspace{-3pt}\{\; \Varid{useMapPresComp}\;\}\hspace{-3pt}={}\<[E]%
\\[\blanklineskip]%
\>[3]{}\hsindent{2}{}\<[5]%
\>[5]{}((\Varid{meas}\mathbin{\circ}\Varid{map}\;(\Varid{const}\;(\Varid{r}\;\Varid{x'}) \mathbin{\medoplus} \Varid{id})\mathbin{\circ}\Varid{map}\;\Varid{s}\mathbin{\circ}\Varid{trj}\;\Varid{ps})\;\Varid{x'}){}\<[66]%
\>[66]{}=\hspace{-3pt}\{\; \Conid{Refl}\;\}\hspace{-3pt}={}\<[E]%
\\[\blanklineskip]%
\>[3]{}\hsindent{2}{}\<[5]%
\>[5]{}((\Varid{meas}\mathbin{\circ}\Varid{map}\;((\Varid{r}\;\Varid{x'}) \mathbin{\oplus} )\mathbin{\circ}\Varid{map}\;\Varid{s}\mathbin{\circ}\Varid{trj}\;\Varid{ps})\;\Varid{x'}){}\<[66]%
\>[66]{}=\hspace{-3pt}\{\; \Varid{useMeasPlusSpec}\;\}\hspace{-3pt}={}\<[E]%
\\[\blanklineskip]%
\>[3]{}\hsindent{2}{}\<[5]%
\>[5]{}((((\Varid{r}\;\Varid{x'}) \mathbin{\oplus} )\mathbin{\circ}\Varid{meas}\mathbin{\circ}\Varid{map}\;\Varid{s}\mathbin{\circ}\Varid{trj}\;\Varid{ps})\;\Varid{x'}){}\<[66]%
\>[66]{}=\hspace{-3pt}\{\; \Conid{Refl}\;\}\hspace{-3pt}={}\<[E]%
\\[\blanklineskip]%
\>[3]{}\hsindent{2}{}\<[5]%
\>[5]{}(((\Varid{const}\;(\Varid{r}\;\Varid{x'}) \mathbin{\medoplus} \Varid{id})\mathbin{\circ}\Varid{meas}\mathbin{\circ}\Varid{map}\;\Varid{s}\mathbin{\circ}\Varid{trj}\;\Varid{ps})\;\Varid{x'}){}\<[66]%
\>[66]{}=\hspace{-3pt}\{\; \Conid{Refl}\;\}\hspace{-3pt}={}\<[E]%
\\[\blanklineskip]%
\>[3]{}\hsindent{2}{}\<[5]%
\>[5]{}((\Varid{r} \mathbin{\medoplus} \Varid{meas}\mathbin{\circ}\Varid{map}\;\Varid{s}\mathbin{\circ}\Varid{trj}\;\Varid{ps})\;\Varid{x'})\;{}\<[66]%
\>[66]{}\hfill\Box{}\<[E]%
\ColumnHook
\end{hscode}\resethooks
The \ensuremath{\Varid{trj}} function never produces an empty \ensuremath{\Conid{M}}-structure of trajectories:
\begin{hscode}\SaveRestoreHook
\column{B}{@{}>{\hspre}l<{\hspost}@{}}%
\column{3}{@{}>{\hspre}l<{\hspost}@{}}%
\column{21}{@{}>{\hspre}c<{\hspost}@{}}%
\column{21E}{@{}l@{}}%
\column{24}{@{}>{\hspre}l<{\hspost}@{}}%
\column{E}{@{}>{\hspre}l<{\hspost}@{}}%
\>[3]{}\Varid{trjNotEmptyLemma}{}\<[21]%
\>[21]{} \mathop{:} {}\<[21E]%
\>[24]{}\{\mskip1.5mu \Varid{t},\Varid{n} \mathop{:} \mathbb{N}\mskip1.5mu\} \to (\Varid{ps} \mathop{:} \Conid{PolicySeq}\;\Varid{t}\;\Varid{n}) \to (\Varid{x} \mathop{:} \Conid{X}\;\Varid{t}) \to \Conid{NotEmpty}\;(\Varid{trj}\;\Varid{ps}\;\Varid{x}){}\<[E]%
\ColumnHook
\end{hscode}\resethooks
\begin{hscode}\SaveRestoreHook
\column{B}{@{}>{\hspre}l<{\hspost}@{}}%
\column{3}{@{}>{\hspre}l<{\hspost}@{}}%
\column{4}{@{}>{\hspre}l<{\hspost}@{}}%
\column{17}{@{}>{\hspre}l<{\hspost}@{}}%
\column{28}{@{}>{\hspre}l<{\hspost}@{}}%
\column{45}{@{}>{\hspre}l<{\hspost}@{}}%
\column{48}{@{}>{\hspre}c<{\hspost}@{}}%
\column{48E}{@{}l@{}}%
\column{51}{@{}>{\hspre}l<{\hspost}@{}}%
\column{60}{@{}>{\hspre}l<{\hspost}@{}}%
\column{E}{@{}>{\hspre}l<{\hspost}@{}}%
\>[3]{}\Varid{trjNotEmptyLemma}\;{}\<[28]%
\>[28]{}(\Conid{Nil})\;{}\<[45]%
\>[45]{}\Varid{x}{}\<[48]%
\>[48]{}\mathrel{=}{}\<[48E]%
\>[51]{}\Varid{pureNotEmpty}\;(\Conid{Last}\;\Varid{x}){}\<[E]%
\\
\>[3]{}\Varid{trjNotEmptyLemma}\;\{\mskip1.5mu \Varid{t}\mathrel{=}\Varid{t}\mskip1.5mu\}\;\{\mskip1.5mu \Varid{n}\mathrel{=}\Conid{S}\;\Varid{n}\mskip1.5mu\}\;(\Varid{p}\mathbin{::}\Varid{ps})\;{}\<[45]%
\>[45]{}\Varid{x}{}\<[48]%
\>[48]{}\mathrel{=}{}\<[48E]%
\\[\blanklineskip]%
\>[3]{}\hsindent{1}{}\<[4]%
\>[4]{}\mathbf{let}\;\Varid{y}{}\<[17]%
\>[17]{}\mathrel{=}\Varid{p}\;\Varid{x}\;{}\<[60]%
\>[60]{}\mathbf{in}{}\<[E]%
\\
\>[3]{}\hsindent{1}{}\<[4]%
\>[4]{}\mathbf{let}\;\Varid{trjps}{}\<[17]%
\>[17]{}\mathrel{=}\Varid{trj}\;\Varid{ps}\;{}\<[60]%
\>[60]{}\mathbf{in}{}\<[E]%
\\
\>[3]{}\hsindent{1}{}\<[4]%
\>[4]{}\mathbf{let}\;\Varid{nxpx}{}\<[17]%
\>[17]{}\mathrel{=}\Varid{next}\;\Varid{t}\;\Varid{x}\;\Varid{y}\;{}\<[60]%
\>[60]{}\mathbf{in}{}\<[E]%
\\
\>[3]{}\hsindent{1}{}\<[4]%
\>[4]{}\mathbf{let}\;\Varid{consxy}{}\<[17]%
\>[17]{}\mathrel{=}((\Varid{x}  \mathbin{*\!*} \Varid{y}) \mathbin{\#\!\#} )\;{}\<[60]%
\>[60]{}\mathbf{in}{}\<[E]%
\\
\>[3]{}\hsindent{1}{}\<[4]%
\>[4]{}\mathbf{let}\;\Varid{nne}{}\<[17]%
\>[17]{}\mathrel{=}\Varid{nextNotEmpty}\;\Varid{x}\;\Varid{y}\;{}\<[60]%
\>[60]{}\mathbf{in}{}\<[E]%
\\
\>[3]{}\hsindent{1}{}\<[4]%
\>[4]{}\mathbf{let}\;\Varid{netrjps}{}\<[17]%
\>[17]{}\mathrel{=}\Varid{trjNotEmptyLemma}\;\Varid{ps}\;{}\<[60]%
\>[60]{}\mathbf{in}{}\<[E]%
\\
\>[3]{}\hsindent{1}{}\<[4]%
\>[4]{}\mathbf{let}\;\Varid{bne}{}\<[17]%
\>[17]{}\mathrel{=}\Varid{bindPresNotEmpty}\;\Varid{trjps}\;\Varid{nxpx}\;\Varid{nne}\;\Varid{netrjps}\;{}\<[60]%
\>[60]{}\mathbf{in}{}\<[E]%
\\[\blanklineskip]%
\>[3]{}\hsindent{1}{}\<[4]%
\>[4]{}\Varid{mapPresNotEmpty}\;\Varid{consxy}\;(\Varid{nxpx} \mathbin{>\!\!>\!\!=} \Varid{trjps})\;\Varid{bne}{}\<[E]%
\ColumnHook
\end{hscode}\resethooks
A technical lemma to lift equalities into the right component of \ensuremath{ \mathbin{\medoplus} }:
\begin{hscode}\SaveRestoreHook
\column{B}{@{}>{\hspre}l<{\hspost}@{}}%
\column{3}{@{}>{\hspre}l<{\hspost}@{}}%
\column{21}{@{}>{\hspre}c<{\hspost}@{}}%
\column{21E}{@{}l@{}}%
\column{24}{@{}>{\hspre}l<{\hspost}@{}}%
\column{E}{@{}>{\hspre}l<{\hspost}@{}}%
\>[3]{}\Varid{oplusLiftEERight}{}\<[21]%
\>[21]{} \mathop{:} {}\<[21E]%
\>[24]{}\{\mskip1.5mu \Conid{A} \mathop{:} \Conid{Type}\mskip1.5mu\} \to (\Varid{f},\Varid{g},\Varid{h} \mathop{:} \Conid{A} \to \Conid{Val}) \to (\Varid{g}\doteq\Varid{h}) \to (\Varid{f} \mathbin{\medoplus} \Varid{g})\doteq(\Varid{f} \mathbin{\medoplus} \Varid{h}){}\<[E]%
\ColumnHook
\end{hscode}\resethooks
\begin{hscode}\SaveRestoreHook
\column{B}{@{}>{\hspre}l<{\hspost}@{}}%
\column{3}{@{}>{\hspre}l<{\hspost}@{}}%
\column{36}{@{}>{\hspre}c<{\hspost}@{}}%
\column{36E}{@{}l@{}}%
\column{39}{@{}>{\hspre}l<{\hspost}@{}}%
\column{E}{@{}>{\hspre}l<{\hspost}@{}}%
\>[3]{}\Varid{oplusLiftEERight}\;\{\mskip1.5mu \Conid{A}\mskip1.5mu\}\;\Varid{f}\;\Varid{g}\;\Varid{h}\;\Varid{ee}\;\Varid{a}{}\<[36]%
\>[36]{}\mathrel{=}{}\<[36E]%
\>[39]{}\Varid{cong}\;(\Varid{ee}\;\Varid{a}){}\<[E]%
\ColumnHook
\end{hscode}\resethooks

\subsection{Properties of monad algebras}
Condition for a function \ensuremath{\Varid{f}} to be an \ensuremath{\Conid{M}}-algebra homomorphism:
\begin{hscode}\SaveRestoreHook
\column{B}{@{}>{\hspre}l<{\hspost}@{}}%
\column{3}{@{}>{\hspre}l<{\hspost}@{}}%
\column{15}{@{}>{\hspre}c<{\hspost}@{}}%
\column{15E}{@{}l@{}}%
\column{18}{@{}>{\hspre}l<{\hspost}@{}}%
\column{E}{@{}>{\hspre}l<{\hspost}@{}}%
\>[3]{}\Varid{algMorSpec}{}\<[15]%
\>[15]{} \mathop{:} {}\<[15E]%
\>[18]{}\{\mskip1.5mu \Conid{A},\Conid{B} \mathop{:} \Conid{Type}\mskip1.5mu\} \to (\alpha \mathop{:} \Conid{M}\;\Conid{A} \to \Conid{A}) \to (\beta \mathop{:} \Conid{M}\;\Conid{B} \to \Conid{B}) \to (\Varid{f} \mathop{:} \Conid{A} \to \Conid{B}) \to \Conid{Type}{}\<[E]%
\\
\>[3]{}\Varid{algMorSpec}\;\{\mskip1.5mu \Conid{A}\mskip1.5mu\}\;\{\mskip1.5mu \Conid{B}\mskip1.5mu\}\;\alpha\;\beta\;\Varid{f}\mathrel{=}(\beta\mathbin{\circ}\Varid{map}\;\Varid{f})\doteq(\Varid{f}\mathbin{\circ}\alpha){}\<[E]%
\ColumnHook
\end{hscode}\resethooks
Structure maps of \ensuremath{\Conid{M}}-algebras are left inverses of \ensuremath{\Varid{pure}}:
\begin{hscode}\SaveRestoreHook
\column{B}{@{}>{\hspre}l<{\hspost}@{}}%
\column{3}{@{}>{\hspre}l<{\hspost}@{}}%
\column{16}{@{}>{\hspre}c<{\hspost}@{}}%
\column{16E}{@{}l@{}}%
\column{19}{@{}>{\hspre}l<{\hspost}@{}}%
\column{E}{@{}>{\hspre}l<{\hspost}@{}}%
\>[3]{}\Varid{algPureSpec}{}\<[16]%
\>[16]{} \mathop{:} {}\<[16E]%
\>[19]{}\{\mskip1.5mu \Conid{A} \mathop{:} \Conid{Type}\mskip1.5mu\} \to (\alpha \mathop{:} \Conid{M}\;\Conid{A} \to \Conid{A}) \to \Conid{Type}{}\<[E]%
\\
\>[3]{}\Varid{algPureSpec}\;\{\mskip1.5mu \Conid{A}\mskip1.5mu\}\;\alpha\mathrel{=}\alpha\mathbin{\circ}\Varid{pure}\doteq\Varid{id}{}\<[E]%
\ColumnHook
\end{hscode}\resethooks
Structure maps of \ensuremath{\Conid{M}}-algebras are themselves \ensuremath{\Conid{M}}-algebra homomorphisms:
\begin{hscode}\SaveRestoreHook
\column{B}{@{}>{\hspre}l<{\hspost}@{}}%
\column{3}{@{}>{\hspre}l<{\hspost}@{}}%
\column{16}{@{}>{\hspre}c<{\hspost}@{}}%
\column{16E}{@{}l@{}}%
\column{19}{@{}>{\hspre}l<{\hspost}@{}}%
\column{28}{@{}>{\hspre}l<{\hspost}@{}}%
\column{57}{@{}>{\hspre}l<{\hspost}@{}}%
\column{E}{@{}>{\hspre}l<{\hspost}@{}}%
\>[3]{}\Varid{algJoinSpec}{}\<[16]%
\>[16]{} \mathop{:} {}\<[16E]%
\>[19]{}\{\mskip1.5mu \Conid{A} \mathop{:} \Conid{Type}\mskip1.5mu\} \to (\alpha \mathop{:} \Conid{M}\;\Conid{A} \to \Conid{A}) \to \Conid{Type}{}\<[E]%
\\
\>[3]{}\Varid{algJoinSpec}\;\{\mskip1.5mu \Conid{A}\mskip1.5mu\}\;\alpha\mathrel{=}{}\<[28]%
\>[28]{}\Varid{algMorSpec}\;\Varid{join}\;\alpha\;\alpha{}\<[57]%
\>[57]{}\mbox{\onelinecomment  \ensuremath{(\alpha\mathbin{\circ}\Varid{map}\;\alpha)\doteq(\alpha\mathbin{\circ}\Varid{join})}}{}\<[E]%
\ColumnHook
\end{hscode}\resethooks
A lemma about computation with \ensuremath{\Conid{M}}-algebras:
\begin{hscode}\SaveRestoreHook
\column{B}{@{}>{\hspre}l<{\hspost}@{}}%
\column{3}{@{}>{\hspre}l<{\hspost}@{}}%
\column{13}{@{}>{\hspre}c<{\hspost}@{}}%
\column{13E}{@{}l@{}}%
\column{16}{@{}>{\hspre}l<{\hspost}@{}}%
\column{E}{@{}>{\hspre}l<{\hspost}@{}}%
\>[3]{}\Varid{algLemma}{}\<[13]%
\>[13]{} \mathop{:} {}\<[13E]%
\>[16]{}\{\mskip1.5mu \Conid{A},\Conid{B},\Conid{C} \mathop{:} \Conid{Type}\mskip1.5mu\} \to (\alpha \mathop{:} \Conid{M}\;\Conid{C} \to \Conid{C}) \to (\Varid{ee} \mathop{:} \Varid{algJoinSpec}\;\alpha) \to {}\<[E]%
\\
\>[16]{}(\Varid{f} \mathop{:} \Conid{B} \to \Conid{C}) \to (\Varid{g} \mathop{:} \Conid{A} \to \Conid{M}\;\Conid{B}) \to {}\<[E]%
\\
\>[16]{}(\alpha\mathbin{\circ}\Varid{map}\;(\alpha\mathbin{\circ}\Varid{map}\;\Varid{f}\mathbin{\circ}\Varid{g}))\doteq(\alpha\mathbin{\circ}\Varid{map}\;\Varid{f}\mathbin{\circ}\Varid{join}\mathbin{\circ}\Varid{map}\;\Varid{g}){}\<[E]%
\ColumnHook
\end{hscode}\resethooks
\begin{hscode}\SaveRestoreHook
\column{B}{@{}>{\hspre}l<{\hspost}@{}}%
\column{3}{@{}>{\hspre}l<{\hspost}@{}}%
\column{5}{@{}>{\hspre}l<{\hspost}@{}}%
\column{55}{@{}>{\hspre}l<{\hspost}@{}}%
\column{E}{@{}>{\hspre}l<{\hspost}@{}}%
\>[3]{}\Varid{algLemma}\;\{\mskip1.5mu \Conid{A}\mskip1.5mu\}\;\{\mskip1.5mu \Conid{B}\mskip1.5mu\}\;\{\mskip1.5mu \Conid{C}\mskip1.5mu\}\;\alpha\;\Varid{ee}\;\Varid{f}\;\Varid{g}\;\Varid{ma}\mathrel{=}{}\<[E]%
\\[\blanklineskip]%
\>[3]{}\hsindent{2}{}\<[5]%
\>[5]{}((\alpha\mathbin{\circ}\Varid{map}\;(\alpha\mathbin{\circ}\Varid{map}\;\Varid{f}\mathbin{\circ}\Varid{g}))\;\Varid{ma}){}\<[55]%
\>[55]{}=\hspace{-3pt}\{\; \Varid{cong}\;(\Varid{mapPresComp}\;\Varid{g}\;(\alpha\mathbin{\circ}\Varid{map}\;\Varid{f})\;\Varid{ma})\;\}\hspace{-3pt}={}\<[E]%
\\[\blanklineskip]%
\>[3]{}\hsindent{2}{}\<[5]%
\>[5]{}((\alpha\mathbin{\circ}\Varid{map}\;(\alpha\mathbin{\circ}\Varid{map}\;\Varid{f})\mathbin{\circ}\Varid{map}\;\Varid{g})\;\Varid{ma}){}\<[55]%
\>[55]{}=\hspace{-3pt}\{\; \Varid{cong}\;(\Varid{mapPresComp}\;(\Varid{map}\;\Varid{f})\;\alpha\;(\Varid{map}\;\Varid{g}\;\Varid{ma}))\;\}\hspace{-3pt}={}\<[E]%
\\[\blanklineskip]%
\>[3]{}\hsindent{2}{}\<[5]%
\>[5]{}((\alpha\mathbin{\circ}\Varid{map}\;\alpha\mathbin{\circ}\Varid{map}\;(\Varid{map}\;\Varid{f})\mathbin{\circ}\Varid{map}\;\Varid{g})\;\Varid{ma}){}\<[55]%
\>[55]{}=\hspace{-3pt}\{\; \Varid{ee}\;(\Varid{map}\;(\Varid{map}\;\Varid{f})\;(\Varid{map}\;\Varid{g}\;\Varid{ma}))\;\}\hspace{-3pt}={}\<[E]%
\\[\blanklineskip]%
\>[3]{}\hsindent{2}{}\<[5]%
\>[5]{}((\alpha\mathbin{\circ}\Varid{join}\mathbin{\circ}\Varid{map}\;(\Varid{map}\;\Varid{f})\mathbin{\circ}\Varid{map}\;\Varid{g})\;\Varid{ma}){}\<[55]%
\>[55]{}=\hspace{-3pt}\{\; \Varid{cong}\;(\Varid{sym}\;(\Varid{joinNatTrans}\;\Varid{f}\;(\Varid{map}\;\Varid{g}\;\Varid{ma})))\;\}\hspace{-3pt}={}\<[E]%
\\[\blanklineskip]%
\>[3]{}\hsindent{2}{}\<[5]%
\>[5]{}((\alpha\mathbin{\circ}\Varid{map}\;\Varid{f}\mathbin{\circ}\Varid{join}\mathbin{\circ}\Varid{map}\;\Varid{g})\;\Varid{ma})\;{}\<[55]%
\>[55]{}\hfill\Box{}\<[E]%
\ColumnHook
\end{hscode}\resethooks

\subsection{Measure specifications}.
The measure needs to be the structure map of an \ensuremath{\Conid{M}}-algebra on \ensuremath{\Conid{Val}}.
This means:

\begin{itemize}

\item It is a left inverse to \ensuremath{\Varid{pure}}:
\begin{hscode}\SaveRestoreHook
\column{B}{@{}>{\hspre}l<{\hspost}@{}}%
\column{3}{@{}>{\hspre}l<{\hspost}@{}}%
\column{17}{@{}>{\hspre}c<{\hspost}@{}}%
\column{17E}{@{}l@{}}%
\column{20}{@{}>{\hspre}l<{\hspost}@{}}%
\column{41}{@{}>{\hspre}l<{\hspost}@{}}%
\column{E}{@{}>{\hspre}l<{\hspost}@{}}%
\>[3]{}\Varid{measPureSpec}{}\<[17]%
\>[17]{} \mathop{:} {}\<[17E]%
\>[20]{}\Varid{algPureSpec}\;\Varid{meas}{}\<[41]%
\>[41]{}\mbox{\onelinecomment  \ensuremath{\Varid{meas}\mathbin{\circ}\Varid{pure}\doteq\Varid{id}}}{}\<[E]%
\ColumnHook
\end{hscode}\resethooks

\item It is an \ensuremath{\Conid{M}}-algebra homomorphism from \ensuremath{\Varid{join}} to itself:
\begin{hscode}\SaveRestoreHook
\column{B}{@{}>{\hspre}l<{\hspost}@{}}%
\column{3}{@{}>{\hspre}l<{\hspost}@{}}%
\column{17}{@{}>{\hspre}c<{\hspost}@{}}%
\column{17E}{@{}l@{}}%
\column{20}{@{}>{\hspre}l<{\hspost}@{}}%
\column{41}{@{}>{\hspre}l<{\hspost}@{}}%
\column{E}{@{}>{\hspre}l<{\hspost}@{}}%
\>[3]{}\Varid{measJoinSpec}{}\<[17]%
\>[17]{} \mathop{:} {}\<[17E]%
\>[20]{}\Varid{algJoinSpec}\;\Varid{meas}{}\<[41]%
\>[41]{}\mbox{\onelinecomment   \ensuremath{\Varid{meas}\mathbin{\circ}\Varid{join}\doteq\Varid{meas}\mathbin{\circ}\Varid{map}\;\Varid{meas}}}{}\<[E]%
\ColumnHook
\end{hscode}\resethooks

\end{itemize}
Moreover, for all \ensuremath{\Varid{v} \mathop{:} \Conid{Val}}, the function \ensuremath{(\Varid{v} \mathbin{\oplus} )} needs to be an
  \ensuremath{\Conid{M}}-algebra homomorphism:
\begin{hscode}\SaveRestoreHook
\column{B}{@{}>{\hspre}l<{\hspost}@{}}%
\column{3}{@{}>{\hspre}l<{\hspost}@{}}%
\column{17}{@{}>{\hspre}c<{\hspost}@{}}%
\column{17E}{@{}l@{}}%
\column{20}{@{}>{\hspre}l<{\hspost}@{}}%
\column{E}{@{}>{\hspre}l<{\hspost}@{}}%
\>[3]{}\Varid{measPlusSpec}{}\<[17]%
\>[17]{} \mathop{:} {}\<[17E]%
\>[20]{}(\Varid{v} \mathop{:} \Conid{Val}) \to (\Varid{mv} \mathop{:} \Conid{M}\;\Conid{Val}) \to (\Conid{NotEmpty}\;\Varid{mv}) \to {}\<[E]%
\\
\>[20]{}(\Varid{meas}\mathbin{\circ}\Varid{map}\;(\Varid{v} \mathbin{\oplus} ))\;\Varid{mv}\mathrel{=}((\Varid{v} \mathbin{\oplus} )\mathbin{\circ}\Varid{meas})\;\Varid{mv}{}\<[E]%
\ColumnHook
\end{hscode}\resethooks
We can omit the non-emptiness condition but this means we implicitly
restrict the class of monads that can be used for \ensuremath{\Conid{M}}.
The condition could then be expressed as
\begin{hscode}\SaveRestoreHook
\column{B}{@{}>{\hspre}l<{\hspost}@{}}%
\column{3}{@{}>{\hspre}l<{\hspost}@{}}%
\column{18}{@{}>{\hspre}c<{\hspost}@{}}%
\column{18E}{@{}l@{}}%
\column{21}{@{}>{\hspre}l<{\hspost}@{}}%
\column{E}{@{}>{\hspre}l<{\hspost}@{}}%
\>[3]{}\Varid{measPlusSpec'}{}\<[18]%
\>[18]{} \mathop{:} {}\<[18E]%
\>[21]{}(\Varid{v} \mathop{:} \Conid{Val}) \to \Varid{algMorSpec}\;\Varid{meas}\;\Varid{meas}\;( \mathbin{\oplus} \Varid{v}){}\<[E]%
\ColumnHook
\end{hscode}\resethooks


\section{Bellman's principle of optimality}
\label{appendix:Bellman}





Basic requirements for monadic backward induction:
\begin{hscode}\SaveRestoreHook
\column{B}{@{}>{\hspre}l<{\hspost}@{}}%
\column{3}{@{}>{\hspre}l<{\hspost}@{}}%
\column{10}{@{}>{\hspre}l<{\hspost}@{}}%
\column{E}{@{}>{\hspre}l<{\hspost}@{}}%
\>[3]{}( \,\sqsubseteq\, ){}\<[10]%
\>[10]{} \mathop{:} \Conid{Val} \to \Conid{Val} \to \Conid{Type}{}\<[E]%
\ColumnHook
\end{hscode}\resethooks
\begin{hscode}\SaveRestoreHook
\column{B}{@{}>{\hspre}l<{\hspost}@{}}%
\column{3}{@{}>{\hspre}l<{\hspost}@{}}%
\column{17}{@{}>{\hspre}c<{\hspost}@{}}%
\column{17E}{@{}l@{}}%
\column{20}{@{}>{\hspre}l<{\hspost}@{}}%
\column{E}{@{}>{\hspre}l<{\hspost}@{}}%
\>[3]{}\Varid{lteRefl}{}\<[17]%
\>[17]{} \mathop{:} {}\<[17E]%
\>[20]{}\{\mskip1.5mu \Varid{a} \mathop{:} \Conid{Val}\mskip1.5mu\} \to \Varid{a} \,\sqsubseteq\, \Varid{a}{}\<[E]%
\\
\>[3]{}\Varid{lteTrans}{}\<[17]%
\>[17]{} \mathop{:} {}\<[17E]%
\>[20]{}\{\mskip1.5mu \Varid{a},\Varid{b},\Varid{c} \mathop{:} \Conid{Val}\mskip1.5mu\} \to \Varid{a} \,\sqsubseteq\, \Varid{b} \to \Varid{b} \,\sqsubseteq\, \Varid{c} \to \Varid{a} \,\sqsubseteq\, \Varid{c}{}\<[E]%
\\
\>[3]{}\Varid{plusMonSpec}{}\<[17]%
\>[17]{} \mathop{:} {}\<[17E]%
\>[20]{}\{\mskip1.5mu \Varid{a},\Varid{b},\Varid{c},\Varid{d} \mathop{:} \Conid{Val}\mskip1.5mu\} \to \Varid{a} \,\sqsubseteq\, \Varid{b} \to \Varid{c} \,\sqsubseteq\, \Varid{d} \to (\Varid{a} \mathbin{\oplus} \Varid{c}) \,\sqsubseteq\, (\Varid{b} \mathbin{\oplus} \Varid{d}){}\<[E]%
\ColumnHook
\end{hscode}\resethooks
\begin{hscode}\SaveRestoreHook
\column{B}{@{}>{\hspre}l<{\hspost}@{}}%
\column{3}{@{}>{\hspre}l<{\hspost}@{}}%
\column{17}{@{}>{\hspre}c<{\hspost}@{}}%
\column{17E}{@{}l@{}}%
\column{20}{@{}>{\hspre}l<{\hspost}@{}}%
\column{E}{@{}>{\hspre}l<{\hspost}@{}}%
\>[3]{}\Varid{measMonSpec}{}\<[17]%
\>[17]{} \mathop{:} {}\<[17E]%
\>[20]{}\{\mskip1.5mu \Conid{A} \mathop{:} \Conid{Type}\mskip1.5mu\} \to (\Varid{f},\Varid{g} \mathop{:} \Conid{A} \to \Conid{Val}) \to ((\Varid{a} \mathop{:} \Conid{A}) \to (\Varid{f}\;\Varid{a}) \,\sqsubseteq\, (\Varid{g}\;\Varid{a})) \to {}\<[E]%
\\
\>[20]{}(\Varid{ma} \mathop{:} \Conid{M}\;\Conid{A}) \to \Varid{meas}\;(\Varid{map}\;\Varid{f}\;\Varid{ma}) \,\sqsubseteq\, \Varid{meas}\;(\Varid{map}\;\Varid{g}\;\Varid{ma}){}\<[E]%
\ColumnHook
\end{hscode}\resethooks
Optimality of policy sequences:
\begin{hscode}\SaveRestoreHook
\column{B}{@{}>{\hspre}l<{\hspost}@{}}%
\column{3}{@{}>{\hspre}l<{\hspost}@{}}%
\column{17}{@{}>{\hspre}c<{\hspost}@{}}%
\column{17E}{@{}l@{}}%
\column{20}{@{}>{\hspre}l<{\hspost}@{}}%
\column{28}{@{}>{\hspre}c<{\hspost}@{}}%
\column{28E}{@{}l@{}}%
\column{31}{@{}>{\hspre}l<{\hspost}@{}}%
\column{E}{@{}>{\hspre}l<{\hspost}@{}}%
\>[3]{}\Conid{OptPolicySeq}{}\<[17]%
\>[17]{} \mathop{:} {}\<[17E]%
\>[20]{}\{\mskip1.5mu \Varid{t},\Varid{n} \mathop{:} \mathbb{N}\mskip1.5mu\} \to \Conid{PolicySeq}\;\Varid{t}\;\Varid{n} \to \Conid{Type}{}\<[E]%
\\
\>[3]{}\Conid{OptPolicySeq}\;\{\mskip1.5mu \Varid{t}\mskip1.5mu\}\;\{\mskip1.5mu \Varid{n}\mskip1.5mu\}\;\Varid{ps}{}\<[28]%
\>[28]{}\mathrel{=}{}\<[28E]%
\>[31]{}(\Varid{ps'} \mathop{:} \Conid{PolicySeq}\;\Varid{t}\;\Varid{n}) \to (\Varid{x} \mathop{:} \Conid{X}\;\Varid{t}) \to \Varid{val}\;\Varid{ps'}\;\Varid{x} \,\sqsubseteq\, \Varid{val}\;\Varid{ps}\;\Varid{x}{}\<[E]%
\ColumnHook
\end{hscode}\resethooks
Optimality of extensions of policy sequences:
\begin{hscode}\SaveRestoreHook
\column{B}{@{}>{\hspre}l<{\hspost}@{}}%
\column{3}{@{}>{\hspre}l<{\hspost}@{}}%
\column{11}{@{}>{\hspre}c<{\hspost}@{}}%
\column{11E}{@{}l@{}}%
\column{14}{@{}>{\hspre}l<{\hspost}@{}}%
\column{20}{@{}>{\hspre}c<{\hspost}@{}}%
\column{20E}{@{}l@{}}%
\column{23}{@{}>{\hspre}l<{\hspost}@{}}%
\column{56}{@{}>{\hspre}l<{\hspost}@{}}%
\column{E}{@{}>{\hspre}l<{\hspost}@{}}%
\>[3]{}\Conid{OptExt}{}\<[11]%
\>[11]{} \mathop{:} {}\<[11E]%
\>[14]{}\{\mskip1.5mu \Varid{t},\Varid{n} \mathop{:} \mathbb{N}\mskip1.5mu\} \to \Conid{PolicySeq}\;(\Conid{S}\;\Varid{t})\;\Varid{n} \to \Conid{Policy}\;\Varid{t} \to \Conid{Type}{}\<[E]%
\\
\>[3]{}\Conid{OptExt}\;\{\mskip1.5mu \Varid{t}\mskip1.5mu\}\;\Varid{ps}\;\Varid{p}{}\<[20]%
\>[20]{}\mathrel{=}{}\<[20E]%
\>[23]{}(\Varid{p'} \mathop{:} \Conid{Policy}\;\Varid{t}) \to (\Varid{x} \mathop{:} \Conid{X}\;\Varid{t}) \to {}\<[56]%
\>[56]{}\Varid{val}\;(\Varid{p'}\mathbin{::}\Varid{ps})\;\Varid{x} \,\sqsubseteq\, \Varid{val}\;(\Varid{p}\mathbin{::}\Varid{ps})\;\Varid{x}{}\<[E]%
\ColumnHook
\end{hscode}\resethooks
Bellman's principle of optimality:
\begin{hscode}\SaveRestoreHook
\column{B}{@{}>{\hspre}l<{\hspost}@{}}%
\column{3}{@{}>{\hspre}l<{\hspost}@{}}%
\column{4}{@{}>{\hspre}l<{\hspost}@{}}%
\column{12}{@{}>{\hspre}c<{\hspost}@{}}%
\column{12E}{@{}l@{}}%
\column{13}{@{}>{\hspre}c<{\hspost}@{}}%
\column{13E}{@{}l@{}}%
\column{15}{@{}>{\hspre}l<{\hspost}@{}}%
\column{16}{@{}>{\hspre}l<{\hspost}@{}}%
\column{29}{@{}>{\hspre}c<{\hspost}@{}}%
\column{29E}{@{}l@{}}%
\column{33}{@{}>{\hspre}l<{\hspost}@{}}%
\column{39}{@{}>{\hspre}c<{\hspost}@{}}%
\column{39E}{@{}l@{}}%
\column{42}{@{}>{\hspre}l<{\hspost}@{}}%
\column{43}{@{}>{\hspre}c<{\hspost}@{}}%
\column{43E}{@{}l@{}}%
\column{58}{@{}>{\hspre}l<{\hspost}@{}}%
\column{62}{@{}>{\hspre}l<{\hspost}@{}}%
\column{82}{@{}>{\hspre}c<{\hspost}@{}}%
\column{82E}{@{}l@{}}%
\column{86}{@{}>{\hspre}l<{\hspost}@{}}%
\column{E}{@{}>{\hspre}l<{\hspost}@{}}%
\>[3]{}\Conid{Bellman}{}\<[12]%
\>[12]{} \mathop{:} {}\<[12E]%
\>[15]{}\{\mskip1.5mu \Varid{t},\Varid{n} \mathop{:} \mathbb{N}\mskip1.5mu\}{}\<[29]%
\>[29]{} \to {}\<[29E]%
\>[33]{}(\Varid{ps}{}\<[39]%
\>[39]{} \mathop{:} {}\<[39E]%
\>[42]{}\Conid{PolicySeq}\;(\Conid{S}\;\Varid{t})\;\Varid{n}){}\<[62]%
\>[62]{} \to \Conid{OptPolicySeq}\;\Varid{ps}{}\<[82]%
\>[82]{} \to {}\<[82E]%
\\
\>[33]{}(\Varid{p}{}\<[39]%
\>[39]{} \mathop{:} {}\<[39E]%
\>[42]{}\Conid{Policy}\;\Varid{t}){}\<[62]%
\>[62]{} \to \Conid{OptExt}\;\Varid{ps}\;\Varid{p}{}\<[82]%
\>[82]{} \to {}\<[82E]%
\>[86]{}\Conid{OptPolicySeq}\;(\Varid{p}\mathbin{::}\Varid{ps}){}\<[E]%
\\[\blanklineskip]%
\>[3]{}\Conid{Bellman}\;\{\mskip1.5mu \Varid{t}\mskip1.5mu\}\;\Varid{ps}\;\Varid{ops}\;\Varid{p}\;\Varid{oep}\;(\Varid{p'}\mathbin{::}\Varid{ps'})\;\Varid{x}{}\<[43]%
\>[43]{}\mathrel{=}{}\<[43E]%
\\[\blanklineskip]%
\>[3]{}\hsindent{1}{}\<[4]%
\>[4]{}\mathbf{let}\;\Varid{y'}{}\<[13]%
\>[13]{}\mathrel{=}{}\<[13E]%
\>[16]{}\Varid{p'}\;\Varid{x}\;{}\<[58]%
\>[58]{}\mathbf{in}{}\<[E]%
\\
\>[3]{}\hsindent{1}{}\<[4]%
\>[4]{}\mathbf{let}\;\Varid{mx'}{}\<[13]%
\>[13]{}\mathrel{=}{}\<[13E]%
\>[16]{}\Varid{next}\;\Varid{t}\;\Varid{x}\;\Varid{y'}\;{}\<[58]%
\>[58]{}\mathbf{in}{}\<[E]%
\\
\>[3]{}\hsindent{1}{}\<[4]%
\>[4]{}\mathbf{let}\;\Varid{f'}{}\<[13]%
\>[13]{}\mathrel{=}{}\<[13E]%
\>[16]{}\Varid{reward}\;\Varid{t}\;\Varid{x}\;\Varid{y'} \mathbin{\medoplus} \Varid{val}\;\Varid{ps'}\;{}\<[58]%
\>[58]{}\mathbf{in}{}\<[E]%
\\
\>[3]{}\hsindent{1}{}\<[4]%
\>[4]{}\mathbf{let}\;\Varid{f}{}\<[13]%
\>[13]{}\mathrel{=}{}\<[13E]%
\>[16]{}\Varid{reward}\;\Varid{t}\;\Varid{x}\;\Varid{y'} \mathbin{\medoplus} \Varid{val}\;\Varid{ps}\;{}\<[58]%
\>[58]{}\mathbf{in}{}\<[E]%
\\
\>[3]{}\hsindent{1}{}\<[4]%
\>[4]{}\mathbf{let}\;s_0{}\<[13]%
\>[13]{}\mathrel{=}{}\<[13E]%
\>[16]{}\lambda \Varid{x'}\Rightarrow \Varid{plusMonSpec}\;\Varid{lteRefl}\;(\Varid{ops}\;\Varid{ps'}\;\Varid{x'})\;{}\<[58]%
\>[58]{}\mathbf{in}{}\<[E]%
\\
\>[3]{}\hsindent{1}{}\<[4]%
\>[4]{}\mathbf{let}\;\Varid{s1}{}\<[13]%
\>[13]{}\mathrel{=}{}\<[13E]%
\>[16]{}\Varid{measMonSpec}\;\Varid{f'}\;\Varid{f}\;s_0\;\Varid{mx'}\;{}\<[58]%
\>[58]{}\mathbf{in}\;\mbox{\onelinecomment  \ensuremath{\Varid{val}\;(\Varid{p'}\mathbin{::}\Varid{ps'})\;\Varid{x} \,\sqsubseteq\, \Varid{val}\;(\Varid{p'}\mathbin{::}\Varid{ps})\;\Varid{x}}}{}\<[E]%
\\
\>[3]{}\hsindent{1}{}\<[4]%
\>[4]{}\mathbf{let}\;\Varid{s2}{}\<[13]%
\>[13]{}\mathrel{=}{}\<[13E]%
\>[16]{}\Varid{oep}\;\Varid{p'}\;\Varid{x}\;{}\<[58]%
\>[58]{}\mathbf{in}\;\mbox{\onelinecomment  \ensuremath{\Varid{val}\;(\Varid{p'}\mathbin{::}\Varid{ps})\;\Varid{x} \,\sqsubseteq\, \Varid{val}\;(\Varid{p}\mathbin{::}\Varid{ps})\;\Varid{x}}}{}\<[E]%
\\[\blanklineskip]%
\>[3]{}\hsindent{1}{}\<[4]%
\>[4]{}\Varid{lteTrans}\;\Varid{s1}\;\Varid{s2}{}\<[E]%
\ColumnHook
\end{hscode}\resethooks

\section{Verification with respect to \ensuremath{\Varid{val}}}
\label{appendix:bilemma}


The empty policy sequence is optimal:
\begin{hscode}\SaveRestoreHook
\column{B}{@{}>{\hspre}l<{\hspost}@{}}%
\column{3}{@{}>{\hspre}l<{\hspost}@{}}%
\column{20}{@{}>{\hspre}c<{\hspost}@{}}%
\column{20E}{@{}l@{}}%
\column{23}{@{}>{\hspre}l<{\hspost}@{}}%
\column{26}{@{}>{\hspre}c<{\hspost}@{}}%
\column{26E}{@{}l@{}}%
\column{29}{@{}>{\hspre}l<{\hspost}@{}}%
\column{E}{@{}>{\hspre}l<{\hspost}@{}}%
\>[3]{}\Varid{nilOptPolicySeq}{}\<[20]%
\>[20]{} \mathop{:} {}\<[20E]%
\>[23]{}\Conid{OptPolicySeq}\;\Conid{Nil}{}\<[E]%
\\
\>[3]{}\Varid{nilOptPolicySeq}\;\Conid{Nil}\;\Varid{x}{}\<[26]%
\>[26]{}\mathrel{=}{}\<[26E]%
\>[29]{}\Varid{lteRefl}{}\<[E]%
\ColumnHook
\end{hscode}\resethooks
Now, provided that we can implement
\begin{hscode}\SaveRestoreHook
\column{B}{@{}>{\hspre}l<{\hspost}@{}}%
\column{3}{@{}>{\hspre}l<{\hspost}@{}}%
\column{15}{@{}>{\hspre}c<{\hspost}@{}}%
\column{15E}{@{}l@{}}%
\column{18}{@{}>{\hspre}l<{\hspost}@{}}%
\column{E}{@{}>{\hspre}l<{\hspost}@{}}%
\>[3]{}\Varid{optExt}{}\<[15]%
\>[15]{} \mathop{:} {}\<[15E]%
\>[18]{}\{\mskip1.5mu \Varid{t},\Varid{n} \mathop{:} \mathbb{N}\mskip1.5mu\} \to \Conid{PolicySeq}\;(\Conid{S}\;\Varid{t})\;\Varid{n} \to \Conid{Policy}\;\Varid{t}{}\<[E]%
\\
\>[3]{}\Varid{optExtSpec}{}\<[15]%
\>[15]{} \mathop{:} {}\<[15E]%
\>[18]{}\{\mskip1.5mu \Varid{t},\Varid{n} \mathop{:} \mathbb{N}\mskip1.5mu\} \to (\Varid{ps} \mathop{:} \Conid{PolicySeq}\;(\Conid{S}\;\Varid{t})\;\Varid{n}) \to \Conid{OptExt}\;\Varid{ps}\;(\Varid{optExt}\;\Varid{ps}){}\<[E]%
\ColumnHook
\end{hscode}\resethooks
then
\begin{hscode}\SaveRestoreHook
\column{B}{@{}>{\hspre}l<{\hspost}@{}}%
\column{3}{@{}>{\hspre}l<{\hspost}@{}}%
\column{7}{@{}>{\hspre}c<{\hspost}@{}}%
\column{7E}{@{}l@{}}%
\column{9}{@{}>{\hspre}l<{\hspost}@{}}%
\column{10}{@{}>{\hspre}l<{\hspost}@{}}%
\column{15}{@{}>{\hspre}c<{\hspost}@{}}%
\column{15E}{@{}l@{}}%
\column{18}{@{}>{\hspre}l<{\hspost}@{}}%
\column{39}{@{}>{\hspre}l<{\hspost}@{}}%
\column{E}{@{}>{\hspre}l<{\hspost}@{}}%
\>[3]{}\Varid{bi}{}\<[7]%
\>[7]{} \mathop{:} {}\<[7E]%
\>[10]{}(\Varid{t} \mathop{:} \mathbb{N}) \to (\Varid{n} \mathop{:} \mathbb{N}) \to \Conid{PolicySeq}\;\Varid{t}\;\Varid{n}{}\<[E]%
\\
\>[3]{}\Varid{bi}\;\Varid{t}\;{}\<[9]%
\>[9]{}\Conid{Z}{}\<[15]%
\>[15]{}\mathrel{=}{}\<[15E]%
\>[18]{}\Conid{Nil}{}\<[E]%
\\
\>[3]{}\Varid{bi}\;\Varid{t}\;(\Conid{S}\;\Varid{n}){}\<[15]%
\>[15]{}\mathrel{=}{}\<[15E]%
\>[18]{}\mathbf{let}\;\Varid{ps}\mathrel{=}\Varid{bi}\;(\Conid{S}\;\Varid{t})\;\Varid{n}\;{}\<[39]%
\>[39]{}\mathbf{in}\;\Varid{optExt}\;\Varid{ps}\mathbin{::}\Varid{ps}{}\<[E]%
\ColumnHook
\end{hscode}\resethooks
is correct with respect to \ensuremath{\Varid{val}}:
\begin{hscode}\SaveRestoreHook
\column{B}{@{}>{\hspre}l<{\hspost}@{}}%
\column{3}{@{}>{\hspre}l<{\hspost}@{}}%
\column{13}{@{}>{\hspre}c<{\hspost}@{}}%
\column{13E}{@{}l@{}}%
\column{16}{@{}>{\hspre}l<{\hspost}@{}}%
\column{E}{@{}>{\hspre}l<{\hspost}@{}}%
\>[3]{}\Varid{biOptVal}{}\<[13]%
\>[13]{} \mathop{:} {}\<[13E]%
\>[16]{}(\Varid{t} \mathop{:} \mathbb{N}) \to (\Varid{n} \mathop{:} \mathbb{N}) \to \Conid{OptPolicySeq}\;(\Varid{bi}\;\Varid{t}\;\Varid{n}){}\<[E]%
\ColumnHook
\end{hscode}\resethooks
\begin{hscode}\SaveRestoreHook
\column{B}{@{}>{\hspre}l<{\hspost}@{}}%
\column{3}{@{}>{\hspre}l<{\hspost}@{}}%
\column{4}{@{}>{\hspre}l<{\hspost}@{}}%
\column{13}{@{}>{\hspre}c<{\hspost}@{}}%
\column{13E}{@{}l@{}}%
\column{15}{@{}>{\hspre}l<{\hspost}@{}}%
\column{16}{@{}>{\hspre}l<{\hspost}@{}}%
\column{21}{@{}>{\hspre}c<{\hspost}@{}}%
\column{21E}{@{}l@{}}%
\column{24}{@{}>{\hspre}l<{\hspost}@{}}%
\column{34}{@{}>{\hspre}l<{\hspost}@{}}%
\column{E}{@{}>{\hspre}l<{\hspost}@{}}%
\>[3]{}\Varid{biOptVal}\;\Varid{t}\;{}\<[15]%
\>[15]{}\Conid{Z}{}\<[21]%
\>[21]{}\mathrel{=}{}\<[21E]%
\>[24]{}\Varid{nilOptPolicySeq}{}\<[E]%
\\
\>[3]{}\Varid{biOptVal}\;\Varid{t}\;(\Conid{S}\;\Varid{n}){}\<[21]%
\>[21]{}\mathrel{=}{}\<[21E]%
\\[\blanklineskip]%
\>[3]{}\hsindent{1}{}\<[4]%
\>[4]{}\mathbf{let}\;\Varid{ps}{}\<[13]%
\>[13]{}\mathrel{=}{}\<[13E]%
\>[16]{}\Varid{bi}\;(\Conid{S}\;\Varid{t})\;\Varid{n}\;{}\<[34]%
\>[34]{}\mathbf{in}{}\<[E]%
\\
\>[3]{}\hsindent{1}{}\<[4]%
\>[4]{}\mathbf{let}\;\Varid{ops}{}\<[13]%
\>[13]{}\mathrel{=}{}\<[13E]%
\>[16]{}\Varid{biOptVal}\;(\Conid{S}\;\Varid{t})\;\Varid{n}\;{}\<[34]%
\>[34]{}\mathbf{in}{}\<[E]%
\\
\>[3]{}\hsindent{1}{}\<[4]%
\>[4]{}\mathbf{let}\;\Varid{p}{}\<[13]%
\>[13]{}\mathrel{=}{}\<[13E]%
\>[16]{}\Varid{optExt}\;\Varid{ps}\;{}\<[34]%
\>[34]{}\mathbf{in}{}\<[E]%
\\
\>[3]{}\hsindent{1}{}\<[4]%
\>[4]{}\mathbf{let}\;\Varid{oep}{}\<[13]%
\>[13]{}\mathrel{=}{}\<[13E]%
\>[16]{}\Varid{optExtSpec}\;\Varid{ps}\;{}\<[34]%
\>[34]{}\mathbf{in}{}\<[E]%
\\[\blanklineskip]%
\>[3]{}\hsindent{1}{}\<[4]%
\>[4]{}\Conid{Bellman}\;\Varid{ps}\;\Varid{ops}\;\Varid{p}\;\Varid{oep}{}\<[E]%
\ColumnHook
\end{hscode}\resethooks


\section{Optimal extension}
\label{appendix:optimal_extension}


The generic implementation of backward induction \ensuremath{\Varid{bi}}
naturally raises the question under which conditions one can
implement \ensuremath{\Varid{optExt}} such that \ensuremath{\Varid{optExtSpec}} holds.\\
To this end, consider the function
\begin{hscode}\SaveRestoreHook
\column{B}{@{}>{\hspre}l<{\hspost}@{}}%
\column{3}{@{}>{\hspre}l<{\hspost}@{}}%
\column{9}{@{}>{\hspre}c<{\hspost}@{}}%
\column{9E}{@{}l@{}}%
\column{12}{@{}>{\hspre}l<{\hspost}@{}}%
\column{20}{@{}>{\hspre}c<{\hspost}@{}}%
\column{20E}{@{}l@{}}%
\column{23}{@{}>{\hspre}l<{\hspost}@{}}%
\column{E}{@{}>{\hspre}l<{\hspost}@{}}%
\>[3]{}\Varid{cval}{}\<[9]%
\>[9]{} \mathop{:} {}\<[9E]%
\>[12]{}\{\mskip1.5mu \Varid{t},\Varid{n} \mathop{:} \mathbb{N}\mskip1.5mu\} \to \Conid{PolicySeq}\;(\Conid{S}\;\Varid{t})\;\Varid{n} \to (\Varid{x} \mathop{:} \Conid{X}\;\Varid{t}) \to \Conid{Y}\;\Varid{t}\;\Varid{x} \to \Conid{Val}{}\<[E]%
\\
\>[3]{}\Varid{cval}\;\{\mskip1.5mu \Varid{t}\mskip1.5mu\}\;\Varid{ps}\;\Varid{x}\;\Varid{y}{}\<[20]%
\>[20]{}\mathrel{=}{}\<[20E]%
\>[23]{}\mathbf{let}\;\Varid{mx'}\mathrel{=}\Varid{next}\;\Varid{t}\;\Varid{x}\;\Varid{y}\;\mathbf{in}{}\<[E]%
\\
\>[23]{}\Varid{meas}\;(\Varid{map}\;(\Varid{reward}\;\Varid{t}\;\Varid{x}\;\Varid{y} \mathbin{\medoplus} \Varid{val}\;\Varid{ps})\;\Varid{mx'}){}\<[E]%
\ColumnHook
\end{hscode}\resethooks
By definition of \ensuremath{\Varid{val}} and \ensuremath{\Varid{cval}}, one has
\begin{hscode}\SaveRestoreHook
\column{B}{@{}>{\hspre}l<{\hspost}@{}}%
\column{4}{@{}>{\hspre}l<{\hspost}@{}}%
\column{6}{@{}>{\hspre}c<{\hspost}@{}}%
\column{6E}{@{}l@{}}%
\column{E}{@{}>{\hspre}l<{\hspost}@{}}%
\>[4]{}\Varid{val}\;(\Varid{p}\mathbin{::}\Varid{ps})\;\Varid{x}{}\<[E]%
\\
\>[4]{}\hsindent{2}{}\<[6]%
\>[6]{}\mathrel{=}{}\<[6E]%
\\
\>[4]{}\Varid{meas}\;(\Varid{map}\;(\Varid{reward}\;\Varid{t}\;\Varid{x}\;(\Varid{p}\;\Varid{x}) \mathbin{\medoplus} \Varid{val}\;\Varid{ps})\;(\Varid{next}\;\Varid{t}\;\Varid{x}\;(\Varid{p}\;\Varid{x}))){}\<[E]%
\\
\>[4]{}\hsindent{2}{}\<[6]%
\>[6]{}\mathrel{=}{}\<[6E]%
\\
\>[4]{}\Varid{cval}\;\Varid{ps}\;\Varid{x}\;(\Varid{p}\;\Varid{x}){}\<[E]%
\ColumnHook
\end{hscode}\resethooks
This suggests that, if we can maximise \ensuremath{\Varid{cval}}, i.e.\ implement
\begin{hscode}\SaveRestoreHook
\column{B}{@{}>{\hspre}l<{\hspost}@{}}%
\column{3}{@{}>{\hspre}l<{\hspost}@{}}%
\column{15}{@{}>{\hspre}c<{\hspost}@{}}%
\column{15E}{@{}l@{}}%
\column{18}{@{}>{\hspre}l<{\hspost}@{}}%
\column{E}{@{}>{\hspre}l<{\hspost}@{}}%
\>[3]{}\Varid{cvalmax}{}\<[15]%
\>[15]{} \mathop{:} {}\<[15E]%
\>[18]{}\{\mskip1.5mu \Varid{t},\Varid{n} \mathop{:} \mathbb{N}\mskip1.5mu\} \to \Conid{PolicySeq}\;(\Conid{S}\;\Varid{t})\;\Varid{n} \to (\Varid{x} \mathop{:} \Conid{X}\;\Varid{t}) \to \Conid{Val}{}\<[E]%
\\[\blanklineskip]%
\>[3]{}\Varid{cvalargmax}{}\<[15]%
\>[15]{} \mathop{:} {}\<[15E]%
\>[18]{}\{\mskip1.5mu \Varid{t},\Varid{n} \mathop{:} \mathbb{N}\mskip1.5mu\} \to \Conid{PolicySeq}\;(\Conid{S}\;\Varid{t})\;\Varid{n} \to (\Varid{x} \mathop{:} \Conid{X}\;\Varid{t}) \to \Conid{Y}\;\Varid{t}\;\Varid{x}{}\<[E]%
\ColumnHook
\end{hscode}\resethooks
that fulfil
\begin{hscode}\SaveRestoreHook
\column{B}{@{}>{\hspre}l<{\hspost}@{}}%
\column{3}{@{}>{\hspre}l<{\hspost}@{}}%
\column{16}{@{}>{\hspre}c<{\hspost}@{}}%
\column{16E}{@{}l@{}}%
\column{19}{@{}>{\hspre}l<{\hspost}@{}}%
\column{22}{@{}>{\hspre}l<{\hspost}@{}}%
\column{70}{@{}>{\hspre}l<{\hspost}@{}}%
\column{E}{@{}>{\hspre}l<{\hspost}@{}}%
\>[3]{}\Varid{cvalmaxSpec}{}\<[16]%
\>[16]{} \mathop{:} {}\<[16E]%
\>[19]{}\{\mskip1.5mu \Varid{t},\Varid{n} \mathop{:} \mathbb{N}\mskip1.5mu\} \to (\Varid{ps} \mathop{:} \Conid{PolicySeq}\;(\Conid{S}\;\Varid{t})\;\Varid{n}) \to (\Varid{x} \mathop{:} \Conid{X}\;\Varid{t}) \to {}\<[E]%
\\
\>[19]{}(\Varid{y} \mathop{:} \Conid{Y}\;\Varid{t}\;\Varid{x}) \to \Varid{cval}\;\Varid{ps}\;\Varid{x}\;\Varid{y} \,\sqsubseteq\, \Varid{cvalmax}\;\Varid{ps}\;\Varid{x}{}\<[E]%
\\[\blanklineskip]%
\>[3]{}\Varid{cvalargmaxSpec}{}\<[19]%
\>[19]{} \mathop{:} {}\<[22]%
\>[22]{}\{\mskip1.5mu \Varid{t},\Varid{n} \mathop{:} \mathbb{N}\mskip1.5mu\} \to (\Varid{ps} \mathop{:} \Conid{PolicySeq}\;(\Conid{S}\;\Varid{t})\;\Varid{n}) \to (\Varid{x}{}\<[70]%
\>[70]{} \mathop{:} \Conid{X}\;\Varid{t}) \to {}\<[E]%
\\
\>[22]{}\Varid{cvalmax}\;\Varid{ps}\;\Varid{x}\mathrel{=}\Varid{cval}\;\Varid{ps}\;\Varid{x}\;(\Varid{cvalargmax}\;\Varid{ps}\;\Varid{x}){}\<[E]%
\ColumnHook
\end{hscode}\resethooks
then we can implement optimal extensions of arbitrary policy
sequences. As it turns out, this intuition is correct. With
\begin{hscode}\SaveRestoreHook
\column{B}{@{}>{\hspre}l<{\hspost}@{}}%
\column{3}{@{}>{\hspre}l<{\hspost}@{}}%
\column{E}{@{}>{\hspre}l<{\hspost}@{}}%
\>[3]{}\Varid{optExt}\mathrel{=}\Varid{cvalargmax}{}\<[E]%
\ColumnHook
\end{hscode}\resethooks
one has
\begin{hscode}\SaveRestoreHook
\column{B}{@{}>{\hspre}l<{\hspost}@{}}%
\column{3}{@{}>{\hspre}l<{\hspost}@{}}%
\column{6}{@{}>{\hspre}l<{\hspost}@{}}%
\column{11}{@{}>{\hspre}l<{\hspost}@{}}%
\column{17}{@{}>{\hspre}c<{\hspost}@{}}%
\column{17E}{@{}l@{}}%
\column{20}{@{}>{\hspre}l<{\hspost}@{}}%
\column{42}{@{}>{\hspre}l<{\hspost}@{}}%
\column{87}{@{}>{\hspre}l<{\hspost}@{}}%
\column{E}{@{}>{\hspre}l<{\hspost}@{}}%
\>[3]{}\Varid{optExtSpec}\;\{\mskip1.5mu \Varid{t}\mskip1.5mu\}\;\{\mskip1.5mu \Varid{n}\mskip1.5mu\}\;\Varid{ps}\;\Varid{p'}\;\Varid{x}\mathrel{=}{}\<[E]%
\\
\>[3]{}\hsindent{3}{}\<[6]%
\>[6]{}\mathbf{let}\;{}\<[11]%
\>[11]{}\Varid{p}{}\<[17]%
\>[17]{}\mathrel{=}{}\<[17E]%
\>[20]{}\Varid{optExt}\;\Varid{ps}\;{}\<[42]%
\>[42]{}\mathbf{in}{}\<[E]%
\\
\>[3]{}\hsindent{3}{}\<[6]%
\>[6]{}\mathbf{let}\;{}\<[11]%
\>[11]{}\Varid{y}{}\<[17]%
\>[17]{}\mathrel{=}{}\<[17E]%
\>[20]{}\Varid{p}\;\Varid{x}\;{}\<[42]%
\>[42]{}\mathbf{in}{}\<[E]%
\\
\>[3]{}\hsindent{3}{}\<[6]%
\>[6]{}\mathbf{let}\;{}\<[11]%
\>[11]{}\Varid{y'}{}\<[17]%
\>[17]{}\mathrel{=}{}\<[17E]%
\>[20]{}\Varid{p'}\;\Varid{x}\;{}\<[42]%
\>[42]{}\mathbf{in}{}\<[E]%
\\
\>[3]{}\hsindent{3}{}\<[6]%
\>[6]{}\mathbf{let}\;{}\<[11]%
\>[11]{}\Varid{s1}{}\<[17]%
\>[17]{}\mathrel{=}{}\<[17E]%
\>[20]{}\Varid{cvalmaxSpec}\;\Varid{ps}\;\Varid{x}\;\Varid{y'}\;{}\<[42]%
\>[42]{}\mathbf{in}{}\<[E]%
\\
\>[3]{}\hsindent{3}{}\<[6]%
\>[6]{}\mathbf{let}\;{}\<[11]%
\>[11]{}\Varid{s2}{}\<[17]%
\>[17]{}\mathrel{=}{}\<[17E]%
\>[20]{}\Varid{replace}\;\{\mskip1.5mu \Conid{P}\mathrel{=}\lambda \Varid{z}\Rightarrow (\Varid{cval}\;\Varid{ps}\;\Varid{x}\;\Varid{y'} \,\sqsubseteq\, \Varid{z})\mskip1.5mu\}\;(\Varid{cvalargmaxSpec}\;\Varid{ps}\;\Varid{x})\;\Varid{s1}\;{}\<[87]%
\>[87]{}\mathbf{in}{}\<[E]%
\\
\>[3]{}\hsindent{3}{}\<[6]%
\>[6]{}\Varid{s2}{}\<[E]%
\ColumnHook
\end{hscode}\resethooks
The observation that functions \ensuremath{\Varid{cvalmax}} and \ensuremath{\Varid{cvalargmax}}
that fulfil \ensuremath{\Varid{cvalmaxSpec}} and \ensuremath{\Varid{cvalargmaxSpec}}
are sufficient to implement an optimal extension \ensuremath{\Varid{optExt}} that fulfils
\ensuremath{\Varid{optExtSpec}} naturally raises the question of what are necessary and
sufficient conditions for \ensuremath{\Varid{cvalmax}} and \ensuremath{\Varid{cvalargmax}}.
Answering this question necessarily requires discussing properties of
\ensuremath{\Varid{cval}} and goes well beyond the scope of formulating a theory of SDPs.
Here, we limit ourselves to remark that if \ensuremath{\Conid{Y}\;\Varid{t}\;\Varid{x}} is finite and
non-empty one can implement the functions \ensuremath{\Varid{cvalmax}} and \ensuremath{\Varid{cvalargmax}} by
linear search.
A generic implementation of \ensuremath{\Varid{cvalmax}} and \ensuremath{\Varid{cvalargmax}} can be found under
\citep{IdrisLibsValVal}.

For the original BJI-theory, tabulated backward induction and several example
applications can be found in the \ensuremath{\Conid{SequentialDecisionProblems}} folder
of \citep{botta20162018}.


\label{lastpage01}

\end{document}

%% file: macros.TeX
\theoremstyle{definition}
\newtheorem{exm}{Example}

\newcommand{\medoplus}{\ensuremath{\scalebox{0.9}{$\bigoplus$}}}

\def\commentbegin{\quad\{\ }
\def\commentend{\}}

\newcommand{\bottaetal}{Botta \emph{et al.\ }}

%% file: PIKPalette.tex
\definecolor{PIKorange} {RGB}{243,90,40}
\definecolor{PIKorangeLight1}{RGB}{245,119,68}
\definecolor{PIKorangeLight2}{RGB}{248,150,103}
\definecolor{PIKorangeLight3}{RGB}{250,183,146}
\definecolor{PIKorangeDark1}{RGB}{222,82,36}
\definecolor{PIKorangeDark2}{RGB}{201,74,32}
\definecolor{PIKorangeDark3}{RGB}{180,66,28}

\definecolor{PIKgray}        {RGB}{104,108,112}
\definecolor{PIKgrayLight1} {RGB}{131,136,139}
\definecolor{PIKgrayLight2} {RGB}{160,164,167}
\definecolor{PIKgrayLight3} {RGB}{190,193,195}
\definecolor{PIKgrayDark1} {RGB}{85,88,91}
\definecolor{PIKgrayDark2} {RGB}{67,67,70}
\definecolor{PIKgrayDark3} {RGB}{48,47,50}

\definecolor{PIKcyan} {RGB}{0,173,239}
\definecolor{PIKcyanLight1} {RGB}{51,190,243}
\definecolor{PIKcyanLight2} {RGB}{102,206,246}
\definecolor{PIKcyanLight3} {RGB}{153,223,249}
\definecolor{PIKcyanDark1} {RGB}{0,141,199}
\definecolor{PIKcyanDark2} {RGB}{0,110,158}
\definecolor{PIKcyanDark3} {RGB}{0,78,117}

\definecolor{PIKgreen} {RGB}{129,190,99}
\definecolor{PIKgreenLight1} {RGB}{154,202,124}
\definecolor{PIKgreenLight2} {RGB}{179,214,152}
\definecolor{PIKgreenLight3} {RGB}{204,227,183}
\definecolor{PIKgreenDark1} {RGB}{106,156,81}
\definecolor{PIKgreenDark2} {RGB}{83,122,63}
\definecolor{PIKgreenDark3} {RGB}{61,88,45}

%% file: main.bbl
\begin{thebibliography}{}

\bibitem[\protect\citename{Audebaud \& Paulin{-}Mohring,
  }2009]{DBLP:journals/scp/AudebaudP09}
Audebaud, P. and Paulin{-}Mohring, C. (2009)
\newblock Proofs of randomized algorithms in {C}oq.
\newblock {\em Sci. Comput. Program.} {\bf 74}(8):568--589.

\bibitem[\protect\citename{Bellman, }1957]{bellman1957}
Bellman, R. (1957)
\newblock {\em Dynamic Programming}.
\newblock Princeton University Press.

\bibitem[\protect\citename{Bertsekas \& Shreve, }1996]{bertsekasShreve96}
Bertsekas, D.~P. and Shreve, S.~E. (1996)
\newblock {\em Stochastic Optimal Control: The Discrete Time Case}.
\newblock Athena Scientific.

\bibitem[\protect\citename{Bertsekas {\em et~al.}\relax,
  }2003]{bertsekas2003convex}
Bertsekas, D., Nedi{\'c}, A. and Ozdaglar, A. (2003)
\newblock {\em Convex Analysis and Optimization}.
\newblock Athena Scientific optimization and computation series.
\newblock Athena Scientific.

\bibitem[\protect\citename{Bertsekas, }1995]{bertsekas1995}
Bertsekas, D., P. (1995)
\newblock {\em Dynamic Programming and Optimal Control}.
\newblock Athena Scientific.

\bibitem[\protect\citename{Bird, }2014]{bird2014thinking}
Bird, R. (2014)
\newblock {\em Thinking {F}unctionally with {H}askell}.
\newblock Cambridge University Press.

\bibitem[\protect\citename{Bird \& Gibbons, }2020]{adwh}
Bird, R. and Gibbons, J. (2020)
\newblock {\em Algorithm {D}esign with {H}askell}.
\newblock Cambridge University Press.

\bibitem[\protect\citename{Botta {\em et~al.}\relax, }2013]{botta+al2013b}
Botta, N., Mandel, A., Hofmann, M., Schupp, S. and Ionescu, C. (2013)
\newblock Mathematical specification of an agent-based model of exchange.
\newblock  {\em Proceedings of the {AISB} Convention 2013, ``{Do-Form}:
  Enabling Domain Experts to use Formalized Reasoning'' Symposium}.

\bibitem[\protect\citename{Botta {\em et~al.}\relax, }2018]{esd-9-525-2018}
Botta, N., Jansson, P. and Ionescu, C. (2018)
\newblock The impact of uncertainty on optimal emission policies.
\newblock {\em Earth System Dynamics} {\bf 9}(2):525--542.

\bibitem[\protect\citename{Botta, }2016\mbox{--}2021]{botta20162018}
Botta, N. (2016\mbox{--}2021)
\newblock {\em {I}dris{L}ibs}.
\newblock \url{https://gitlab.pik-potsdam.de/botta/IdrisLibs}.

\bibitem[\protect\citename{Botta {\em et~al.}\relax,
  }n.d.]{botta2020extensional}
Botta, N., Brede, N., Jansson, P. and Richter, T.
\newblock (in press) {E}xtensional equality preservation and verified generic
  programming.
\newblock {\em J. Funct. Program.}
\newblock (Accepted for publication August 2021).
  \url{https://arxiv.org/abs/2008.02123}.

\bibitem[\protect\citename{Botta {\em et~al.}\relax,
  }2017a]{2017_Botta_Jansson_Ionescu}
Botta, N., Jansson, P. and Ionescu, C. (2017a)
\newblock Contributions to a computational theory of policy advice and
  avoidability.
\newblock {\em J. Funct. Program.} {\bf 27}:e23.

\bibitem[\protect\citename{Botta {\em et~al.}\relax, }2017b]{2014_Botta_et_al}
Botta, N., Jansson, P., Ionescu, C., Christiansen, D.~R. and Brady, E. (2017b)
\newblock Sequential decision problems, dependent types and generic solutions.
\newblock {\em Logical Methods in Computer Science} {\bf 13}(1).

\bibitem[\protect\citename{Brady, }2013]{JFP:9060502}
Brady, E. (2013)
\newblock Idris, a general-purpose dependently typed programming language:
  Design and implementation.
\newblock {\em J. Funct. Program.} {\bf 23}(9):552--593.

\bibitem[\protect\citename{Brady, }2017]{idrisbook}
Brady, E. (2017)
\newblock {\em Type-{D}riven {D}evelopment in {I}dris}.
\newblock Manning Publications Co.

\bibitem[\protect\citename{Brede \& Botta, }2021]{IdrisLibsValVal}
Brede, N. and Botta, N. (2021)
\newblock {\em On the Correctness of Monadic Backward Induction}.
\newblock
  \href{https://gitlab.pik-potsdam.de/botta/papers/-/tree/master/2021.On\%20the\%20Correctness\%20of\%20Monadic\%20Backward\%20Induction}{Git
  repository}.

\bibitem[\protect\citename{{De Moor}, }1995]{de_moor1995}
{De Moor}, O. (1995)
\newblock A generic program for sequential decision processes.
\newblock  {\em {PLILPS} '95 Proceedings of the 7th International Symposium on
  Programming Languages: Implementations, Logics and Programs} pp.  1--23.
\newblock Springer.

\bibitem[\protect\citename{{De Moor}, }1999]{de_moor1999}
{De Moor}, O. (1999)
\newblock Dynamic programming as a software component.
\newblock {\em Proc. 3rd WSEAS Int. Conf. Circuits, Systems, Communications and
  Computers ({CSCC} 1999)}  4--8.

\bibitem[\protect\citename{Diederich, }2001]{diederich01}
Diederich, A. (2001)
\newblock Sequential decision making.
\newblock  Smelser, N.~J. and Baltes, P.~B. (eds), {\em International
  Encyclopedia of the Social \& Behavioral Sciences}, pp.  13917--13922.
\newblock Pergamon.

\bibitem[\protect\citename{Erwig \& Kollmansberger,
  }2006]{DBLP:journals/jfp/ErwigK06}
Erwig, M. and Kollmansberger, S. (2006)
\newblock Functional {P}earls: {P}robabilistic functional programming in
  {H}askell.
\newblock {\em J. Funct. Program.} {\bf 16}(1):21--34.

\bibitem[\protect\citename{Finus {\em et~al.}\relax, }2003]{finus+al2003}
Finus, M., van Ierland, E. and Dellink, R. (2003)
\newblock {\em Stability of Climate Coalitions in a Cartel Formation Game}.
\newblock FEEM Working Paper No. 61.2003.

\bibitem[\protect\citename{Gintis, }2007]{gintis2007}
Gintis, H. (2007)
\newblock The dynamics of general equilibrium.
\newblock {\em Economic Journal} {\bf 117}:1280--1309.

\bibitem[\protect\citename{Giry, }1981]{giry1981}
Giry, M. (1981)
\newblock A categorial approach to probability theory.
\newblock  Banaschewski, B. (ed), {\em Categorical Aspects of Topology and
  Analysis}.
\newblock Lecture Notes in Mathematics 915, pp.  68--85.
\newblock Springer.

\bibitem[\protect\citename{Heitzig, }2012]{heitzig2012}
Heitzig, J. (2012)
\newblock {\em Bottom-Up Strategic Linking of Carbon Markets: Which Climate
  Coalitions Would Farsighted Players Form?}
\newblock SSRN Environmental Economics eJournal.

\bibitem[\protect\citename{Helm, }2003]{helm2003}
Helm, C. (2003)
\newblock International emissions trading with endogenous allowance choices.
\newblock {\em Journal of Public Economics} {\bf 87}:2737--2747.

\bibitem[\protect\citename{Ionescu, }2009]{ionescu2009}
Ionescu, C. (2009)
\newblock {\em Vulnerability Modelling and Monadic Dynamical Systems}.
\newblock PhD thesis, Freie Universit{\"a}t Berlin.

\bibitem[\protect\citename{Jacobs, }2011]{DBLP:journals/tcs/Jacobs11}
Jacobs, B. (2011)
\newblock Probabilities, distribution monads, and convex categories.
\newblock {\em Theor. Comput. Sci.} {\bf 412}(28):3323--3336.

\bibitem[\protect\citename{MacLane, }1978]{maclane}
MacLane, S. (1978)
\newblock {\em Categories for the {W}orking {M}athematician}. 2nd edn.
\newblock Graduate Texts in Mathematics.
\newblock Springer.

\bibitem[\protect\citename{Mercure {\em et~al.}\relax, }2020]{mercure2020risk}
Mercure, J.-F., Sharpe, S., Vinuales, J., Ives, M., Grubb, M., Pollitt, H.,
  Knobloch, F. and Nijsse, F. (2020)
\newblock Risk-opportunity analysis for transformative policy design and
  appraisal.
\newblock {\em C-EENRG Working Papers} {\bf 2020-4}:1--40.

\bibitem[\protect\citename{Puterman, }2014]{puterman2014markov}
Puterman, M.~L. (2014)
\newblock {\em Markov {D}ecision {P}rocesses: {D}iscrete {S}tochastic {D}ynamic
  {P}rogramming}.
\newblock John Wiley \& Sons.

\bibitem[\protect\citename{TiPES, }2019\mbox{--}2023]{TiPES::Website}
TiPES. (2019\mbox{--}2023)
\newblock {\em {T}i{PES} {H}2020 {P}roject {W}ebsite}.
\newblock \url{https://www.tipes.dk/}.

\bibitem[\protect\citename{Wadler, }2015]{DBLP:journals/cacm/Wadler15}
Wadler, P. (2015)
\newblock Propositions as types.
\newblock {\em Commun. {ACM}} {\bf 58}(12):75--84.

\end{thebibliography}
